\documentclass[letterpaper,superscriptaddress,twocolumn,accepted=2020-06-25]{quantumarticle}
\pdfoutput=1

\usepackage{float}
\usepackage{amsthm}
\usepackage{amssymb,epic,eepic}
\usepackage{framed2006}
\usepackage{xcolor}
\usepackage{colortbl}
\usepackage{bm}

\usepackage[citebordercolor=red,urlbordercolor=red,linkbordercolor=blue]{hyperref}

\hypersetup{colorlinks=false,linkbordercolor=blue,citebordercolor=blue,linkcolor=blue,pdfborderstyle={/S/U/W 1}}

\usepackage{graphicx}
\usepackage{fancybox,amssymb,color}

\definecolor{mblue}{rgb}{0,0,0.7}

\hypersetup{
   colorlinks=true,
   linkcolor=mblue,
   citecolor=mblue,
   urlcolor=mblue
}

\makeatletter
\renewcommand*\l@section{\@dottedtocline{1}{1.5em} {2.3em}}
\renewcommand*\l@subsection{\@dottedtocline{2}{3.8 em}{3.2em}}
\renewcommand*\l@subsubsection{\@dottedtocline{3}{ 7.0em}{4.1em}}
\renewcommand*\l@paragraph{\@dottedtocline{4}{10em }{5em}}
\renewcommand*\l@subparagraph{\@dottedtocline{5}{1 2em}{6em}}
\renewcommand*\l@figure{\@dottedtocline{1}{1.5em}{ 2.3em}}
\makeatother

\theoremstyle{definition}

\newcommand{\s}{{\mathcal{S}}}

\newcommand{\R}{{\mathbb R}}
\newcommand{\N}{{\mathbb N}}

\newcommand{\Km}{{\rm Km}}
\newcommand{\p}{{\mathbf{P}}}

\newtheorem{theorem}{Theorem}[section]
\newtheorem{lemma}[theorem]{Lemma}
\newtheorem{corollary}[theorem]{Corollary}
\newtheorem{example}[theorem]{Example}
\newtheorem{definition}[theorem]{Definition}

\newtheorem{fixing}[theorem]{Informal definition}

\newtheorem{thoughtexp}[theorem]{Thought experiment}
\newtheorem{observation}[theorem]{Observation}
\newtheorem{assumption}[theorem]{Assumption}

\newcommand{\lineclear}{{\rule{0pt}{0pt}\nopagebreak\par\nopagebreak\noindent}}

\newcommand{\sdefinition}[3]{\noindent\enspace\begin{shaded}$\strut$\vskip -3em$\strut$\begin{definition}[#1]\label{#2}#3\end{definition}\vskip -0.7em\end{shaded}}

\newcommand{\stheorem}[3]{\noindent\enspace\begin{shaded}$\strut$\vskip -3em$\strut$\begin{theorem}[#1]\label{#2}#3\end{theorem}\vskip -0.7em\end{shaded}}

\newcommand{\slemma}[3]{\noindent\enspace\begin{shaded}$\strut$\vskip -3em$\strut$\begin{lemma}[#1]\label{#2}#3\end{lemma}\vskip -0.7em\end{shaded}}
\newcommand{\emptyslemma}[2]{\noindent\enspace\begin{shaded}$\strut$\vskip -3em$\strut$\begin{lemma}\label{#1}#2\end{lemma}\vskip -0.7em\end{shaded}}
\newcommand{\sexample}[3]{\noindent\enspace\begin{shaded}$\strut$\vskip -3em$\strut$\begin{example}[#1]\label{#2}#3\end{example}\vskip -0.7em\end{shaded}}
\newcommand{\sobservation}[3]{\noindent\enspace\begin{shaded}$\strut$\vskip -3em$\strut$\begin{observation}[#1]\label{#2}#3\end{observation}\vskip -0.7em\end{shaded}}
\newcommand{\emptysassumption}[2]{\noindent\enspace\begin{shaded}$\strut$\vskip -3em$\strut$\begin{assumption}\label{#1}#2\end{assumption}\vskip -0.7em\end{shaded}}

\definecolor{xblue}{rgb}{.9,.9,.9}
\definecolor{shadecolor}{named}{xblue}
\definecolor{xred}{rgb}{1,.95,.9}
\definecolor{xgreen}{rgb}{1,1,.88}
\newtheorem{postulating}[theorem]{Postulates}
\newcommand{\postulates}[3]{\noindent\enspace\begin{shaded}\begin{postulating}[#1]\label{#2}#3\end{postulating}\end{shaded}}

\newcommand\ci{\mathbin{\perp\kern-9.4pt\perp\,}}

\begin{document}
\title{Law without law: from observer states to physics via\newline algorithmic information theory}
\author{Markus P. M\"uller}
\affiliation{Institute for Quantum Optics and Quantum Information, Austrian Academy of Sciences, Boltzmanngasse 3, A-1090 Vienna, Austria}
\affiliation{Perimeter Institute for Theoretical Physics, Waterloo, ON N2L 2Y5, Canada}

\begin{abstract}
According to our current conception of physics, any valid physical theory is supposed to describe the objective evolution of a unique external world. However, this condition is challenged by quantum theory, which suggests that physical systems should not always be understood as having objective properties which are simply revealed by measurement. Furthermore, as argued below, several other conceptual puzzles in the foundations of physics and related fields point to limitations of our current perspective and motivate the exploration of an alternative: to start with the first-person (the observer) rather than the third-person perspective (the world).

In this work, I propose a rigorous approach of this kind on the basis of algorithmic information theory. It is based on a single postulate: that \emph{universal induction} determines the chances of what any observer sees next. That is, instead of a world or physical laws, it is the local state of the observer alone that determines those probabilities. Surprisingly, despite its solipsistic foundation, I show that the resulting theory recovers many features of our established physical worldview: it predicts that it appears to observers \emph{as if there was an external world} that evolves according to simple, computable, probabilistic laws. In contrast to the standard view, objective reality is not assumed on this approach but rather provably emerges as an asymptotic statistical phenomenon. The resulting theory dissolves puzzles like cosmology's Boltzmann brain problem, makes concrete predictions for thought experiments like the computer simulation of agents, and suggests novel phenomena such as ``probabilistic zombies'' governed by observer-dependent probabilistic chances. It also suggests that some basic phenomena of quantum theory (Bell inequality violation and no-signalling) might be understood as consequences of this framework.
\end{abstract}

\maketitle

\renewcommand \thesection{\arabic{section}}
\renewcommand{\thesubsection}{\arabic{section}.\arabic{subsection}}
\makeatletter
\def\p@subsection{}\makeatother

{
\hypersetup{linkcolor=black}
\onecolumngrid
\begin{center}
\begin{minipage}[t]{0.7\textwidth}
\setcounter{tocdepth}{2}
\begin{shaded}
\footnotesize
\begin{center}
$\strut$\vskip -3em$\strut$
\tableofcontents
\end{center}
\end{shaded}
\end{minipage}
\end{center}
}

\pagebreak

\newgeometry{margin=0.55in, voffset=0.2in, bottom=1in}
\fancyheadoffset{0pt}
\fancyfootoffset{0pt}
\twocolumngrid

\renewcommand{\sectionmark}[1]{\markboth{#1}{}}
\renewcommand{\headrule}{\vbox to 0pt{\hbox to\headwidth{\color{quantumgray}\leaders\hrule\hfil}\vss}}
\chead{}
\rhead{\fancyplain{}{\textit{\thesection. \leftmark}}}

\section{Introduction}
\label{SecIntroduction}
\sectionmark{Introduction}
Theoretical physics is more than just a fixed framework that allows us to predict measurable quantities. Ever since the first philosophers have wondered what our universe is made of, the very nature of the questions that we ask in physics has been continuously evolving. Novel discoveries and problems have led to completely new concepts that did not even make sense within earlier theories. For example, the problem of ether and of the Lorentz transformations in electrodynamics have ultimately led us to a framework (relativity) in which the structure of spacetime itself is dynamical, which is an idea that could not even have been formulated within Newtonian mechanics.

The starting point of this work is the hypothesis that we are perhaps at a point where we may want to consider another substantial revision of some traditional aspects of our worldview, at least in certain contexts. We are facing several conceptual problems, some of them of enormous importance, for which systematic problems and difficulties arise when we try to address them with standard approaches. While some of these questions are simply free-floating expressions of human curiosity (like \emph{``Why are there simple laws of physics at all?''}), others have emerged as notorious and persistent problems in physics and related areas. They seem to show us in a rather annoying way that there is something that we fundamentally do not understand (see Table~\ref{fig_motivation} for an overview).

For example, consider some questions that are currently being discussed in the context of cosmology: what if the universe is really large (as in eternal inflation) and contains a multitude of copies of every observer~\cite{AguirreTegmark}? How can we assign probabilities to properties of ``possible worlds''~\cite{Linde}? What if thermal fluctuations produce a massive amount of randomly appearing ``Boltzmann brains''~\cite{Albrecht2002,AlbrechtSorbo2004,Nomura} --- can we use the assumption that we are \emph{not} the result of such fluctuations to constrain our cosmological models? Independently, philosophers are discussing questions related to agents or observers that seem at first sight to be of a very different category, like: What happens if we simulate an intelligent agent on a computer --- would the simulation be ``alive''~\cite{Bostrom}?

Even though these puzzles seem to be of quite different nature at first sight, they do have a common core --- they are all specific instances of the question: \textbf{``What will I see next?''} In the empirical regime, \emph{physics} allows us to answer this question, at least probabilistically. For example, if we send a photon to a half-silvered mirror in the laboratory, then quantum physics tells us that we will see the photon being transmitted (or rather hear a specific detector click) with $50\%$ probability. But we can ask this question also in exotic situations, some of which are listed in Table~\ref{fig_motivation}. For example, if we are promised to be scanned in all detail to a computer, and then to be simulated in one (or even many different) virtual worlds, will we ``wake up'' in a simulation (and, if so, in which one)? In this context, it seems inappropriate to try to predict what happens to us solely on the basis of information about the detailed physical composition of body or computer. Instead, the question now seems to fall into the realm of the philosophy of mind.

Similarly, if we assume the validity of a cosmological model predicting a universe with a large number of Boltzmann brains, does it make sense for \emph{me} to hold a degree of belief on whether \emph{I am actually one of them}? Can we assign a meaningful probability to the possibility that \emph{what I see next} is the strange experience of one of those fluctuating beings, perhaps suddenly realizing that something is very strange before disappearing? Conversely, can we use the empirical fact that \emph{this is not what we see} to rule out some cosmological models? The very existence of controversy among cosmologists regarding these questions tells us that we have no idea how to approach them in a conclusively coherent way.

\noindent\enspace\begin{shaded}$\strut$\vskip -2.5em$\strut$\begin{itemize}\item \textbf{Quantum theory.} ``Unperformed experiments have no results''~\cite{PeresNoResults,FuchsSchack}; measurement problem~\cite{FuchsQFQIT}; no-go results about observer-independent facts~\cite{BruknerNogo,Bong}.
\item \textbf{Cosmology.} Boltzmann brain problem~\cite{Albrecht2002,AlbrechtSorbo2004}; self-locating uncertainty~\cite{AguirreTegmark}; measure problem~\cite{Linde}.
\item \textbf{Philosophy of mind / future technology.} \emph{``Are you living in a computer simulation?''}~\cite{Bostrom}, puzzles of personal identity like \emph{``A Conversation with Einstein's Brain''} in Hofstadter's and Dennett's ``The Mind's I''~\cite{HofstadterDennett}, or Parfit's \emph{teletransportation paradox}~\cite{Parfit}.
\item \textbf{Fundamental curiosity.} Why is there a ``world'' with ``laws of nature'' in the first place?
\end{itemize}
\vskip -0.5em
\end{shaded}
\label{fig_motivation}
{\small TABLE 1. Some enigmas that motivate the approach of this paper. As explained in more detail in the main text, even though these conceptual puzzles are rooted in different fields, they have a common feature: they can all in principle be reformulated in terms of the question of \emph{what is the probability of my future state, given my current state} (including my momentary observations and memory, conscious or not). This motivates the attempt to formulate a framework for which these first-person conditional probabilities are fundamental, and which does not assume that they come from an external world.
}\\

From this perspective, it seems odd that a single unifying question has to be approached with so different methods in the different regimes --- physics, philosophy, or outright speculation. But is this actually a fair comparison? Isn't physics, after all, more concerned with the question of ``What is the world like?'' rather than ``What will I see next?'' Not if we live in a quantum world. Ultimately, the formalism of quantum theory tells us the probabilities of outcomes of experiments, i.e.\ \emph{the chances of what to see next}, given the physical context. In particular, due to results like Bell's theorem~\cite{Bell1964,Bell1966}, it is provably inconsistent to assume that measurements simply reveal preexisting unknown facts of the world, without sacrifizing other important principles of physics like locality. We should not think of the wave function as the ``configuration of the world'' in a naive sense, but rather as a catalogue of expectations about what an agent will see next. Therefore, quantum theory gives us a physical motivation to regard the question in boldface above as more fundamental than the question of what the world is like.

Given that this single question appears in so many instances in different fields --- could there be a single, unified approach or theory that \emph{answers this question in all contexts uniformly}? Such a theory would have an important advantage: while most ad hoc claims about problems like the brain emulation question above do not seem to be directly amenable to empirical testing\footnote{After all, we cannot directly empirically test any predictions of the form ``Yes, if we do a simulation of this or that kind, then the simulated mind \emph{really has} an inner life in the same way that we do''. Simply observing the simulated mind, or asking it, will not allow us to draw any ultimate conclusions; see e.g.\ the philosophical discussion of ``zombies''~\cite{SEP}. Of course we can (and should) study other aspects of this problem empirically, e.g.\ via neuroscience.}, the hypotheses of such a unified approach about these exotic phenomena could be put to an indirect test. Namely, if that theory made in principle successful empirical predictions in the regime of physics, then this would justifiably increase our trust in its predictions in the more speculative regime.

The goal of this work is to provide a proof of principle that we can indeed have a theory\footnote{Note that this is \emph{not} supposed to be a ``theory of everything''; in fact, the theory predicts its own limitations. By construction, it will have to say nothing about most things. As an obvious example, it will not be useful for the search for a theory of quantum gravity.} of this kind --- one that is simple, rigorous, and well-motivated.  We arrive at such a theory quite naturally by following a few well-motivated assumptions. Our first assumption is to committ to the \emph{first-person perspective of observers}\footnote{In line with Rovelli~\cite{Rovelli}, here the word ``observer'' is by no means restricted to \emph{human} observers, and it is not (at least not directly) related to the notion of ``consciousness''. The question of consciousness is irrelevant for this paper; my notion of ``first-person perspective'' is not meant to be equivalent to consciousness. The former (but probably not the latter) describes a very general, technically formalizable notion that is agnostic about the question ``what that perspective really feels like''. As a rough analogy, note that computer science can reason about the information content of a painting (say, after it is digitized and saved on a hard drive) without the need to decide what it is supposed to depict, or whether it is ``beautiful''.} as being fundamental. In more detail, we start with what we call the ``observer state'': a mathematical formalization of the information-theoretic state of the observer, including its current observations and its memory (conscious and unconscious). This will be our primitive notion, and we will drop all assumptions of an ``external world''. A moment's thought shows that such a move is unavoidable if we want to address questions like those mentioned above. For example, if we ask \emph{``why is there a ``world'' with ``laws of nature'' at all?}, then we must have a starting point that does not assume the existence of such a world from the outset. Similarly, if we do not think that detailed insights into physical properties of the world can help us resolve puzzles like Parfit's teletransportation paradox, then we must be able to argue without these ingredients.

Given such a notion of ``observer state'', we can formulate a possible answer to the question of ``what the observer will see next'': namely, we would like to write down some notion of propensity, or chance,
\begin{equation}
   \p(\mbox{next observer state}\,|\,\mbox{current observer state}).
   \label{eqP}
\end{equation}
Our second assumption is that this chance~\cite{Myrvold} always exists, and that there is a mathematical object $\p$ that formalizes it. For the moment, think of $\p$ as a probability distribution; later on, its role will in fact be played by a more general object (a countable set of asymptotically equivalent distributions). Consider the following example. Suppose that $x$ describes the state of an observer who knows that she will now be put to sleep, scanned, and simulated in a computer. Let $y$ be the observer state that she would have at the first moment of computer simulation. Then what we assume here is that there is in fact an ``objective chance'' $\p(y|x)$ that the observer will ``wake up'' in the simulation. Moreover, this notation implies that this chance is \emph{independent of all other ``facts of the world''} --- it really only depends on the state of the observer.

It is important to understand that $\p$ is \emph{not} meant to represent the observer's degree of belief. As a colourful and imprecise example, suppose that $x$ describes the state of a little insect that is crawling across the edge of a table. Then (we think that) there is a large chance $\p(y|x)$ of transitioning into a state that experiences falling, even if the insect is too stupid to hold beliefs (let alone to compute probabilities). Moreover, the observer state should be interpreted as encompassing all information ``contained in'' the observer, not just what the observer is consciously aware of. In this example, $x$ could contain enough information from the insect's nervous system to indicate \emph{in principle} the presence of the table's edge, even if the insect is not aware of it.

Finally, to obtain a complete theory, we have to concretely postulate what $\p$ should be. As mentioned above, $\p$ will be something like (but not quite) a probability distribution. In order to obtain a meaningful, mathematically formalized, objective theory, it should \emph{not} be necessary to determine what it ``feels like'' to be in a particular observer state $x$ in order to determine $\p(y|x)$. Instead, $\p(y|x)$ should only depend on the abstract information content of $x$ and $y$, and not on questions of qualia. As we will explain in Section~\ref{SecAlgorithmicProbability} and motivate in detail in Section~\ref{SecPostulates}, we will here postulate that $\p$ should express some form of ``universal induction'': $\p(y|x)$ is large if an external rational agent with complete knowledge of $x$ would be led to predict $y$. This will lead us to claim that $\p$ is some version of \emph{algorithmic probability}. Such $\p$ is related to description length: the more \emph{compressible} the conceivable future state $y$ (given the current state $x$), the more likely. Thus, in the approach of this paper, answering the brain emulation question above boils down to estimating the algorithmic complexities of the simulated observer states. We study this problem in detail in Subsection~\ref{SubsecBrainEmulation}.

The theory is introduced in two successive steps, distinguished by their color shading:

\begin{tabular}{cl}
\rowcolor{xred} \textbf{I.} & \textbf{Mathematical formulation} \\
\rowcolor{xred} & Sections~\ref{SecAlgorithmicProbability} (algorithmic probability) and \ref{SecPostulates}.\\
\rowcolor{xgreen} \textbf{II.} & \textbf{Predictions of the theory} \\
\rowcolor{xgreen} & Sections~\ref{SecSimpleLaws}--\ref{SecQuantum}.
\end{tabular}
$\strut$\\

Section~\ref{SecAlgorithmicProbability} will introduce the notions of observer states and algorithmic probability. Section~\ref{SecPostulates} spells out the postulates of this paper's approach, and motivates why algorithmic probability is our measure of choice.

The second part reconstructs aspects of physics from the postulates, and uses them to address some of the puzzles of Table~\ref{fig_motivation}. While our methodological starting point is in some sense solipsistic, Section~\ref{SecSimpleLaws} shows how we can nevertheless understand the existence of an external world with simple computable probabilistic laws of physics as a \emph{consequence} of this framework. Furthermore, Section~\ref{SecObjectivity} proves that we also obtain an emergent notion of objective reality. Subsections~\ref{SubsecZombies} and \ref{SubsecImmortality} argue, however, that there are extreme situations in which objective reality breaks down, leading to the phenomena of ``probabilistic zombies'' and ``subjective immortality''. Section~\ref{SecApplication} describes how the Boltzmann brain problem gets dissolved, and what we can say about the computer simulation of agents. Finally, Section~\ref{SecQuantum} argues that some basic phenomena of quantum theory can perhaps be understood as consequences of this paper's approach, before we conclude in Section~\ref{SecConclusions}.

\section{Algorithmic probability}
\label{SecAlgorithmicProbability}
\sectionmark{Algorithmic probability}
\colorlet{shadecolor}{xred}
There are two main notions mentioned in the introduction that we have to discuss in all mathematical details: the \emph{state} of an observer, and the chance $\p$. We will begin by stipulating that observer states shall be modelled by the \emph{finite binary strings}:
\[
   \s=\{\varepsilon,0,1,00,01,10,11,000,\ldots\}.
\]
The \emph{length} of a string $x\in\s$ will be denoted $\ell(x)$; for example $\ell(11)=2$. The symbol $\varepsilon$ denotes the empty string of length zero. We will assume that \emph{every possible observer state corresponds to a binary string; and, vice versa, to every binary string there is a corresponding observer state, i.e.\ ``state of being'' of a conceivable observer}. As explained in the introduction, we should think of an observer state as an exhaustive description of an observer's memory (conscious or unconscious) and momentary observations --- all her ``locally accessible information''. Naively, think of a raw dump of the information-theoretic content of a human brain at some moment in time, scanned up to all functionally relevant detail. Now, this is not an exact interpretation. To what detail exactly are we supposed to scan? Where exactly do we put the boundaries of the brain? In the following, we will see that we do not need to answer these questions to construct our theory and to extract its predictions. Moreover, the \emph{interpretation} of an observer state will gradually become more clear in the course of construction of the theory.

It is important to understand that most observer states are completely unrelated to states of humans or animals. (This is a truism as obvious as stating that ``almost all theoretically possible digital pictures do not show anything that you are familiar with''.) We have to, may, and will ignore questions of qualia like ``what does it feel like to be in state $x$''? Moreover, the actual zeroes and ones in an observer state do not carry any meaning in isolation. This is comparable to, say, the theory of general relativity, where coordinates of spacetime points like $x=(0,0.3,-0.14,1.25)$ do not carry any meaning in themselves, but only \emph{relative to a choice of coordinate system}. While general relativity allows for a mostly arbitrary choice of coordinate system, we will see in Section~\ref{SecPostulates} that we also have a mostly \emph{arbitrary choice of encoding}, and changing the encoding will change the bit string.

Since observer states are discrete, it makes sense that the state of the observer changes in discrete (subjective) time steps. That is, every observer will be in some state now, in another state next, and so on. This leads us to study transition probabilities of the form $\p(y|x)$ (as indicated in~(\ref{eqP})).

What are those probabilities? We would like to postulate a probability measure $\p$ that determines the chances of what observers see. How can we do so, without making arbitrary choices or smuggling known facts of physics into the definition? I will argue in the following that a version of \emph{algorithmic probability} is a good candidate, since it uses only structure that is unavoidably available once we start to reason logically: the computability structure of axiomatic systems. While a more detailed discussion of the motivation will be deferred to Section~\ref{SecPostulates}, this subsection will now appeal to intuition, and at the same time derive and present a definition of algorithmic probability.

Let us step back for a moment and recall some basic ideas from probability theory. When students start to learn stochastics at school, often the first example they discuss is that of an ``urn'', containing balls that have different colors, and of some experimenter drawing one of the balls at random. In our case, the differently colored balls correspond to the observer states, i.e.\ the finite binary strings. Also, bit strings are purely mathematical objects, so in some sense, mathematics itself represents the analog of the urn, or, say, the formal axiomatic system that is used to define the notion of a ``binary string''. But what corresponds to the act of ``drawing'' such a mathematical object at random?

Intuitively, the concrete mechanism of drawing determines the resulting probability distribution. If the urn contains two red balls and a green ball, say, then the chance of drawing the green ball will only be close to $\frac 1 3$ if the experimenter has equal access to all the balls (for example, none of the balls lies at the bottom of the urn and cannot be reached by the experimenter), if she moves her arm uniformly inside the urn in a pseudo-random fashion, and if she does not see, feel or detect the color of the balls in any way\footnote{What I write here has only motivational value; I do not claim to say anything profound about the foundations of probability theory.}. On the other hand, if one of the balls is
in some sense ``easier to draw'' (say, there are $3$ balls, and $2$ of them are hidden in the urn's corners), then the chance of drawing that ball will be higher.

How can we ``draw'' a finite binary string? If we are looking for a ``natural'' mechanism that is not just chosen arbitrarily from all conceivable mechanisms, then we should only use structures that are given to us a priori --- that is, ones that are supplied by mathematics itself. Mathematics constitutes the ``urn'' that contains the finite binary strings, and supplies mechanisms for drawing them. In a mathematical formal system, we can ``draw'' a mathematical object by {\em describing} it. That is, we can write down a {\em definition}, based on the axioms of our formal system, and thereby selecting
a mathematical object from the ``urn'' of all mathematical objects.

Thus, our random experiment might be performed by a mathematician, equipped with paper and pencil, who draws finite binary strings by describing them. Some strings are much easier to describe than others, even if they contain more bits. For example, the binary string
\[
   x:=\underbrace{00000\ldots 0}_{\mbox{a million zeroes}}
\]
is easy to describe --- in fact, we have just described it (and it remains easily describable even if we demand
a more formally sound way, say, a definition according to the rules of a fixed formal system). Similarly, it is easy to describe the string
\[
   x_\pi:=0010001000011111101\ldots
\]
containing the first $10^6$ binary digits of $\pi$. Some strings are much more difficult to describe, like
\[
   x_c:=010010100010000011110\ldots
\]
which is a concrete structureless string of $1000$ bits, generated by a thousand tosses of a fair coin. The simplest way to describe the string by mathematical means seems to be to write it down bitwise, which arguably needs more effort (and more paper space) than the previous two strings.

So the strings $x$ and $x_\pi$ seem to be easier to describe, and, according to our urn metaphor, easier to ``draw'' than $x_c$, for example. Hence they should have larger probability with respect to the distribution that we are looking for.

But how can we formalize this idea? How can we ``describe a string at random'' and get a meaningful probability distribution? The idea of a mathematician, randomly writing down definitions on a piece of paper, is clearly not formal enough to determine a well-defined distribution.

It turns out that there is a precise formal definition of this very idea, which is known as \emph{algorithmic probability}. The main insight is as follows: every step of formal manipulation performed by the mathematician can also be done by a universal computing machine. Thus, instead of asking how easy it is for a mathematician to write down a definition of a binary string, we can ask how easy it is to program a universal computer to output the corresponding string.

I will now briefly summarize a few key concepts from algorithmic information theory as they are relevant for this work. I will mainly focus on the book by Hutter~\cite{Hutter} and a subsequent paper~\cite{Wood}, and assume that the reader is familiar with some basic notions of theoretical computer science (e.g.\ the Turing machine, the halting problem, and computability). A more detailed and pedagogical introduction can be found in the book by Li and Vit\'anyi~\cite{LiVitanyi}, see also~\cite{Chaitin}.

One of the basic models of computation is the Turing machine~\cite{CoverThomas},
consisting of several (input, work and output) {\em tapes} carrying some data given by bits, a finite state
machine, and some {\em read-write-heads} pointing to a single cell on each tape and giving the position
where to read or write next. In accordance with~\cite{Hutter}, we shall only consider Turing machines
with one unidirectional input tape, one unidirectional output tape (to be generalized later), and several
bidirectional work tapes. {\em ``Input tapes are read only, output tapes are write only, unidirectional
tapes are those where the head can only move from left to right. All tapes are binary (no blank symbol),
work tapes initially filled with zeros.''}

Now we distinguish two different possible events: first, the Turing machine $T$ might halt and output
a fixed, finite binary string $x\in\s$. Second, the Turing machine $T$ might compute a possibly infinite
bit string without ever halting; in this case, we may still observe that the output string starts
with a finite bit sequence $x\in\s$. This is due to the fact that the output tape is assumed to be
unidirectional. We use the definition given in~\cite{Hutter}:

\emph{{\bf Monotone TM.} We say $T$ outputs/computes a string starting with $x\in\s$ on input $p\in\s$, and write $T(p)=x*$ if $p$ is to the left of the input head when the last bit of $x$ is output ($T$ reads all of $p$ but no more). $T$ may continue operation and need not halt. For given $x$, the set of such $p$ forms a prefix code. We call such codes $p$ minimal programs.}

This allows us to define the concepts of {\em Kolmogorov complexity} and {\em algorithmic probability}:
\sdefinition{Algorithmic probability and complexity}{DefAlgProb}{
Let $T$ be any monotone Turing machine.
The \emph{monotone complexity} or {\em (monotone) Kolmogorov complexity} of a string $x\in\s$ with respect to $T$ is given by
\[
   {\rm Km}_T(x):=\min\{\ell(p)\,\,|\,\, T(p)=x*\}
\]
or by $\infty$ if no such program $p$ exists. Moreover, define the \emph{algorithmic probability}
that $T$ outputs some string that starts with $x\in\s$ by
\[
   \mathbf{M}_T(x):=\sum_{p:T(p)=x*} 2^{-\ell(p)}.
\]}

Since the set of programs $p$ such that $T(p)=x*$ is prefix-free, it follows from the Kraft inequality that $\mathbf{M}_T(x)\leq 1$ for all $x$. This expression can be interpreted as the probability that $T$ outputs a string that starts with $x$ if the input is chosen by tossing a fair coin. In more detail, $\mathbf{M}_T$ is a \emph{semimeasure} in the sense of the following definition:
\sdefinition{Measures and semimeasures~\cite{Hutter}}{DefSemimeasures}{\lineclear
A function $m:\s\to\R_0^+$ is called a
\emph{semimeasure} if $m(\varepsilon)\leq 1$ and $m(x)\geq m(x0)+m(x1)$, and a \emph{probability
measure} if equality holds in both cases.

We define the conditional (semi)measure as
\[
   m(y|x):=\frac{m(xy)}{m(x)} \quad (x,y\in\s)
\]
if $m(x)\neq 0$, where $xy$ denotes the concatenation of $x$ and $y$.
}

One of the most important facts in computer science is the existence of ``universal computers'' that are capable of simulating every other computer. The following theorem defines what we mean by a ``universal monotone Turing machine'', and claims the existence of such machines~\cite{Wood}:
\stheorem{Universal monotone Turing machine \cite{Hutter,Wood}}{TheUniversalTM}{
There exist monotone Turing machines $U$ which simulate every (other) monotone Turing machine $T$ in the following sense.
There is an enumeration $\{T_i\}_{i\in\N}$ of all monotone Turing machines, and a computable uniquely decodable self-delimiting code $I:\N\to\s$, such that
\[
   U\left(\strut I(i)p\right) = T_i(p)\mbox{ for all }i\in\N, p\in\s,
\]
where $I(i) p$ denotes the binary string obtained by concatenating the strings $I(i)$ and $p$.
}

Intuitively, the string $I(i)$ is a program that makes $U$ emulate the machine $T_i$. Since a universal Turing machine $U$ can simulate every other machine, its monotone complexity measure ${\rm Km}_U$ is ``optimal'' in the sense that ${\rm Km}_U(x)\leq {\rm Km}_T(x)+c_T$ for every Turing machine $T$, where $c_T\in\N$ is a constant that does not depend on $x$. In particular, if $U$ and $V$ are both universal, then there are constants $c,C\in\N$ such that
\[
   \Km_U(x)+c\leq \Km_V(x)\leq \Km_U(x)+C\mbox{ for all }x\in\s.
\]
In other words, $\Km_U$ and $\Km_V$ agree up to an additive constant, which is sometimes denoted $\Km_U(x)=\Km_V(x)+\mathcal{O}(1)$. Similarly, we will find that $\mathbf{M}_U(x)=\mathbf{M}_V(x)\cdot\mathcal{O}(1)$. This kind of ``weak'' machine-independence will be of high relevance for the theory of this paper, as we will discuss in Section~\ref{SecPostulates}.

\begin{figure}[!hbt]
\begin{center}
\includegraphics[angle=0, width=9cm]{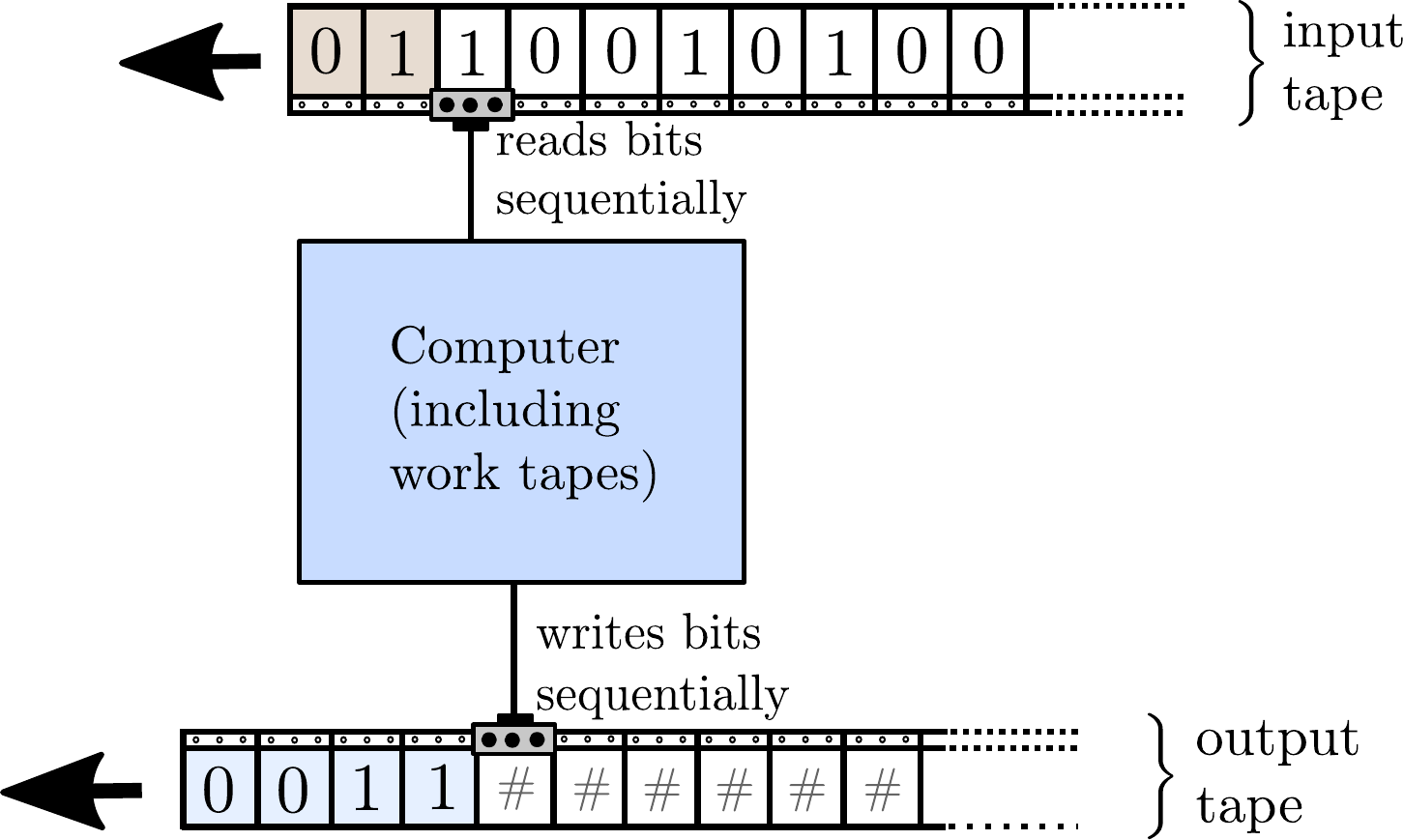}
\caption{Sketch of a monotone Turing machine $T$. The machine reads some (possibly infinite) binary input string, here starting with $0110010100$, and the snapshot depicts the output relation $T(01)=0011*$.}
\label{FigGraphMachine}
\end{center}
\end{figure}

In this paper, we will make extensive use of the following property of universal monotone Turing machines.
\stheorem{Universal enumerable semimeasure~\cite{Hutter}}{TheUniversalityM}{\lineclear If $T$ is a monotone Turing machine then $\mathbf{M}_T$ is an enumerable semimeasure. Vice versa, for every enumerable semimeasure $m$ there exists a monotone Turing machine $T$ with $\mathbf{M}_T(x)=m(x)$ for all non-empty strings $x\in\s$. Moreover, if $U$ is universal, then $\mathbf{M}_U$ is a universal enumerable semimeasure; that is, for every enumerable semimeasure $m$, it holds
\[
   \mathbf{M}_U(x)\geq 2^{-\mathrm{K}_U(m)}\cdot m(x)
\]
for all $x\in\s$, where $\mathrm{K}_U(m)$ denotes the length of the shortest binary string that makes $U$ emulate any monotone Turing machine which has $m$ as its semimeasure, i.e.
\[
   \mathrm{K}_U(m)=\min\{\ell(x)\,\,|\,\,\forall p: U(xp)=T(p)\mbox{ and }\mathbf{M}_T=m\}.
\]
}
This definition uses the notion of \emph{enumerability} of a function $f:\s\to\R$. Suppose we have a computable
function $\Phi:\s\times\N\to\R$ such that $\lim_{n\to\infty} \Phi(x,n)=f(x)$ and $\Phi(x,n)\leq \Phi(x,n+1)$ for all $x\in\s$ and $n\in\N$.
Then $f$ can be approximated from below by a single computer program (computing $\Phi$), without necessarily knowing how
close the approximation will be to the true value $f(x)$.
In this case $f$ is called \emph{enumerable}.
If additionally $(-f)$ is enumerable as well, then we can estimate the error of approximation for finite $n$
by computably determining a finite interval that contains $f(x)$. If this is the case, $f$ is called \emph{computable}.

The semimeasures $\mathbf{M}_U$ will be the subject of the key claims of the postulates of this paper. So far, it seems as if the $\mathbf{M}_U$ represent properties of our specific choice of computational model, the monotone Turing machine. Since this model was chosen somewhat arbitrarily, doesn't this undermine our motivation from above to find a natural (class of) probabilities for which such a choice does not have to be made?

We will now see that the semimeasures $\mathbf{M}_U$ have an alternative definition \emph{that does not refer to monotone Turing machines}. This shows that the $\mathbf{M}_U$ represent natural mathematical structure independent of our favorite choice of computational model.
\sdefinition{Universal mixture~\cite{Wood}}{DefUniversalMixture}{
A \emph{universal mixture} $\mathbf{M}$ is a mixture with non-zero positive weights over an enumeration $\{\nu_i\}_{i\in\N}$ of all enumerable semimeasures:
\[
   \mathbf{M}(x)=\sum_{i\in\N} w_i \nu_i(x),\qquad \R\ni w_i>0,\enspace \sum_{i\in\N} w_i\leq 1,
\]
where, in addition, $i\mapsto w_i$ is an enumerable function.}
It turns out that the universal mixtures are exactly the semimeasures $\mathbf{M}_U$ that we have defined above via monotone Turing machines (MTMs):
\slemma{Universal mixtures and the $\mathbf{M}_U$~\cite{Wood}}{LemUniv2}{$\strut$\newline
Up to their value at the empty string\footnotemark\enspace $\varepsilon$, we have
\[
   \{\mathbf{M}_U\,\,|\,\, U\mbox{ universal MTM}\}= \{\mathbf{M}\,\,|\,\,\mathbf{M}\mbox{ universal mixture}\}.
\]
In other words, for every universal mixture $\mathbf{M}$ there is a universal MTM $U$ such that $\mathbf{M}(x)=\mathbf{M}_U(x)$ for all $x\in\s\setminus\{\varepsilon\}$, and vice versa.
}
This gives a \emph{model-independent} characterization of the $\mathbf{M}_U$: the definition of universal mixtures uses only the notion of computability, without referring specifically to the monotone Turing machine. Since the notion of computability is identical for all models, including quantum computation (more on this in Subsection~\ref{SubsecWhyAlgProb}), the approach of this paper is independent of the choice of model of computation.
\addtocounter{footnote}{-1}
\footnotetext{The special role of the empty string $\varepsilon$ follows from the fact that $\mathbf{M}_U(\varepsilon)=1$ for all $U$ by construction, but $\mathbf{M}(\varepsilon)<1$ for all universal mixtures $\mathbf{M}$~\cite{Wood}.
}

While any given universal mixture $\mathbf{M}$ is only a semimeasures, we can define its \emph{Solomonoff normalization}~\cite{LiVitanyi}
\[
   \p(\varepsilon):=1,\quad \p(xa):=\p(x)\cdot\frac{\mathbf{M}(xa)}{\sum_b \mathbf{M}(xb)}\quad (a\in\{0,1\})
\]
to obtain a \emph{measure} $\p$ that shares many (but not all) desirable properties with $\mathbf{M}$. Universal mixtures $\mathbf{M}$ and their Solomonoff normalizations $\p$ are related by the inequalities
\[
   \p(x)\geq \mathbf{M}(x), \quad
   \p(y|x)\geq \mathbf{M}(y|x).
\]
Every $\p$ that is derived from a universal mixture $\mathbf{M}$ in this way will be called an \textbf{algorithmic prior}.

\section{Postulates of an incomplete theory}
\label{SecPostulates}
\sectionmark{Postulates of an incomplete theory}
Let me clarify right away that the theory of this paper will not satisfy all the desiderata that have been formulated in the introduction. Namely, what we would \emph{like} to have is a theory that satisfies the following Postulates:
\colorlet{shadecolor}{xblue}
\postulates{Desired postulates; not used in that form}{DesiredPostulates}{
$\strut$
\vskip -2em $\strut$
\begin{itemize}
\item[(i)] \textbf{Observer states.} Having a first-person perspective means to be in some observer state at any given (subjective) moment. The observer states are in computable one-to-one correspondence with the finite binary strings.
\item[(ii)] \textbf{Dynamics.} Being in some observer state $x$ now, there is a well-defined chance of being in some other observer state $y$ next. It is denoted
\begin{equation}
   \p(y|x),
   \label{eqDesired}
\end{equation}
where $\p$ is an algorithmic prior.
\item[(iii)] \textbf{Predictions.} The predictions of the theory are those that are identical for every choice of algorithmic prior. They follow from (i) and (ii) alone; no underlying physical world is assumed to ``cause'' those probabilities. ``Now'' and ``next'' are understood as purely first-person notions, not related to any external notion of time or clock.
\end{itemize}}
Before giving the actual form of the postulates that we use in this paper, let me give some more intuition on the worldview that they express. The formulation itself seems solipsistic in some sense: it talks about what it means to ``have a first-person perspective''. In this sense, it talks about the ``I'': I am currently in some state $x$, and then I will be in some other state $y$. So who is this ``I''? What about ``you'' or ``them'', i.e.\ other observers? Or is there only ever one observer?

We will address these questions in more detail in the following sections, when the ontology of the theory will become gradually more clear while working out the postulates' consequences. A preliminary answer is that the postulates describe \emph{everybody}: they allow to determine the chances of what happens to \emph{any} observer next, given their current observer states. This is somewhat similar to the tenets of Bayesianism, which can be used by \emph{everybody} to make rational bets on their future.

However, in the approach of this paper, the probabilities $\p(y|x)$ are \emph{not} betting probabilities. They are interpreted as \emph{private, but objective chances}, not as degrees of belief. For an observer in state $x$, they are meant to say which states $y$ are more or less likely to be actualized next for this observer. In particular, these probabilities are seen as \emph{fundamental}: neither do they represent missing knowledge about an underlying state of the world (as in statistical mechanics), nor do they arise from some kind of fundamental quantum state. In particular, there is \emph{no} claim of any actual underlying computation or Turing machine which would justify the appearance of these probabilities: monotone Turing machines have only been used in the \emph{mathematical definition} of $\p$, but they are not part of any ontological claim.

In the theory described by these postulates, there is no fundamental notion of an ``observer'', but only of ``observer \emph{states}''. That is, observers are not material objects in some universe which could be distinguished and counted, at least not fundamentally. This is perhaps not so surprising, given that we are about to construct a theory that intends to say something meaningful about puzzles like Parfit's teletransportation paradox or about ``copying'' observers. For a colorful example, think of the ``Back to the Future'' movie series, in which the main protagonist (Marty McFly) meets his older self in the future. Is this now one observer, or two? Here, we view this as a fundamentally meaningless question. Nonetheless, our theory will admit situations that can best be interpreted as ``observers encountering other observers'', and we will discuss these situations in detail in Section~\ref{SecObjectivity}.

So when did the observer start? When did she hold her first observer state? Will she not die some time? Indeed, the postulates above say that the answer to the latter question is negative --- the observer will follow a never-ending Markovian process. It also doesn't make sense to talk of a ``beginning''. Any notions of this kind --- and of an external world that seems to have begun in a Big Bang a long time ago --- will have to be reconstructed from the postulates alone. The exciting news is that this can in fact be done to some extent. That is, we will understand ``why'' observers will see something like an external world ``around them'' as a \emph{consequence} of these postulates.
\footnotetext{This notion of ``forgetting'' should be interpreted in purely technical terms: it refers to a situation in which an exhaustive description of an observer at some moment (given by its observer state) does not admit the reconstruction of its earlier observer states in principle. This is \emph{not} the same as the colloquial notion of ``forgetting'' that we use for human observers, in the sense that some information in the brain becomes consciously unavailable (``when again is my wife's birthday?''). Observer states include much more than just consciously accessible information (indeed, most types of observers will anyway not be ``conscious'' in any meaningful sense of the word).}
\addtocounter{footnote}{-1}

Before discussing Postulate (iii) in more detail, let me explain why our notion of algorithmic probability does not quite satisfy all the desiderata of Postulates~\ref{DesiredPostulates} above. Intuitively, if an observer is in state $x$, then she can transition into another state $y$ \emph{that may hold more or less information than $x$}. In particular, it is possible for observers to ``forget'' information: sometimes, memory is erased, and our next observer state $y$ does not contain full information on the previous state $x$. It seems overly restrictive to disallow this possibility. On the other hand, conditional algorithmic probability $\p(y|x)=\p(xy)/\p(x)$ is defined as the probability that the next observer state will be $xy$, given that it is now $x$. In other words, algorithmic probability defines a situation in which \emph{an observer's state will in principle always contain full information on its previous states}. This will define the postulates that we are actually working with --- and, as expressed below, it will therefore be an approximation to our desired postulates which applies whenever memory erasure can be neglected:
\colorlet{shadecolor}{xred}
\postulates{Simplified postulates as actually used}{Postulates}{
$\strut$
\begin{itemize}
\item[(i)] \textbf{Observer states.} Having a first-person perspective means to be in some observer state at any given (subjective) moment. The observer states are in computable one-to-one correspondence with the finite binary strings.
\item[(ii)] \textbf{Dynamics.} Being in some observer state $x$ now, there is a well-defined chance of being in some other observer state $xa$ next, where $a\in\{0,1\}$ is a bit. It is given by
\begin{equation}
   \p(a|x),
   \label{eqActually}
\end{equation}
where $\p$ is an algorithmic prior which can be chosen arbitrarily, but has to be fixed.
\item[(iii)] \textbf{Predictions.} The predictions of the theory are those that are identical for every choice of algorithmic prior. ``Now'' and ``next'' are understood as purely first-person notions, not related to any external notion of time or clock.
\end{itemize}
\textbf{Interpretation:} These postulates will make similar predictions as the ``desired theory'' (expressed in Postulates~\ref{DesiredPostulates}) in those cases where the observer holds a large amount of memory on her previous states; they will fail to do so, however, when ``forgetting''\footnotemark \enspace(information erasure) becomes relevant.
}
Since the only predictions that we will make are those that agree for every choice of algorithmic prior, we will now fix an arbitrary reference universal monotone Turing machine $U$, and use $\mathbf{M}:=\mathbf{M}_U$ and $\mathbf{P}$ its Solomonoff normalization in all the calculations of this paper. It is no loss of generality to assume that observer states grow \emph{one} bit at a time: the probability of receiving more than one bit can be expressed via the chain rule. If $y=y_1 y_2 \ldots y_m\in\{0,1\}^m$, then
\[
\   \p(y|x)=\p(y_1|x)\p(y_2|xy_1)\ldots \p(y_m|xy_1\ldots y_{m-1}).
\]
In the following, we will work with Postulates~\ref{Postulates}. For more comments on the relation between those and the desired postulates~\ref{DesiredPostulates}, see the beginning of the appendix. In the remainder of this section, we will motivate why an algorithmic prior $\p$ is assumed in the postulates, and describe in more detail how its non-uniqueness should be understood.

\subsection{Algorithmic priors and levels of indeterminism}
Let us now turn to Postulate (iii), which is closely related to the following question: \emph{which universal mixture $\mathbf{M}$ should we choose? That is, which universal machine $U$ should we use in~(\ref{eqActually}) to define $\mathbf{P}\equiv \mathbf{P}_U$?} At first sight, it seems as if the infinitude of universal monotone Turing machines $U$, and the corresponding infinitude of algorithmic priors $\p_U$, would make the theory underdetermined. Is there a ``correct'' choice of $U$? Or can we somehow average over all $\p_U$ in a clever way to fix the probabilities? Unfortunately, it can be shown that the answer to this question must be ``no'' as a matter of principle~\cite{StatAlg,HutterOpenProblems,Schack97,SterkenburgThesis}, which can be related to well-known insights from philosophy like ``Goodman's new riddle of induction''~\cite{Goodman}.

Instead, Postulate (iii) says that it is arbitrary which $U$ to choose --- the predictions of the theory are exactly those that are the same for every $U$. In fact, as we shall see, there are \emph{many interesting} predictions that are identical for all $\p_U$. This is due to the invariance property: if $U$ and $V$ are universal monotone Turing machines, then there are constants $0<c<C$ such that $c\, \mathbf{M}_V(x)\leq \mathbf{M}_U(x)\leq C\, \mathbf{M}_V(x)$ for all $x\in\s$. Hence $\mathbf{M}_U$ and $\mathbf{M}_V$ will agree on many asymptotic statistical properties, and so will their Solomonoff normalizations $\p_U$ and $\p_V$.

But shouldn't there be an \emph{actual probability} $\p(a|x)$ of the next bit $a$, given the current observer state $x$? How can we have any predictive power whatsoever if we can make $\p_U(a|x)$ equal to basically any real number between zero and one? First of all, note that while one is allowed to choose an arbitrary $U$, one is also \emph{obliged to keep it fixed}. That is, $U$ is not allowed to be changed after learning a new bit $a$, so we cannot simply make our probabilistic predictions attain arbitrary desired numerical values, independently over time.

To better understand the meaning of Postulate (iii), let me argue by analogy. Imagine a physicist (the Conservative) who is familiar with Newtonian mechanics (and perhaps its relativistic version), but nothing else. Newtonian mechanics is a fully deterministic theory --- exact knowledge of the initial conditions allows us to predict the future behavior of any physical system exactly. Let us denote this property by ``level-0 indeterminism''. That is, Newtonian mechanics, as a physical theory, is level-0 indeterministic (namely, deterministic). If our world was actually exactly governed by Newtonian mechanics, then we would live in a level-0 indeterministic world.

Now suppose that the Conservative is shocked to meet another physicist (the Liberal) who has just proposed a physical theory which is \emph{irreducibly probabilistic}: quantum theory. As the Liberal proposes, quantum theory does not allow us to predict the outcomes of experiments with certainty, but instead tells us \emph{the probabilities} of the outcomes. For example, if we send a single photon to a half-silvered mirror, then there is a $50\%$ chance of detecting the transmitted versus the reflected photon. In other words: the Liberal claims that quantum theory is \emph{level-1 indeterministic}. This leads to the following conversation.

\textbf{Conservative:} ``Of course, every good physical theory should tell us \emph{what the particles are going to do}, and not just give us probabilities! Doesn't this just mean that you don't know exactly all the details of your mirror? That you just have to look at it more closely, and then you can actually predict whether the photon will be reflected or transmitted?''

\textbf{Liberal:} ``No. If my approach is correct, then it is in principle impossible to predict this --- the process is \emph{intrinsically random}. In other words, I claim that \emph{nature simply does not carry enough structure} to determine the outcome of the experiment. This should not come as a surprise: for example, you have already learnt from relativity that the notion of \emph{simultaneity} is, surprisingly, not part of the furniture of the world, even though it intuitively should be. I claim that quantum theory tells us that \emph{determination of measurement results} is yet another piece of furniture that has to go.''

\textbf{Conservative:} ``But \emph{what exactly} is the content of the claim that the transmission probability is $50\%$? When I repeat the experiment ten times, does this tell me that the photon will be transmitted five times? You explained to me that it does not. It could be six times, or actually all ten times. Nothing is excluded and nothing predicted.''

\textbf{Liberal:} ``Well, if you repeat the experiment many times (say, $n$ times), then the law of large numbers tells you that there is a high probability that the fraction of times $m$ it was transmitted (that is, $m/n$) is $\varepsilon$-close to $1/2$.''

\textbf{Conservative:} ``... there is a \emph{high probability} of that happening... You just explained the meaning of $50\%$ probability to me with yet another reference to probability. This is circular reasoning! I'm still waiting for your explanation of what a probabilistic claim actually means.''

\textbf{Liberal:} ``Yes, this reasoning is circular --- the notion of probability \emph{cannot} ultimately be grounded in any deterministic notion. Nevertheless, probabilistic claims are not meaningless. They resemble structure of the world --- though weaker structure than determinism --- that allows us to place successful bets. We have learned by experience how to use probabilities to act rationally in the face of indeterminism. This is an empirical fact that's hard to deny.''

Half convinced, the conservative leaves and downloads a copy of the Liberal's lecture notes. In the meantime, the Liberal meets yet another physicist (the Gambler) who has just proposed Postulates~\ref{Postulates}. These postulates describe a \emph{level-2 indeterministic theory}: that is, one in which there is not a \emph{single} probability distribution, but an \emph{infinite set} of possible distributions that are supposed to describe the chances. The Liberal is shocked.

\textbf{Liberal:} ``Of course, every good physical theory should tell us \emph{what the probabilities are}, and not just give us an infinite set of priors! Doesn't this just mean that you don't yet know what the correct distribution is? I guess you have to work a bit harder, until you can improve the postulates to tell us the actual values of the probabilities!''

\textbf{Gambler:} ``No. If my theory is correct, then it is in principle impossible to say this --- \emph{nature simply does not carry enough structure} to determine the actual numerical values of probabilities. You must be well aware that we had to let go of other beloved furniture of the world before.''

\textbf{Liberal:} ``But \emph{what exactly} is the content of the claim that observer states are described by the infinite set of algorithmic priors $\p_U$? Does this mean that when I pick my favorite universal machine $U$, my future states will be distributed according to $\p_U$? You explained to me that it does not.''

\textbf{Gambler:} ``Well, it tells you that if there is a computable regularity that you have observed often enough, then for every $\p_U$ there will ultimately be a high probability that this regularity remains. This is the nature of algorithmic probability, or universal induction.''

\textbf{Liberal:} ``... for \emph{every $\p_U$ it will ultimately...} But for some $\p_U$ it will happen faster than for others. So, suppose I have seen a regularity $n$ times, then what is \emph{the actual} probability that this regularity will remain? I'm still waiting for your concrete explanation of what the set of $\p_U$ is supposed to mean.''

\textbf{Gambler:} ``The notion of level-2 indeterminism cannot ultimately be grounded in any level-1 notion. Nevertheless, level-2 claims are not meaningless. They resemble structure of the world --- though weaker structure than level-1 --- that allows us to place rational bets. In fact, I claim that this is \emph{how we actually bet in this world anyway}: we can start with an arbitrary prior, and use new data to update it. When you say that there is a $50\%$ chance that a photon will be reflected or transmitted, you are basing this claim on a strong belief that this is \emph{really a half-silvered mirror}, and that repetitions of the experiment can be treated as exchangeable. But these are \emph{also} only statements with certain probabilities attached to them (even if you think that these probabilities are close to one), and so on, ad infinitum. The best a physical theory can do is to tell us which \emph{kinds of priors} are admissible, and which ones are not (for example, ones with a belief in the violation of conservation laws).''

Convinced by some amount that is hard to quantify (perhaps half convinced, perhaps $30\%$), the Liberal leaves... and starts to feel really old.

\noindent\rule{.48\textwidth}{1pt}

The position of the Gambler resembles some arguments that are also often heard in another camp: by supporters of \emph{QBism}~\cite{Fuchs,FuchsSchack,Timpson}, i.e.\ of (subjective) Quantum Bayesianism (or nowadays called ``Quantum Betabiliterianism'' by its founders). QBists argue that the actual numerical values of probabilities (or, for that matter, the concrete entries of a quantum density matrix) are not themselves properties of the world, but represent subjective beliefs. As the argument goes, it is instead \emph{the update rules (e.g.\ the Born rule) and the structure of the state space (e.g.\ the Hilbert space dimension)} that resemble actual ``facts of the world''. In this sense, QBists would perhaps agree that ``quantum states represent beliefs which are level-2'', whereas Postulates~\ref{Postulates} should be understood as saying that ``objective chances are level-2''.

The idea of a \emph{set} of priors is not new. It has been studied in many different forms under the name of \emph{imprecise probability}~\cite{Walley}, and it appears in physics under the name of \emph{equivalence of ensembles}~\cite{Lima1,Lima2,MuellerAdlam}. That is, in many situations, thermodynamics is understood as a theory with predictions that agree for the canonical and microcanonical ensembles, and it is in some sense arbitrary which ensemble to select.

There is another important advantage of postulating a set of priors over picking a single distribution: it allows for a strong notion of \emph{encoding invariance}.

\subsection{Encoding invariance}
What do the bits in an observer state $x$ actually \emph{mean}? Postulates~\ref{Postulates} tells us that the length $\ell(x)$ says \emph{how many} state transition the observer has suffered since its state of no information. Moreover, the order of bits is relevant as well: if $x=x_1 x_2 \ldots x_n$, then the first bit is the one that has been acquired first, the second bit next, and so on.

But what about the meaning of ``zero'' versus ``one''? Isn't it naive to claim that an observer's first-person perspective is described by a bunch of bits in the first place? This conceptual confusion can be clarified by taking a more abstract point of view. According to this view, observer states themselves \emph{are} not binary strings, but they can be \emph{encoded} into binary strings. Think of an unspecified countably-infinite set $\mathcal{O}$ of observer states, and a bijective map $\alpha:\mathcal{O}\to\s$ that tells us which observer state is represented by which binary string. We assume that observer states have additional structure: there is a distinguished \emph{empty} observer state $o_\varepsilon$, and for every observer state $o$, there are \emph{two} different distinguished observer states $o'$ and $o''$ that we can think of as ``continuations'' of $o$, encoding an additional answer to some yes-no-question. Then $\alpha$ should be structure-preserving: it should satisfy $\alpha(o_\varepsilon)=\varepsilon$ and $\ell(\alpha(o'))=\ell(\alpha(o''))=\ell(\alpha(o))+1$.

Consider \emph{another} bijective structure-preserving encoding map $\beta:\mathcal{O}\to\s$. The two maps $\alpha$ and $\beta$ encode any $o\in\mathcal{O}$ into different binary strings, $x_\alpha:=\alpha(o)$ and $x_\beta:=\beta(o)$. Consequently, $x_\beta=\beta\circ \alpha^{-1}(x_\alpha)$. We can think of the bijective map $\beta\circ\alpha^{-1}:\s\to\s$ as an \emph{encoding transformation}: it preserves the length and prefix properties of any string, but switches between two different possible encodings of observer states.

We need one additional piece of structure on the observer states: \emph{computability structure}. Intuitively, we would like to say that both $\alpha$ and $\beta$ should be computable. However, we have no idea what this means. But whatever it \emph{does} mean, it should imply that the composition $\beta\circ\alpha^{-1}$ is computable too --- and, as a map on the binary strings, computability of this function \emph{is} a well-defined notion. This argumentation leads us to the following definition and theorem:
\stheorem{Encoding invariance}{TheEncodingInv}{
Let $\varphi:\s\to\s$ be any \emph{structure-preserving map} on the observer states, i.e.\ a computable bijective map whose inverse is computable, and which preserves prefixes in the sense that $\varphi(\varepsilon)=\varepsilon$ and
$\strut$\vskip -0.5em
\[
   \varphi(\{x0,x1\})=\{\varphi(x)0,\varphi(x)1\}.
\]
Then the theory expressed by Postulates~\ref{Postulates} is invariant under every such map; that is,
$\strut$\vskip -0.7em
\[
   \{\p_U\,\,|\,\, U\mbox{ universal}\}=\{\p_U\circ\varphi\,\,|\,\, U\mbox{ universal}\},
\]
i.e.\ the set of algorithmic priors is invariant under structure-preserving maps. As explained above, we can interpret this as a ``freedom of choice of encoding'' of observer states into binary strings.}

An example of a structure-preserving map is given by the \emph{bitwise inversion}: for example, $\varphi(1011)=0100$. This theorem tells us that the predictions of our theory do not change if we decide to switch zeros with ones in the representation of observer states.

\proof
Note that $\p_U\circ \varphi=\p_{\varphi^{-1}\circ U}$, i.e.\ the distribution $\p_U\circ\varphi$ can be obtained as the one coming from a machine that works like $U$, but applies $\varphi^{-1}$ to its outputs. Since $\varphi$ is structure-preserving if and only if $\varphi^{-1}$ is, it only remains to show that
\[
   \{\varphi^{-1}\circ U\,\,|\,\, U\mbox{ universal}\}\subseteq \{U\,\,|\,\, U \mbox{ universal}\}.
\]
In other words, we have to show that $V:=\varphi^{-1}\circ U$ is universal whenever $U$ is universal. To see this, let $\{T_i\}_{i\in\N}$ be the enumeration of all monotone Turing machines, and $I:\N\to\s$ the corresponding computable uniquely decodable self-delimiting code, that makes $U$ universal according to Definition~\ref{TheUniversalTM}. Then $\tilde T_i:=\varphi^{-1}\circ T_i$ defines another enumeration of all monotone Turing machines. We have
\[
   V\left(\strut I(i)p\right)=\varphi^{-1}\circ U\left(\strut I(i)p\right)=\varphi^{-1}\circ T_i(p)=\tilde T_i(p).
\]
This proves that $V$ is universal.
\qed

There is an intuitive analogy of these insights with differential geometry as used, for example, in the theory of general relativity: the set $\mathcal{O}$ is analogous to the manifold of spacetime points, and the maps $\alpha$ and $\beta$ correspond to two different coordinate systems. These coordinate systems should preserve the differentiability structure of the manifold, and the physical laws are invariant under changes of coordinates. Here, the structure to be preserved is the observer states' \emph{computability structure} (together with the prefix structure), and our theory is invariant under changes of encoding.

Now that we have argued that a \emph{set} of priors is the way to proceed, we still have to discuss why this set should be chosen as the set of \emph{algorithmic} priors. In Section~\ref{SecAlgorithmicProbability}, we have given an intuitive motivation for selecting algorithmic probability: it is in some sense a ``natural probability structure'', derived from the structure of mathematics alone, without the need for any further choices. The next subsection will give a second, independent argument.

\subsection{Why use algorithmic probability?}
\label{SubsecWhyAlgProb}
Our choice of algorithmic probability can also be understood in a pragmatic way, namely as the result of extrapolating a successful method of prediction to a larger and more general domain: Solomonoff induction~\cite{Hutter,LiVitanyi}.

In a nutshell, Solomonoff induction is a simple prescription for predicting future data, given previous data. Suppose that a random process generates one bit $x_1,x_2,\ldots$ after the other, according to an unknown and perhaps very complicated measure $\mu$. Think of a scientist who receives these bits, and is supposed to predict \emph{the probability of the next bit}, $\mu(a|x_1\ldots x_n)$. There is one extra promise that may help the scientist to place her guess: that $\mu$ is \emph{computable}. That is, there exists a (potentially extremely inefficient) algorithm that, on input $x=x_1\ldots x_n$ and $n\in\mathbb{N}$, outputs an $n$-digit approximation of $\mu(x)$. Needless to say, neither $\mu$ nor this algorithm are known to the scientist.

Solomonoff induction is the following prescription: \emph{as a good guess for $\mu(a|x)$, use algorithmic probability $\p(a|x)$}. That is all. As we will study in more detail in Section~\ref{TheSimpleLaws}, this guess $\p$ is guaranteed to be close to the actual probability $\mu$ in the limit of $n\to\infty$.

Solomonoff induction works in the context of computable, probabilistic processes --- and there is one particularly relevant process of this kind: \emph{our physical world}, as it presents itself to the observations of a physicist. Given data on initial conditions of some physical system, we can in principle write a computer program that simulates the laws of physics as we know them and produces predictions for all observations that we may perform on the system at later times. While these predictions will in general be probabilistic (as dictated by quantum theory), the statistical inferences that we draw from them are in great agreement with our actual observations. It is a remarkable empirical finding that the notion of ``universal computation'', based on the Turing machine~\cite{Turing,Cooper}, seems to capture every kind of process in our universe that can be subjected to this kind of controlled empirical analysis\footnote{This is a statement of \emph{principle} and not of \emph{practice}. For all practical purposes, it may e.g.\ remain forever impossible to produce an accurate simulation of the statistics of all of planet Earth in any detail (as described by quantum theory or any future theory), even though the physical Church-Turing thesis claims that a corresponding (extremely impractical) algorithm exists in principle.}.

This observation and its extrapolation to all physical experiments that we may perform in the future --- sometimes called the ``physical version'' of the Church-Turing thesis\footnote{While the original Church-Turing thesis does not directly relate to physics, several different versions of this thesis have been formulated over the decades. The version that we refer to here resembles, for example, Wolfram's~\cite{WolframBook} ``principle of computational equivalence''. It has been analyzed in more detail by Gandy~\cite{Gandy}, who calls (something very similar to) it ``Thesis M'', and in the quantum context by Arrighi and Dowek~\cite{Arrighi}. For an overview and discussion of different versions of the Church-Turing thesis, see e.g.~\cite{Hofstadter,PiccininiTuring}.} --- is supported by experience and a variety of arguments.
All theoretical attempts to construct reasonable mathematical models of computation, especially under the constraint to be realizable in our physical world, have so far turned out to be equivalent
to the Turing machine model. This includes quantum computers, which can compute exactly the same class of functions as
classical ones (the fact that they may be superpolynomially faster at some tasks~\cite{Deutsch} does not invalidate the
formulation of the thesis that we are considering here since questions of efficiency are irrelevant for our purpose). Despite some claims in the opposite, no physical system performing
``hypercomputation'' has ever been identified~\cite{Davis}.

This implies a very simple and at the same time surprising consequence: \emph{Solomonoff induction can be used to make successful predictions in our physical world} --- predictions that must agree with those of our best physical theories (or even better future theories) if their regime of applicability is the regime of data collection. Solomonoff induction will automatically ``discover'' the probabilistic laws of nature that we already have (such as quantum theory), or possible future ones. In some sense, Solomonoff induction can thus be seen as a formal analogue of the scientific method itself.

One of the major motivations of this paper is the insight that there are regimes of experience that we are currently entering which go beyond the standard domain of physics, cf.~Table~\ref{fig_motivation}: for example, we are interested in the experience of observers in extremely large universes (cosmology and the Boltzmann brain problem), or we would like to understand what an agent is going to see if her brain is simulated on a computer (philosophy of mind). The problem is that physics as we currently know it is not designed to address these questions, at least not directly, as heralded by the controversial discussions that characterize these fields. It is at this point where the insight above is quite suggestive: \emph{Solomonoff induction agrees perfectly with our best physical theories in the usual regime of physics, but it can also be applied in more exotic domains}. But then, there is an obvious approach that we can take: \emph{Go ahead and apply Solomonoff induction to these new regimes of experience!}

Applying Solomonoff induction means nothing but predicting the future according to conditional algorithmic probability. But this is exactly what Postulates~\ref{Postulates} are claiming: namely, that the chances of our future observations are given by conditional algorithmic probability. In this sense, our postulates can be seen as simply formalizing the prescription that we have derived above --- namely, to apply Solomonoff induction in \emph{all} situations in which we may ask ``what will I see next?''\footnote{In more detail, the prescription is not not that an observer should \emph{actually use Solomonoff induction to predict her future}, because observers will not in general know all details of their observer states, and algorithmic probability is not computable. It is rather the idea that we should \emph{think of algorithmic probability as dictating what happens in these new regimes of experience}, and then do our best to extract concrete predictions from this claim.}.

In the rest of the paper, we are indeed going to do exactly that. For example, we will apply Solomonoff induction to exorcise Boltzmann brains in Subsection~\ref{SubsecBB}, and we will derive predictions on the computer simulation of agents in Subsection~\ref{SubsecBrainEmulation}. By doing so, we will work with Postulates~\ref{Postulates}, but will sometimes speculate what would happen if we instead had a better theory that realizes Postulates~\ref{DesiredPostulates}.

The motivation above is so simple that it may sound almost trivial: all it really says is that \emph{agents who have made observations in agreement with our physical theories in the past should bet on the validity of those theories in the future}. But is there any reason to expect that agents will arrive in such a situation in the first place? In other words, even \emph{before} having made any observations, do our postulates predict that agents will make observations that are consistent with some kind of theory -- observations that correspond to some simple, probabilistic, computable laws of some external physical world? This is the question that we will address next.

\section{Emergence of an external physical world}
\label{SecSimpleLaws}
\definecolor{shadecolor}{named}{xgreen}
\sectionmark{Emergence of an external physical world}
\emph{``The only thing harder to understand than a law of statistical origin would be a law that is not of statistical origin, for then there would be no way for it [...] to come into being. On the other hand, when we view each of the laws of physics [...] as at bottom statistical in character, then we are at last able to forego the idea of a law that endures from everlasting to everlasting.''} (John A.\ Wheeler, ``Law Without Law''~\cite{Wheeler}).\\

Now we are ready to prove the first consequence of our postulates: that observers will, with high probability, see an external world that is governed by \emph{simple, computable, probabilistic} laws (that is, laws that assign probabilities to observations and which have a short description). However, we will not be able to make any claim as to \emph{what these laws actually are}: their specific form will be contingent. In particular, we will in principle not be able to predict the exact form of the laws of physics as they present themselves to us human observers (say, in the form of General Relativity and the Standard Model), only that they have the three structural properties just mentioned (for a possible notable exception, see Section~\ref{SecQuantum}).

Before showing this in full generality, let us start with a first ``warm-up'' that gives us some intuition as to why and how regularities may emerge and stabilize themselves.

\subsection{Warm-up: persistence of regularities}
Suppose that an observer is currently in observer state $x$. Then, her state will subsequently change to a longer strong $xy$, with probability
\[
   \p(y|x)\geq \mathbf{M}(y|x)=\frac{\mathbf{M}(xy)}{\mathbf{M}(x)}
   \geq \frac{2^{-\Km(xy)}}{\mathbf{M}(x)}.
\]
This inequality tells us that transitions to those $xy$ tend to be preferred which are ``more natural continuations'' of the previous state $x$. That is, if $xy$ has a short description, i.e.\ if $\Km(xy)$ is small, then the corresponding $xy$ tends to occur with higher probability than other possible states $xy'$. Thus, simplicity in the sense of compressibility is favored. Intuitively, highly compressible histories (or strings) are those that contain \emph{regularities} which can be used to generate shorter descriptions.

How can we define the notion of ``regularities'' and prove that they are somehow favored by algorithmic probability $\p$? It turns out that an abstract approach is the most simple and powerful one: namely, defining a ``regularity'' of an observer state $x$ as some property for which a computer program can check in finite time whether or not it is present:
\sdefinition{Computable tests}{DefComputableTest}
{A computable function $f:\s\to\{0,1\}$ is called a \emph{computable test}. A computable test is called \emph{sustainable} if $f(\varepsilon)=1$ and if for all $x\in\s$ with $f(x)=1$ there is some bit $a\in\{0,1\}$ with $f(xa)=1$.}
In a nutshell, a computable test is sustainable if whenever it gives the answer ``yes'', it can possibly still give the answer ``yes'' in the next moment.

Imagine an observer in state $x=x_1 x_2\ldots x_n$ (where the $x_i$ are the bits), and suppose that there is a computable test $f$ such that $f(x_1)=f(x_1 x_2)=\ldots =f(x_1 \ldots x_n)=1$. This describes a regularity: all previous observer states (including, for example, observations) had the property that the test $f$ yielded the outcome ``yes''. In this case, the observer may be led to believe that $f$ will yield ``yes'' in the next moment, too. The following theorem, inspired by~\cite{HutterConfirmation}, shows that this guess will asymptotically be correct due to the properties of algorithmic probability. We use the notation $x_1^k:=x_1 x_2\ldots x_k$ if $x=x_1 \ldots x_n$ and $k\leq n$.
\stheorem{Persistence of regularities}{ThePersistence}{Let $f$ be a sustainable computable test. For bits $a_1,\ldots,a_n,b\in\{0,1\}$, define the measure $p$ as
\begin{eqnarray*}
   p(b|a_1 a_2\ldots a_n):=
   \p\{f(x_1^{n+1})=b\,\,&|&\,\,f(x_1^1)=a_1, \ldots,\\
      &&f(x_1^n)=a_n\}.
\end{eqnarray*}
Then we have $p(1|1^n)\stackrel{n\to\infty}\longrightarrow 1$, and the convergence is rapid since $\sum_{n=0}^\infty p(0|1^n)<\infty$. That is, \emph{computable regularities that were holding in the past tend to persist in the future}.
}
\proof
Since $f$ is a sustainable computable test, there is an algorithm that constructs an infinite string $z\in\{0,1\}^\infty$ with the property that $f(z_1^k)=1$ for all $k\in\mathbb{N}$. Namely, the algorithm starts with the empty string, and then picks some bit $z_1$ such that $f(z_1)=1$, and then picks some next bit $z_2$ such that $f(z_1 z_2)=1$, and so forth. Define
\[
   p(a_1 a_2\ldots a_n):=\sum_{x\in\{0,1\}^n:f(x_1)=a_1,\ldots, f(x_n)=a_n} \p(x),
\]
which yields a measure in the sense of Definition~\ref{DefSemimeasures}. Its conditional version $p(b|a)$ is the quantity we are interested in. Let $\mu$ be the computable deterministic measure with $\mu(x)=1$ if $x$ is a prefix of $z$, and $0$ otherwise. Since $\mathbf{M}$ is a universal mixture, there is some constant $c>0$ such that $\mathbf{M}(x)\geq c\,\mu(x)$ for all $x\in\s$ (according to Theorem~\ref{TheUniversalityM}, we can choose $c=2^{-{\rm K}(\mu)}$), thus
\[
   p(1^n)\geq \p(z_1^n)\geq \mathbf{M}(z_1^n)\geq c\, \mu(z_1^n)=c\quad\mbox{for all }n\in\N.
\]
On the other hand, $p(1^n)=\prod_{j=0}^{n-1}p(1|1^j)$, hence
\[
   \log c\leq \sum_{j=0}^{n-1}\log p(1|1^j)\leq\sum_{j=0}^{n-1}\left(p(1|1^j)-1\right).
\]
Since this is true for all $n$, the claim follows.
\qed

As a simple example, consider the ``frequency of ones'' of some string $x$, defined as $\#_1(x)/\ell(x)$, where $\#_1(x)$ is the number of ones in $x$ (for example, $\#_1(1011)=3$). Let us define a computable test $f$ that asks whether the frequency of ones is larger than $90\%$. This is only an interesting question for longer strings. For all $x\in\s$ with $\ell(x)\geq 10$, set
\[
   f(x):=\left\{\begin{array}{cl} 1 & \mbox{if }\#_1(x)/\ell(x)\geq 0.9 \\ 0 & \mbox{otherwise}.
\end{array}\right.
\]
and for strings $x\in\s$ with $\ell(x)\leq 9$, set $f(x):=1$ if and only if $x$ is a prefix of any string $y$ of length $10$ with $f(y)=1$. For example, $f(1101111111)=1$, hence $f(110)=1$ and $f(\varepsilon)=1$.

This is a computable test, but is it sustainable? Suppose that $f(x)=1$ and $\ell(x)\leq 9$, then there exists some $a\in\{0,1\}$ with $f(xa)=1$ by construction. If $\ell(x)\geq 10$, the
\[
   \frac{\#_1(x1)}{\ell(x1)}=\frac{\#_1(x)+1}{\ell(x)+1}\geq \frac{\#_1(x)}{\ell(x)}\geq 0.9,
\]
hence $f$ is indeed sustainable. Thus, Theorem~\ref{ThePersistence} that \emph{an observer that has been in states with more than $90\%$ of ones for long enough will probably continue to be in states with this property in the future}.

However, a moment's thought shows that Theorem~\ref{ThePersistence} doesn't really say very much: facts can change, and the answers to yes-no-questions can flip over time. This simple observation creates a puzzle of relevance far beyond this paper, which is known under the name of ``Goodman's new riddle of induction''~\cite{Goodman}. To illustrate this, consider another sustainable computable test $f$ which simply asks whether the observer state's last bit equals one. Fix some very large number $N\in\N$ (say, one with large Kolmogorov complexity ${\rm K}(N)\gg 1$), and define a modified test $f'$ by
\[
   f'(x):=\left\{
      \begin{array}{cl}
      	   f(x) & \mbox{if }\ell(x)\leq N,\\
      	   1-f(x) & \mbox{if }\ell(x)>N.
      \end{array}
   \right.
\]
The computable test $f$ seems as simple or ``natural'' as properties like ``green'' or ``blue'', whereas $f'$ resembles Goodman's properties ``grue'' or ``bleen'': \emph{``Is the observer young and her last bit equals one, or is she old and her last bit equals zero?''} Now if our observer has seen that $f(x_1)=\ldots=f(x_1 x_2\ldots x_n)=1$, but $n\leq N$, then Theorem~\ref{ThePersistence} applies to both $f$ and $f'$. So what should the observer bet on --- that the last bit switches at some point?

This puzzle is resolved by noting that Theorem~\ref{ThePersistence} gives only an asymptotic statement: it only says that \emph{if $n$ is large enough}, then $f$ (resp.\ $f'$) will yield the answer ``yes`` (i.e.\ $1$) with high probability in the future, if it did in the past. Intuitively, what happens is that the regularity $f=1$ stabilizes itself much faster than the regularity $f'=1$. In particular, if $n=N$, then we expect that $f(x_1 x_2\ldots x_{n+1})=1$ has higher probability than $f'(x_1 x_2\ldots x_{n+1})=1$, since $f$ is a \emph{simpler} computable test, and thus the corresponding regularity statement is preferred by algorithmic probability. Intuitively, the regularity $f'$ would stabilize itself only after $n\gg N$ and if the observer has \emph{in fact} seen her last bit switch. In the notation of the proof of Theorem~\ref{ThePersistence}, we have
\[
   \sum_{n=0}^\infty p(0|1^n)\leq -\log c = \mathrm{K}(\mu)+\mathcal{O}(1)\leq \mathrm{K}(f)+\mathcal{O}(1),
\]
which means that $\sum_{n=0}^\infty p(0|1^n)$ (a measure of the ``total exceptions from the rule'') tends to be smaller if $f$ has a shorter description.

This simple example teaches us two things. First, we should not only look at regularities, but also at their \emph{complexities} --- simple regularities will be more relevant. Second, instead of answering a single yes-no-question, we should try to answer a multitude of questions as parts of some ``web of knowledge''. We would like to show that the regularities fit together to give the observer a coherent notion of a ``world''. This is what we are going to address next.

\subsection{Computable laws and the external process}
\label{SubsecExternalProcess}
Having studied a (rather weak) notion of ``persistence of regularities'' in the previous subsection, it is clear that physics as we know it has much more to offer. For example, regularities in physics are often at the level of the \emph{statistics} rather than in the actual results. In particular in the context of quantum theory, when we talk about ``simple laws of physics'', we have simplicity in a peculiar form: the \emph{probabilistic laws themselves} seem to be simple, but the \emph{individual measurement outcomes} turn out to be complex.

As an example\footnote{This is a rather naive example to illustrate the main idea, not a profound statement about the foundations of quantum mechanics. To study the exact nature of randomness in quantum mechanics, one would have to dive into the subject of interpretations, and also into the field of device-independent randomness amplification~\cite{ColbeckThesis}.}, consider a single quantum spin-$\frac 1 2$-particle (a qubit) and the following experimental setup: the spin is first measured in $Z$-direction, then in $X$-direction, then in $Z$-direction again and so on --- that is, $Z$- and $X$-directions are alternately measured on the single qubit; in total, there are $n$ measurements. Assume for the sake of the argument that the particle starts in a quantum state where the spin points exactly in $X$-direction. Whenever the result is ``spin-up'', it will be denoted by a one, and ``spin-down'' will be denoted by a zero. The result of the measurement is a binary string, consisting of $n$ bits which encode the measurement outcomes.

Denote the eigenstates in $Z$-direction by $|0\rangle$ and $|1\rangle$, and those in
$X$-direction by $|+\rangle=(|0\rangle+|1\rangle)/\sqrt{2}$ and $|-\rangle=(|0\rangle-|1\rangle)/\sqrt{2}$.
The particle starts in state $|+\rangle$. Thus, the first measurement (which is in $Z$-direction)
will yield outcome ``spin-up'' or ``spin-down'' with probability $\frac 1 2$ each.
After that measurement, by the projection postulate, the state of the system will be either $|0\rangle$
or $|1\rangle$. But then, the following measurement in $X$-direction will again yield spin up or
down with probability $\frac 1 2$ each, and so on. According to elementary quantum mechanics,
the resulting string will be completely random (and there are good arguments that this randomness is irreducible and not just ``apparent'' in some sense~\cite{ColbeckThesis,Pironio}).

In the end, the situation is equivalent to $n$ tosses of a fair coin: the $n$ bit values are independently and identically distributed. But a string $x\in\s$ that is generated by such a process is incompressible with high probability~\cite{ZvonkinLevin,Brudno}, i.e.\ its monotone complexity $\Km$ is close to maximal, such that $\Km(x)\approx n =\ell(x)$.

In this example, the rules of quantum mechanics (which yield the outcome probabilities) are very simple, but the outcomes themselves are arbitrarily complex. This is a typical situation in physics. In what follows we will show that this kind of behavior is predicted by the theory of this paper.

Our main technical tool will be the formulation of Solomonoff induction~\cite{LiVitanyi,Hutter} as the following theorem~\cite[Corollary 5.2.1]{LiVitanyi}. It rests on a slightly different interpretation of \emph{measures} $\mu$ as defined in Definition~\ref{DefSemimeasures}: namely, as probability measures in the set of infinite strings $\{0,1\}^\infty$. The idea is that a statement like $\mu(101)=\frac 1 4$ should be understood as saying that the \emph{set of all infinite strings that start with $101$} has measure $\frac 1 4$. Such sets are called \emph{cylinder sets} and denote, as in this example, $[101]$. Then, as subsets of $\{0,1\}^\infty$, we have for example the disjoint union $[10]=[100]\cup [101]$, and consequently $\mu(10)=\mu(100)+\mu(101)$. Formally, the cylinder sets generate a $\sigma$-algebra on $\{0,1\}^\infty$ on which $\mu$ is a probability measure. The measure-theoretic details can be found in~\cite{LiVitanyi}.
\slemma{Solomonoff induction}{LemSolomonoffInduction}{
Let $\mu$ be a computable measure. Then there is a set $S\subseteq\{0,1\}^\infty$ of $\mu$-measure one,
such that for every $x\in S$ and every $b\in\{0,1\}$
\[
   \mathbf{M}(b|x_1^n)\stackrel{n\to\infty}\longrightarrow \mu(b|x_1^n).
\]
Moreover, for all $x\in S$, we have asymptotic normalization:
\[
   \sum_{b\in\{0,1\}} \mathbf{M}(b|x_1^n)\stackrel{n\to\infty}\longrightarrow 1.
\]
}
The arrow is to indicate that the difference between left- and right-hand side turns to zero, not that
both sides converge individually. In this equation, $x=x_1 x_2 x_3\ldots$ is an infinite
binary string, and $x_1^n=x_1 x_2\ldots x_n$ denotes the string of length $n$
that consists of the first $n$ bits of $x$.

What does this mean? Consider a physicist who observes a certain random process which emits random bits
$x_1,x_2,\ldots$. Suppose that the physicist has no idea what random process it is; all she knows
is that there is \emph{some} underlying probability measure $\mu$ that describes the process accurately, and that $\mu$
is computable. Despite her ignorance, she would like to predict the probability of future outcome bits.

This situation is analogous to what scientists are doing when they try to uncover the hidden mechanisms of nature; it is a simple model of science. The lemma above now says that in
the long run, i.e.\ for large $n$, the physicist may simply use the universal semimeasure $\mathbf{M}$ (or, equivalently, the measure $\p$) to predict the probability of the next outcomes. This is Solomonoff induction: algorithmic probability is used as a tool for prediction.

A simple example is given in~\cite{HutterConfirmation,LiVitanyi}: suppose that the unknown process is actually
deterministic and emits only $1$'s, that is, $\mu(1^n)=1$ for all $n\in\N$, where $1^n=111\ldots 1$
is the string of length $n$ consisting only of $1$'s.
Then the probability that Solomonoff induction predicts a $0$ as the next bit
is asymptotically $\mathbf{M}(0|1^{n})=2^{-{\rm K}(n)+\mathcal{O}(1)}$, which is of the order $1/n$
for most $n$. In particular, this converges to the ``correct'' probability zero for large $n$.\label{SomePage}

This formulation of Solomonoff induction will have several applications in the following. As our first application, we will see that it can give us a surprising prediction: that our theory implies the emergence of an abstract notion of ``external world'' in some sense:
\stheorem{Asymptotic computability of statistics}{TheSimpleLaws}{\lineclear
Let $\mu$ be some computable measure. Then,
\[
   \p\left\{\strut \p(b|x_1 \ldots x_n)\stackrel{n\to\infty}\longrightarrow \mu(b|x_1 \ldots x_n)\,\forall b\right\}\geq 2^{-{\rm K}(\mu)};
\]
that is, with probability at least $2^{-{\rm K}(\mu)}$ (which is large if and only if $\mu$ is simple), the actual transition probability $\p$ will in the long run converge\footnotemark\, to the computable measure $\mu$.
}
\addtocounter{footnote}{-1}
\footnotetext{Regarding the speed of convergence, it seems to be a generic phenomenon
that the standard notion of limit (for every $\delta>0$ there is some $N\in\mathbb{N}$ such that
the difference is smaller than $\delta$ for all $n\geq N$) does not yield the strongest or most relevant
notion of convergence in this context. That is, one would expect that even for ``most'' $n<N$, the difference is small already.
This is a consequence of the irregular behavior of Kolmogorov complexity, and can be seen nicely in the
example on page~\pageref{SomePage}: we have $\mathbf{M}(0|1^n)=2^{-{\rm K}(n)+\mathcal{O}(1)}$. This expression
tends to zero, but does so extremely slowly, since there are always astronomically large $n$ with exceptionally small complexity ${\rm K}(n)$.
However, it is close to zero (or, in more detail, to $1/n$) for \emph{most} $n$, since most $n$ have ${\rm K}(n)\approx \log n$. An observer subject to a random process with transition probabilities $\mathbf{M}(b|1^n)$ (or of some other sort where outcomes actually vary) will thus typically not notice the exceptional values of $n$, and see convergence much faster than in the formal $\delta$-criterion. This shows that Theorem~\ref{TheSimpleLaws} features a perhaps too strong form of convergence, and it might be more relevant to ask whether $\mu$-typicality of outcomes might be a ``persistent regularity'' (in the sense of Theorem~\ref{ThePersistence}), for example.
}
That is, in the long run, it happens with probability at least $2^{-\mathrm{K}(\mu)}$ that the conditional version of algorithmic probability converges to the conditional version of the computable measure $\mu$. This probability is larger for \emph{simpler} $\mu$, i.e.\ for ones that have small Kolmogorov complexity $\mathrm{K}(\mu)$ as defined in Theorem~\ref{TheUniversalityM}. Note that despite the computability of the measure $\mu$, the actual infinite sequence of observer states $x_1 x_2 x_3\ldots$ will typically \emph{not} be computable. As a simple example, think of a fair coin tossing process $\mu$: this process has small algorithmic complexity $\mathrm{K}(\mu)$ since it can be implemented with a short program on a monotone Turing machine as described in Theorem~\ref{TheUniversalityM}, but it generates algorithmically complex (random) outcomes with unit probability.

In a nutshell, the correspondence between this theorem and our physical world can be thought of as follows. The string $x_1^n=x_1 x_2\ldots x_n$ denotes the (classical) information held by an observer at some given moment. This is in general an algorithmically complex string of data. The measure $\mu$ describes the probabilities of the different possible data that the observer can acquire successively. This is an algorithmically simple probability measure, because it can in principle be computed by a process with a short computer program: our universe's quantum process (which we believe follows computable laws of short description), supplemented e.g.\ by the Born rule adjusted to the specific observer.

We will discuss the physical interpretation in more detail below. The issue of quantum versus classical probabilities will be discussed in Section~\ref{SecQuantum}.

\proof
Lemma~\ref{LemSolomonoffInduction} and asymptotic normalization imply that there is a Borel set $S\subseteq \s^\infty$ with $\mu(S)=1$ such that $\mathbf{M}(b|x_1^n)-\mu(b|x_1^n)\stackrel{n\to\infty}\longrightarrow 0$ for all $x\in S$. On the other hand, let $\varepsilon_n(x):=1-\sum_{b\in\{0,1\}} \mathbf{M}(b|x_1^n)$, then the same lemma shows that $\varepsilon_n(x)\stackrel{n\to\infty}\longrightarrow 0$ for all $x\in S$, and
\[
   \mathbf{M}(b|x_1^n)\leq \p(b|x_1^n)=\frac{\mathbf{M}(b|x_1^n)}{1-\varepsilon_n(x)},
\]
and so $\p(b|x_1^n)-\mu(b|x_1^n)\stackrel{n\to\infty}\longrightarrow 0$ for all $x\in S$. But for every finite string $x\in\s$, Theorem~\ref{TheUniversalityM} shows that $\p(x)\geq\mathbf{M}(x)\geq 2^{-{\rm K}(\mu)}\mu(x)$. According to elementary measure theory, this inequality must then also be true for $S$, i.e.\ $\p(S)\geq  2^{-{\rm K}(\mu)} \mu(S)$, lower-bounding the probability of the stated event as claimed\footnote{It is tempting to conjecture an alternative proof of Theorem~\ref{TheSimpleLaws} in the following way. Let $p\in\s$ be a minimal program for $\mu$ in the sense of Theorem~\ref{TheUniversalityM}; in particular, $\ell(p)={\rm K}(\mu)$. Consider the set of infinite strings $T:=\{pq\,\,|\,\,q\in\{0,1\}^\infty,\enspace q\mbox{ is Martin-L\"of random}\}$, and the corresponding set $S$ of infinite output strings that are generated by our universal reference machine $U$ if all strings of $T$ are chosen as inputs. Since almost all infinite strings are Martin-L\"of random, we have $\mu(S)\geq 2^{-\ell(p)}=2^{-{\rm K}(\mu)}$, and these output strings should be ``$\mu$-typical'' since the strings in $T$ are, and thus satisfy the property stated in Theorem~\ref{TheSimpleLaws}. However, Exercise 5.2.8 in~\cite{LiVitanyi} shows that the latter assertion is not quite true. Thus, one needs more refined arguments to make this proof idea work. See also the result of~\cite{HutterMuchnik}.
}.
\qed

While I do not know whether the the event in Theorem~\ref{TheSimpleLaws} (i.e.\ convergence to \emph{some} computable measure $\mu$) happens with total probability one, we can interpret it in a way that suggests that some possibly weaker form should always be true. Namely, it seems to say that all regularities which persist according to Theorem~\ref{ThePersistence} will tend to ``fit together'' into a coherent overall lawlike behavior. That is, if the answers to several computable tests $f$ all remain ``yes'', then this can be interpreted as consequences of a single computable statistical law (namely $\mu$) which yields $\mu$-typical outcomes. Since $\mu$-typicality of the sequence of observations is a much weaker statement than the convergence of probabilities in Theorem~\ref{TheSimpleLaws}, it might still hold in all cases even if that strong convergence does not happen with unit probability. We leave a more detailed analysis of this to future work.

In the following, we will assume that the event of Theorem~\ref{TheSimpleLaws} happens, for some $\mu$. This assumption is also natural in the context of the ``desired'' (not yet realized) Postulates~\ref{DesiredPostulates}: if we have a theory that admits processes of ``forgetting'', then the event of Theorem~\ref{TheSimpleLaws} describes a situation in which an observer accumulates more and more information, and fully remembers all its earlier state. Now, the idea is that as long as this process continues, there is a tendency for regularities to stabilize, and for a resulting computable measure $\mu$ to determine the asymptotic transitions. However, if the observer loses most of her memory, then the process will basically restart, and there is yet another chance for asymptotic computability described by some measure $\mu$. If this repeats often enough, then the event of Theorem~\ref{TheSimpleLaws} will eventually happen.

We will now argue that Theorem~\ref{TheSimpleLaws} predicts that observers should expect to see two facts which are features of physics as we know it: first, the fact that there is an \emph{simple, computable, probabilistic external world that evolves in time} (a ``universe''), and second, that this external world seems to have had an absolute beginning in the past (the ``Big Bang'').

Let us start by taking these two features as empirically confirmed facts about our physical world, and look at the ``informational'' consequences
of these facts in the context of algorithmic complexity. A possible analysis (assuming a certain view on the quantum state that we do not need to share)
has been performed by Tegmark~\cite{Tegmark} in a paper with the title \emph{``Does the universe in fact contain almost no information?}

In this paper, Tegmark argues that the universe's quantum state at (or shortly after) the Big Bang has been very simple, in the sense that it had in principle a very short description. Furthermore, there seem to be algorithmically simple laws of nature determining the state's
time evolution. Thus, there should in principle exist a concise complete description of the current quantum state of the universe: simply append a description of the physical laws and a description (in some coordinate system) of the time that has passed since the Big Bang to a description of this initial quantum state. From this, a computationally immensely complex but algorithmically very simple computer program will be able to extract the present quantum state of the universe.

If we continue to accept Tegmark's Everettian view of quantum mechanics for the sake of the argument, then we can argue as follows. If, instead of the full quantum state, we restrict our attention to \emph{observations in typical branches of the wavefunction}, then these observations will nevertheless look very complex, i.e.\ have large Kolmogorov complexity. The reason is very similar to, say a classical coin tossing process. While the process itself has a very short description, the actual sequence of outcomes of, say, $10^9$ coin tosses will typically have large Kolmogorov complexity (about $10^9$).

We can thus reformulate the two empirical facts about our universe in informational terms: observers make observations that are typically complex, but that are nevertheless described by an algorithmically simple evolution of an external world. This external world has the property that its evolution will in general only allow \emph{probabilistic} predictions of future observations.

Let us now see how these observations can be understood as consequences of Theorem~\ref{TheSimpleLaws}. If the event that is described in this theorem happens, then the transition probability $\p$ will converge to a simple computable measure $\mu$,
\[
   \p(b|x_1 \ldots x_n)\stackrel{n\to\infty}\longrightarrow \mu(b|x_1 \ldots x_n),
\]
where ${\rm K}(\mu)$ is likely small. According to the definition in Theorem~\ref{TheUniversalityM} this means the following. Denote the universal reference machine by $U$. Then there is a short computer program (a finite binary string) $q$ of length $\ell(q)={\rm K}(\mu)$ with the property that
\begin{equation}
   \sum_{p:U(qp)=x*} 2^{-\ell(p)} = \mu(x)
   \label{eqSuggest}
\end{equation}
for all finite strings $x\in\s$. That is, the computer program $q$ causes the universal machine $U$ to operate in the following manner:
\begin{itemize}
\item after having read the prefix $q$ from its input, the monotone Turing machine enters a particular mode of computation. In this mode,
it sequentially reads further (random) bits from the input tape (the finite sequence of these bits that have previously been read, at any given time step, is called $p$).
\item The machine does (possibly very complex) computations in its working memory, and
\item sometimes produces a new output $x_i$ on its output tape, building up an output string $x_1 x_2 x_3\ldots$. Since $\mu$ is a measure, the computation will never halt, and produce an infinite sequence of output bits.
\end{itemize}
Attaching the weights $2^{-\ell(p)}$ to the input strings $p$ can be interpreted as supplying independent, identically distributed random bits to the machine $U$ as input. These bits are read by the machine, and processed in a computation which produces outputs from time to time. The outputs are distributed according to $\mu$.
The sequence of output bits $x_1 x_2 x_3,\ldots$ constitutes the observer's (say, Alice's) sequence of observer states, with $x_1^n=x_1\ldots x_n$ her current state. If Alice herself --- or an imaginary bystander --- would like to predict her next state, then there are now two possibilities: first, predict $\p(b|x_1 \ldots x_n)$ directly. But this is certainly hard, given that algorithmic probability is not computable. Second, predict $\mu(b|x_1 \ldots x_n)$ --- which is the probability that the machine $U$ will generate $b$ as its next output bit, in the course of the process that we have just described.

But this allows a much more natural way of prediction: \emph{given all you know (as encoded in $x_1^n$), make a model of the state of the machine $U$ and of its inner workings. Then, use \emph{that} for prediction.} Given that the machine implements a \emph{computable stochastic process}, unfolding according to a simple algorithm of short description, this promises to be a much more plausible endeavor. 
\begin{figure}[!hbt]
\begin{center}
\includegraphics[angle=0, width=9cm]{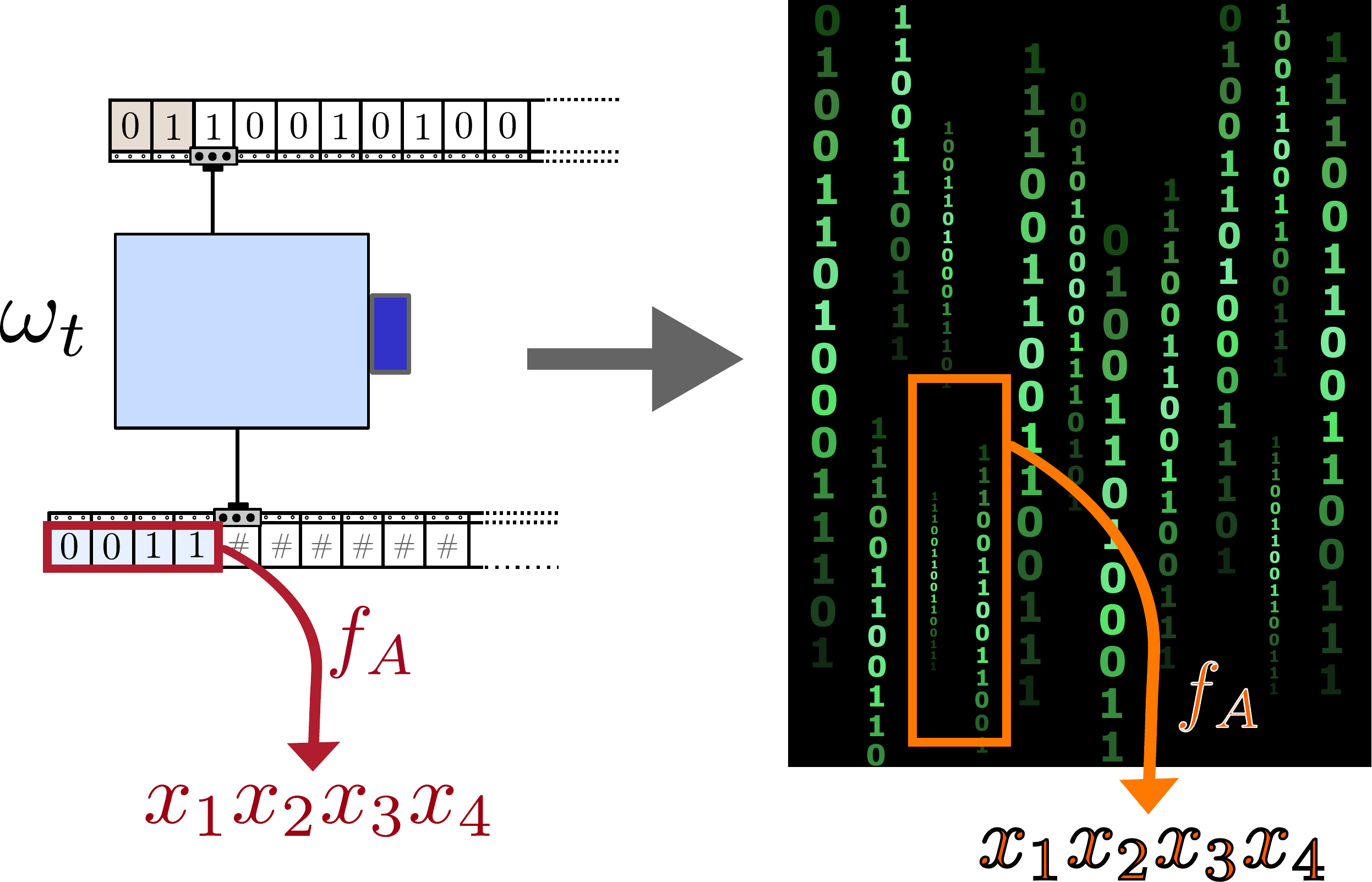}
\caption{As explained in the main text, we can abstract from the concrete monotone Turing machine model. It is irrelevant that we have used this particular model with tapes, internal memory etc.\ in Section~\ref{SecAlgorithmicProbability}, because the set of algorithmic priors is identical for every choice of model of computation. This motivates a more general view: a \emph{computational ontological model} is any computable stochastic process which generates the observer's statistics $\mu$ via some read-out function $f_A$ applied to the process' state $\omega_t$. This includes concrete computations via monotone Turing machines (as on the left), but also other more exotic or abstract computable processes (symbolically sketched on the right).}
\label{fig_gmabstraction}
\end{center}
\end{figure}

Indeed, it is arguably what \emph{human observers} do to predict: given the data that we hold (some of which is collected by our senses), we infer properties of a hypothetical process (the external world) that is not directly ``part of us'', but that turns out to be correlated to our future observations. (``According to the signals that my eyes have just sent me, there seems to be a tiger approaching... I better run...'') 

The situation for Alice is similar: her changes of observer state (her ``experience'') is shaped by the fact that it is consistent to view this state as part of a ``larger'' process, namely of the computational process that would be run if the universal machine $U$ was to produce $\mu$ via the shortest program. Under suitable conditions, Alice may discover an exhaustive or useful approximate description of this external process, and call this her ``external world''.

But does this mean that Alice should see Turing machine tapes in her external world? Is this ``larger'' process unique at all? To see that the answer is ``no'' in both cases, and to discuss what we can expect Alice to see, let us introduce some general terminology.
\sdefinition{Computational ontological model}{DefOntologicalModel}{$\strut$\newline
Consider some computable measure $\mu$ --- for example, the measure $\mu$ that appears in Theorem~\ref{TheSimpleLaws}, as explained above. A \emph{computational ontological model} for $\mu$ is a computable, Markovian, time-discrete stochastic process $\{\omega_t\}_{t\in\N}$, where $\omega_t\in\Omega$ and $\Omega$ is a countable set, together with a computable function $f_A:\Omega\to\s$. We interpret $f_A(\omega_t)$ as the process' \emph{output} at time $t$, and demand that $f_A(\omega_{t+1})$ is either equal to, or one bit longer than, $f_A(\omega_t)$. This gives us a sequence of output bits $x_1=f_A(\omega_{t_1}),x_1 x_2=f_A(\omega_{t_2}),\ldots$, and we demand that these are distributed according to $\mu$.

For example, the monotone Turing machine computation as described above, under~(\ref{eqSuggest}), is a computational ontological model for $\mu$. In this case, the state space $\Omega$ is the set of all possible states of the monotone Turing machine, such that $\omega_t$ includes the input bits that have already been read, the working memory that has already been used, and the output bits that have already been written, and $f_A$ is a function that reads the output tape.
}

The notion of an ``ontological model'' is inspired by Ref.~\cite{HarriganSpekkens,SpekkensContextuality}, who define this notion in the context of quantum mechanics: \textit{``An ontological model is an attempt to offer an explanation of the success of an operational theory by assuming that there exist physical systems that are the subject of the experiment. These systems are presumed to have attributes regardless of whether they are being subjected to experimental test, and regardless of what anyone knows about them.''} When we reason about our physical world, even in every-day life, then we are building an ontological model of a similar sort: given the data in our memory and current observations, our brain creates a model of ``other things'' that are not directly accessible to us, but that may be correlated with our future observations. For example, if we are standing at a street corner, we may reason about a possible car that is about to approach, and use this to predict that we might get hit later on if we move now, even though we do not yet see the car directly.

Moreover, ontological models give us \emph{mechanistic explanations}: for example, we see certain pixels in a picture taken by our telescope \emph{because Jupiter's moon has been moving into our field of sight}. In other words, instead of talking \emph{only} about our observations (or our observer state), an ontological model allows us to understand those observations as consequences of the evolution of variables external to us.

Similarly, a computational ontological model in the sense of Definition~\ref{DefOntologicalModel} allows us (or perhaps the observer herself) to understand the observer's states and their probabilities as consequences of the time evolution of an external computational process. For example, if the model is the monotone Turing machine computation as in the left of Figure~\ref{fig_gmabstraction}, then the arrival of new bits in the observer's state can be mechanistically understood as consequences of the way that the machine's working memory is processed. The observer can thus interpret a suitable computational ontological model as her \emph{external world}:
\sobservation{External world}{ObsExternalWorld}{
If the event of Theorem~\ref{TheSimpleLaws} happens, and a simple computable measure $\mu$ governs the observer states asymptotically, then there exists an algorithmically simple computational ontological model for it in the sense of Definition~\ref{DefOntologicalModel}: the computation of the universal monotone Turing machine, as described under~(\ref{eqSuggest}) above.

The interesting fact is that such a model \emph{exists}, not that it is unique. Indeed, different models of computation (say, monotone Turing machines versus cellular automata, or others) will give different kinds of ontological models. Among those that are algorithmically simple, some will be more ``natural'' or ``useful'' than others. A natural model of this kind can then be interpreted as (an emergent notion of) the observer's \emph{external world}.}
Ontological models cannot be unique. This is even true for our usual understanding of the physical world. For example, we can represent spacetime in many different inequivalent ways (in different coordinates, or via simulation in some computer program), or choose to add pilot waves to our description of quantum theory, without altering any observable predictions. The fact that we describe our world as a $3+1$-dimensional spacetime, and not as a sufficiently accurate simulation on the one-dimensional Turing machine, for example, is merely a matter of utility and not of conflicting predictions. If we have this freedom in orthodox physics, then we should certainly expect to find it in the approach of this paper, too. Whenever there is such a freedom, it makes sense to focus on ontological models that are ``natural'' or ``useful'' (as we do in physics and every-day life), but it may remain difficult to formalize exactly what we mean by that.

In the approach of this paper, it is the set of algorithmic priors, derived from the \emph{set of universal mixtures} (Definition~\ref{DefUniversalMixture}), that determines what observers see. Universal mixtures are defined abstractly, without reference to monotone Turing machines. Lemma~\ref{LemUniv2} shows that this set is equal to the set of semimeasures generated by universal monotone Turing machines, which is ultimately the reason why we can have a simple computational ontological model in terms of such machines. But we could equally well have started with a different computational model (say, cellular automata of some kind). Doing so generates other computational ontological models, admitting the same kind of abstract computational processes, but representing them differently.

Indeed, our definition of a computational ontological model is very broad. It also includes, for example, \emph{quantum computations} like the process generated by a quantum Turing machine. This is explained in Example~\ref{ExQTM} in the appendix, where we also give a few further comments.

Since monotone Turing machines do not necessarily generate the most natural computational ontological model for some given measure $\mu$, our observer Alice should not expect to see Turing tapes or other machine-specific phenomena in her external world. However, there are some model-independent typical properties of abstract computational processes that Alice arguably \emph{can} expect to see. One of them is the fact that computations initially start in a simple state (at least if the computer program has small algorithmic complexity, which is the case here), and compute forever without halting. This is because we are considering measures $\mu(x_1x_2x_3\ldots)$ on one-sided infinite bit strings. To be computable, i.e.\ simulatable by a monotone Turing machine, the process has to start in \emph{some} initial state\footnote{For every model of computation, the initial state $\omega_1$ must be the same for all possible measures $\mu$, since the formulation of Theorem~\ref{TheSimpleLaws} counts all input bits that depend on $\mu$ as contributing to ${\rm K}(\mu)$. But if $\omega_1$ was complex, then we would obtain an arguably very unnatural machine model. For example, it could correspond to the Turing machine model, but always starting with an arbitrary, fixed, finite, very complex binary string written on its working tape. Such a model would not be very useful for the observer, or to reason about the observer, similarly as superdeterministic models of physics which put the outcomes of all quantum experiments into the initial state of the universe are not very useful to us.} $\omega_1$; and to generate all bits, it has to run forever.

Furthermore, let us assume that $\mu$ is in some sense a \emph{typical} simple computable measure. For example, typical measures will not be deterministic (like $\mu(0^n)=1$ for all $n\in\N$), but have entropy of the distribution $\mu(x_1x_2\ldots x_n)$ that grows with $n$. Then, the corresponding process will tend to look more and more complex over the course of the computation. Indeed, the random input bits (as described under eq.~(\ref{eqSuggest}) above) will increase the entropy of the machine's state over time. Moreover, inspection of many simple example, e.g.\ cellular automata~\cite{WolframBook}, shows that even deterministic computations often exhibit increasing apparent complexity over time. For some ideas of how to prove rigorous versions of this observation in terms of complexity measures like Kolmogorov minimal sufficient statistics, see e.g.~\cite{Wolf}.

Hence, these common features of computational processes will also be present in the process that corresponds to Alice's external world. Since this world is governed by a short computer program, its time evolution can in principle be described by concise ``laws of nature''. Suppose that Alice can obtain some sufficiently good approximation to those laws and to her world's current state\footnote{Whether this is \emph{actually} possible for Alice is a question of epistemology that we do not intend to address in this paper. The answer will depend on Alice's characteristics as an observer, on her computational and experimental abilities, on the details of how she is embedded into her external process, and on further characteristics of the process. Even in our physical world, some observers (say, guinea pigs or very simple informational structures that we can view as observer states) will not be able to obtain such information (they may not even try), while others (we human observers) \emph{think} to hold it but can never be completely sure. Indeed, this fact reflects well-known general obstructions to the unambiguous learning of computable laws from a finite amount of data, see e.g.~\cite{ZeugmannZilles}. The aim of this section is not to argue that Alice can discover these laws in practice, but that there \emph{exist laws} waiting for discovery in the first place.}. Not only will this allow Alice to statistically predict her future observations, but she may also ``calculate backwards'' and retrodict her world's earlier states. This may include internal process times that have been before the machine has produced any output; in this sense, before what might be called Alice's ``birth''.

Extrapolating backwards far enough will lead to earlier and earlier stages of an unfolding computation and thus to simpler and more ``compact'' (in particular, less entropic) states. Finally, this will lead to the initial state of the machine's computation (right after the machine $U$ has read the prefix $q$), where simplicity and compactness are maximal. Alice may call this initial state the ``Big Bang'', and hypothesize that the world had its beginning in this moment. This is broadly consistent with our actual physical observations. We will discuss this further in Subsection~\ref{SubsecBB}.

At the same time, our approach claims that this appearance of an external world is ultimately not fundamental: according to Postulates~\ref{Postulates}, what \emph{actually} determines Alice's future observations is conditional algorithmic probability. In particular, her observations do not fundamentally supervene on this ``physical universe''; it is merely a useful tool (an ontological model) to predict her future observations. Nonetheless, this universe will seem perfectly real to her, since its state is correlated with her future experiences.  If the measure $\mu$ that is computed within her computational universe assigns probability close to one to the experience of hitting her head against a brick, then the corresponding experience of pain will probably render all abstract insights into the non-fundamental nature of that brick irrelevant.

This view suggests some sort of effective \emph{pancomputationalism}~\cite{Pancomp}: our (emergent external) world \emph{is} in some sense a computation. However, Section~\ref{SecQuantum} will add an interesting twist to this: if some version of our theory applies that involves memory loss (Postulates~\ref{DesiredPostulates}), then there will be an extra layer ``between us and the world'', hiding some degrees of freedom, and leading us to an effective description that displays some features of quantum theory. This will imply that there is no direct correspondence between the causal structure of our phenomenal world and the causal structure of the computational process. This is in contrast to other pancomputationalist approaches, like the ones by Zuse~\cite{Zuse}, Schmidhuber~\cite{Schmidhuber}, 't Hooft~\cite{tHooft} or Lloyd~\cite{Lloyd}, which claim a one-to-one mapping between modules of the computation (such as ``gates'' or ``tape cells'') and space-time regions.

\section{The rise and fall of objective reality}
\label{SecObjectivity}
\sectionmark{The rise and fall of objective reality}
\textit{``How come `one world' out of many observer-participants?''} (J.\ A.\ Wheeler~\cite{WheelerInformation}).\\

In Section~\ref{SecSimpleLaws}, we have seen that our approach predicts some features of physics as we know it: with high probability, observers see simple probabilistic ``laws of nature'', and find themselves to be part of an external world that they may call ``the universe''. However, there is one further crucial aspect of physics that is \emph{not} a priori true in our theory: namely that \emph{different observers see the same physical world}.

\begin{figure*}[!hbt]
\begin{center}
\includegraphics[angle=0, width=15cm]{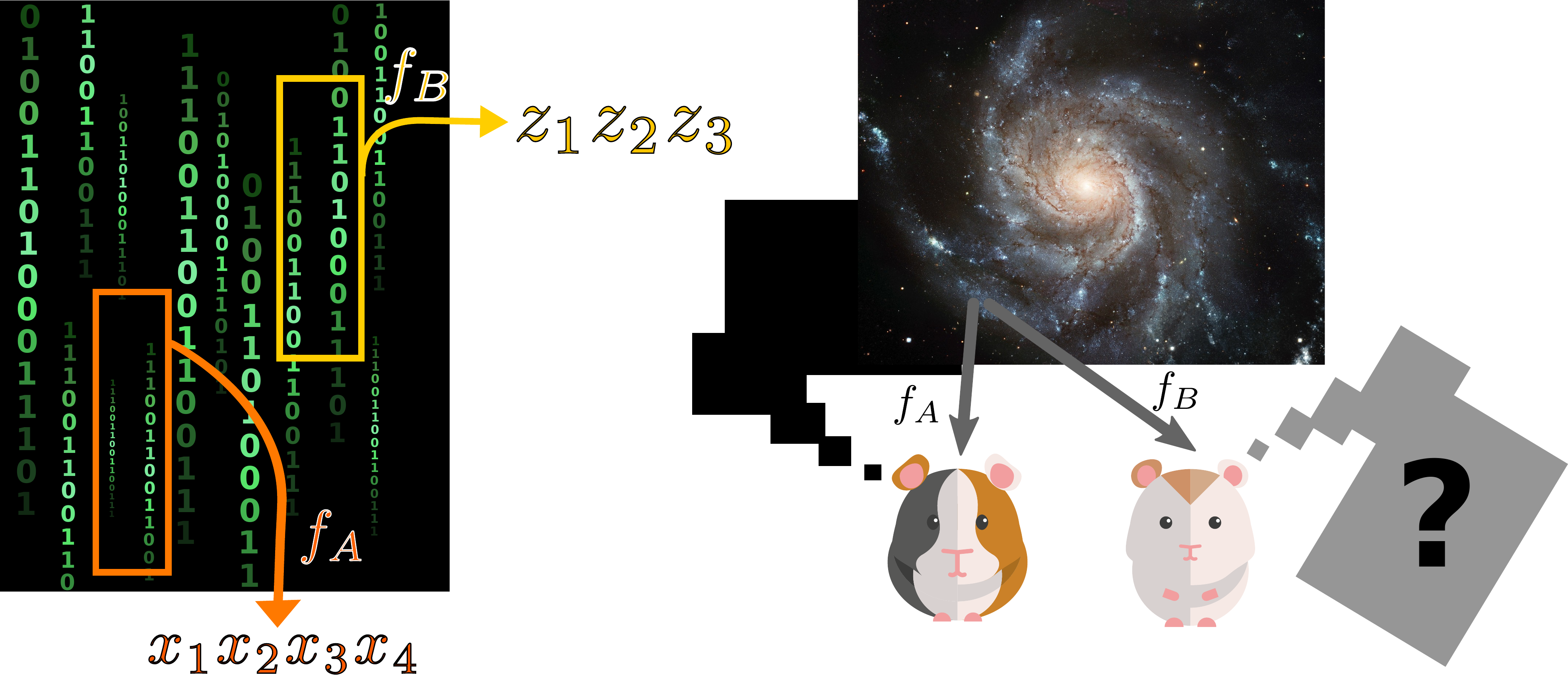}
\caption{Informal illustration of the setup considered in this section. We have an observer $A$ (Alice) who finds herself to be part of a simple computational process which generates some measure $\mu$ (according to Theorem~\ref{TheSimpleLaws}). This means that the computational process is what she may call her ``external world'' as explained in Section~\ref{SecSimpleLaws}; her observer state (here e.g.\ $x=x_1 x_2 x_3 x_4$) is a function $f_A$ of the process' state (see also Figure~\ref{fig_gmabstraction}). Suppose that there is another simple computable function $f_B$, acting on the states of this process, which yields another string of bits that grows over time. Then Alice can interpret this as ``having another observer Bob$_{3\rm rd}$ in her world'', and her world will yield a probability measure $\p_{3\rm rd}$ determining what is going to happen to Bob$_{3\rm rd}$ in her world. However, Bob's first-person perspective, Bob$_{1\rm st}$, is governed by algorithmic probability, $\p_{1\rm st}\equiv\p$; and these probabilities may a priori be completely unrelated. In other words, what Bob$_{1\rm st}$ will probably really see next (symbolized by the grey speech bubble) may be very different from \emph{what Alice will probably be seeing Bob$_{3\rm rd}$ see next}. But as we show in Theorem~\ref{TheNoZombies} below, for a large number of bits learned by Bob, the conditional probabilities $\p_{1\rm st}$ and $\p_{3\rm rd}$ will be very close to each other --- in this sense, Alice and Bob will be ``part of the same world'', and Bob$_{3\rm rd}$ will be a probabilistically faithful ``representation'' of Bob$_{1\rm st}$. This is a probabilistic form of emergent objective reality.}
\label{FigObjectivity}
\end{center}
\end{figure*}

In fact, it is not even clear what it means to talk about ``different observers'' in our context. Postulates~\ref{Postulates} only talk about \emph{observer states}, not about ``observers'' as physical objects of which you could have two or three. In principle, we could have a single observer --- ``the mind'' --- taking a never-ending random walk on the observer states, with transitions determined by algorithmic probability. Or could we? What would this claim even mean? Is it not empirically obvious that there are \emph{other observers} out there, not just \emph{I}?

We certainly believe that there are other observers because we encounter things in our world that seem to hold first-person perspectives, too: other humans, for example. Let us now discuss and formalize what this means in our framework. Suppose that an observer --- say, Alice the guinea pig --- finds herself located in some simple, computable, probabilistic external world (characterized by a computable measure $\mu$ on her observer states), as derived in detail in Section~\ref{SecSimpleLaws}. 
 
Suppose that Alice encounters another guinea pig, called Bob$_{3\rm rd}$, in her external world. The subscript indicates that this describes a ``third-person perspective'': Alice points to something (an object, a pattern) in her world that she calls ``Bob''. Moreover, Alice can in principle consider the information that is stored and processed in Bob$_{3\rm rd}$'s brain, and reason about how this information changes in time. Even if it may be impossible for Alice (or unethical) to open Bob$_{3\rm rd}$'s brain and read out all this information in practice, Alice can still argue that it contains \emph{some} information that changes over time, in accordance with the evolution of Alice's world. If this evolution is probabilistic, then there will be an induced probability distribution $\p_{3\rm rd}$ that describes the distribution of Bob$_{3\rm rd}$'s brain states one moment after the other.

But now, if this momentary information content corresponds to a finite binary string $x$, then our approach enables us to regard this string as an \emph{observer state}. In other words, we can think of the corresponding first-person perspective that describes ``what it is like'' to be in state $x$, and think of a corresponding abstract observer Bob$_{1\rm st}$ --- the actual ``mode of being'' in this state\footnote{Note that Bob$_{1\rm st}$ is not a well-defined ``object'' or variable, but a handy way to talk about the first-person perspective corresponding to $x$.}.

Typically, Alice will expect that Bob$_{3\rm rd}$ and Bob$_{1\rm st}$ are in some sense ``identical''. For example, when she observes Bob$_{3\rm rd}$ expressing a feeling of happiness on seeing Alice (together with the corresponding neural correlates), she will assume that Bob$_{1\rm st}$ really \emph{does} have that feeling in some sense. But is this really true? Given that our theory admits that Bob$_{3\rm rd}$ and Bob$_{1\rm st}$ are in some sense different things, is such an identification possible and meaningful?

We certainly cannot formally analyze what it means to ``feel'' something, but the question just asked has an obvious formal counterpart: namely, we can compare two probability measures. On the one hand, we have $\p_{3\rm rd}$, the probability induced by Alice's world on Bob$_{3\rm rd}$'s fate; on the other hand, we have algorithmic probability, $\p_{1\rm st}:=\p$ that determines Bob$_{1\rm st}$'s actual first-person chances according to Postulates~\ref{Postulates}. Is there any relation between the two?

To put these considerations into a more concrete form, consider the question whether the sun rises tomorrow. Suppose that Alice has gathered enough information about her external world, and about the physical measure $\mu$, to know that there is a probability close to one that the sun is going to rise tomorrow. Thus, Alice will have a close to 100\% chance of \emph{seeing Bob$_{3\rm rd}$ see the sun rise tomorrow}. But what is Bob$_{1\rm st}$'s actual chance to see the sun rise tomorrow, from his first-person perspective?

\subsection{Asymptotic coherence of $\p_{\rm 1st}$ and $\p_{\rm 3rd}$}
\label{SubsecAsymptoticCoherence}
Let us go into more formal detail. Consider Alice's external world, as described in Observation~\ref{ObsExternalWorld}: a computational ontological model in the sense of Definition~\ref{DefOntologicalModel}, and as depicted in Figure~\ref{fig_gmabstraction}. That is, we have a computational process together with a function $f_A$ that ``reads out'' Alice's current state, and subsequently generates her sequence of observer states.

But now, let us consider some other variable in this process that we will call Bob$_{3\rm rd}$, read out by some \emph{other} ``locator function'' $f_B$. It can be any variable whatsoever, as long as it satisfies the minimal requirements to be interpreted as generating a sequence of observer states:
\emptysassumption{AssObjectivity}{Consider observer Alice whose states are asymptotically governed by a computable measure $\mu$ as in Theorem~\ref{TheSimpleLaws}. Fix an arbitrary computational ontological model, i.e.\ a stochastic process $\{\omega_t\}_{t\in\N}$, $\omega_t\in\Omega$, and function $f_A:\Omega\to\s$ that reads out Alice's state, with all the properties given in Definition~\ref{DefOntologicalModel} (for example the model that Alice interprets as her external world).

Let $f_B:\Omega\to\s$ be another computable map; we interpret it as a ``locator function'' that reads out the state of Bob$_{3\rm rd}$. We assume that $f_B$ always yields valid sequences of observer states. That is, $f_B(\omega_{t+1})$ is either equal to or one bit longer than $f_B(\omega_t)$; and, in addition, for every $t$ there will be $t'>t$ such that $f_B(\omega_{t'})\neq f_B(\omega_t)$.
}
If this assumption is satisfied, then the process will generate, via $f_B$, an infinite sequence of bits $z_1,z_2,\ldots$. At any computational time $t$, we interpret $z:=f_B(\omega_t)=z_1 z_2\ldots z_n$ as an observer state, where $n$ grows with $t$. Since this is a random variable, the process generates a well-defined probability measure (in the sense of Definition~\ref{DefSemimeasures}) $\p_{\rm 3rd}$. Intuitively, the distribution $\p_{\rm 3rd}$ tells us the probabilities of what happens to Bob$_{\rm 3rd}$ within Alice's world.

However, there is another probability distribution of relevance in our framework: the algorithmic prior $\p_{\rm 1st}\equiv \p$ for which Postulates~\ref{Postulates} claim that it determines what actually happens to Bob$_{1\rm st}$ from his first-person perspective.

In the example above, if there is an almost 100\% chance that Alice will see that Bob$_{3\rm rd}$ sees the sun rise tomorrow, then this is a probability assignment of $\p_{3\rm rd}\approx 1$. If we ask what Bob$_{1\rm st}$ will actually see, then this asks for the corresponding value of $\p_{1\rm st}$.

A priori, both probabilities can take different values. However, if they \emph{are} in fact different, then we have a quite strange situation, reminiscent of Wittgenstein's philosophical concept of a ``zombie''~\cite{SEP}: Bob$_{1\rm st}$ would in fact not observe what Alice sees Bob$_{3\rm rd}$ observe, but would divert into his own ``parallel world'' with high probability. This does not mean that Alice will subsequently be confronted with a ``soulless'' Bob$_{3\rm rd}$ (since $f_B$ will still produce an observer state, associated with \emph{some} first-person perspective); it would somehow, very roughly, mean that Alice is confronted with some sort of ``very unlikely instance'' of Bob$_{\rm 1st}$, and that the Bob$_{\rm 1st}$ that she knew earlier has somehow subjectively ``fallen out of the universe''. It is probably save to say that we lack both intuition and terminology to describe non-mathematically what that would mean\footnote{Note that this would be much stranger than the simple effect of having different ``computational branches'', following different values that the random variable $\omega_t$ can take. Similarly as in Everettian interpretations of quantum mechanics, a ``many-worlds''-like picture suggests that we should imagine different ``instances'' of Alice and Bob, following the different branches. Nevertheless, if Alice and Bob meet in one branch of an Everettian world, they will both be subject to the same  objective chances of joint future observations (like seeing the sun rise tomorrow). For ``probabilistic zombies'' as just described, this would not be the case.}.

As we will now see, the good news is that the properties of algorithmic probability imply that this strange phenomenon will not typically happen in every-day situations\footnote{Strictly speaking, this kind of consistency (as expressed in Theorem~\ref{TheNoZombies}) would not \emph{necessarily} have to hold in order to have a well-defined theory; physics would still make sense in a solipsistic world in which every observer is surrounded by probabilistic zombies. But such a world would be truly terrifying.} (but see Subsection~\ref{SubsecZombies}). Instead, $\p_{1\rm st}$ and $\p_{3\rm rd}$ \emph{will} be very close to each other under natural circumstances:
\stheorem{Emergence of objective reality}{TheNoZombies}{\lineclear
In the setting of Assumption~\ref{AssObjectivity}, the probabilities $\p_{3\rm rd}$ that determine the fate of Bob$_{3\rm rd}$ within Alice's external world are asymptotically close to the actual chances $\p\equiv \p_{1\rm st}$ of Bob$_{1\rm st}$'s first-person perspective. That is, with $\p_{3\rm rd}$-probability one,
\begin{equation}
   \p_{3\rm rd}(y|z_1 z_2\ldots z_n)\stackrel{n\to\infty}\longrightarrow \p_{1\rm st}(y|z_1 z_2\ldots z_n),
   \label{eqNoZombie}
\end{equation}
i.e.\ the difference between the conditional versions of $\p_{\rm 3rd}$ and $\p_{\rm 1st}$ tends to zero.}
This theorem follows directly from applying Solomonoff induction (Lemma~\ref{LemSolomonoffInduction}), noting that Assumption~\ref{AssObjectivity} implies that the measure $\p_{\rm 3rd}$ is \emph{computable}.

In this sense, Alice and Bob ``inhabit the same world'' --- Bob$_{\rm 3rd}$ as encountered by Alice is a faithful representation of an actual first-person perspective of some Bob$_{\rm 1st}$. Note that this theorem is formulated \emph{from Alice's perspective}: it is Alice who assigns $\p_{3\rm rd}$-probability one to convergence.

At first sight, this seems to resemble the idea of \emph{Bayesian consistency}~\cite{DiaconisFreedman}: if two agents start with different prior distributions, but receive equivalent data, their Bayesian posterior distributions will in many cases converge towards each other. In this view, both agents are by definition part of the ``same world'' such that they receive data which is in principle compatible between the two, and the prior and posterior distributions represent their beliefs. In the approach of this paper, however, this not the case: $\p_{\rm 3rd}$ and $\p_{\rm 1st}$ are not beliefs but \emph{actual chances}, and observers are \emph{not} from the outset assumed to be part of a joint world.

\subsection{Probabilistic zombies}
\label{SubsecZombies}
Theorem~\ref{TheNoZombies} shows in what sense our theory predicts the \emph{emergence of objective reality}: while the fundamental ontology is given
by each observer's first-person perspective, there is nevertheless a tendency for observers to agree that they see a specific objective ``external world'': there is a \emph{single} computational ontological model (up to the locator function) that works for both. However, this theorem relies on two premises as formalized in Assumption~\ref{AssObjectivity}:
\begin{itemize}
\item \textbf{Bob is ``old/complex enough'':} the length $n=\ell(z)$ of Bob's observer state $z$ must be large. 
\item \textbf{Bob ``survives and remembers forever'':} the locator function $f_B$ that is supposed to read out Bob's state from Alice's world will always yield a consistently growing observer state, even in the very distant future.
\end{itemize}
While it seems plausible that both assumptions are satisfied \emph{approximately} in typical situations, they will not hold in all cases. In this subsection and the next, we will have a closer look at what happens if we drop these two assumptions. Let us start by dropping the first of the two:
\sobservation{Probabilistic zombies}{ObsYoungZombies}{
In the notation of Theorem~\ref{TheNoZombies}, the probabilities $\p_{3\rm rd}$ that determine the state of Bob$_{3\rm rd}$ in Alice's world, and $\p_{1\rm st}\equiv \p$ that determine Bob$_{1\rm st}$'s actual first-person chances, will in general be very different if the length $n$ of Bob's observer state $z$ is small.

If this is the case, we will say that Bob$_{3\rm rd}$ is a ``probabilistic zombie'' for Alice. In particular, this will be the case if Bob's current state $z$ is \emph{too simple}, namely if ${\rm K}(z)\ll {\rm K}(\p_{3\rm rd})$.
}
As explained in Section~\ref{SecObjectivity}, this notion of ``probabilistic zombie'' vaguely resembles Wittgenstein's notion of a zombie~\cite{SEP}, but it is on the one hand more precise and on the other hand less intuitive. From Alice's perspective, it means that Bob$_{\rm 3rd}$ does \emph{not} faithfully (in probability) represent the actual first-person perspective of some corresponding Bob$_{\rm 1st}$. See the previous subsection for a more detailed explanation.

Since the complexity ${\rm K}(z)$ tends to grow with $n=\ell(z)$, the statement that ${\rm K}(z)$ is small can roughly be interpreted as saying that ``Bob is still young''. But this intuition should be taken with a grain of salt, since there is no monotonous relationship between length and complexity.

In more detail, ${\rm K}(z)$ has to be compared to ${\rm K}(\p_{\rm 3rd})$, the complexity of the measure $\p_{\rm 3rd}$ as defined in Theorem~\ref{TheUniversalityM}. The latter is the length of the shortest program that generates $\p_{\rm 3rd}$ an a universal computer --- in other words, it is \emph{the description length of the probabilistic laws of Alice's world, together with a description of Bob$_{\rm 3rd}$'s location $f_B$}. We can thus interpret the quantitative statement in Observation~\ref{ObsYoungZombies} as follows: \emph{If Bob$_{\rm 3rd}$'s complexity is much smaller than the number of bits it takes to describe the laws of physics in Alice's world, \emph{and} to locate Bob$_{\rm 3rd}$ inside that world, then Bob$_{\rm 3rd}$ is a probabilistic zombie for Alice.}

We will not formally prove this quantitative statement, but give some intuition as to why it represents a reasonable conjecture. To this end, let us return to Theorem~\ref{TheNoZombies} which proves the asymptotic emergence of objective reality. Intuitively, if $n$ is large, then Bob's state $z=z_1 z_2 \ldots z_n$ contains enough information to \emph{infer via universal induction, without reasonable doubt, that this data has been generated by Alice's world}. In other words, a description of Alice's world, and thus of $\p_{\rm 3rd}$, can be obtained from $z$. If this can be done via some algorithm, then we could conclude that
\[
   {\rm K}(\p_{\rm 3rd}) \lesssim {\rm K}(z).
\]
Therefore, if this inequality is very strongly violated, then we expect that $\p_{\rm 3rd}\not \simeq \p_{\rm 1st}$. Let us go into some more detail. Consider some enumeration $m_1, m_2, m_3,\ldots$ of the enumerable semimeasures, and define
\begin{equation}
   \mathbf{M}_V(w):=\sum_{j=1}^\infty 2^{-{\rm K}(j)} m_j(w)\quad (w\in\s\setminus\{\varepsilon\}).
   \label{eqMPrime}
\end{equation}
This is well-defined since $\sum_j 2^{-{\rm K}(j)}\leq 1$~\cite{LiVitanyi}. Since ${\rm K}(j)$ can be computably estimated from above, $\mathbf{M}_V$ is a universal mixture in the sense of Definition~\ref{DefUniversalMixture}. Hence, by Lemma~\ref{LemUniv2}, there is a universal monotone Turing machine $V$ for which this quantity is indeed equal to $\mathbf{M}_V$, justifying the notation. It follows that
\begin{equation}
   \mathbf{M}_V(y|z)=\sum_{j=1}^\infty 2^{-{\rm K}(j)}\frac{m_j(z)}{\mathbf{M}_V(z)} m_j(y|z).
   \label{eqMPrimeConditional}
\end{equation}
Starting with this equation, emergence of objectivity as in Theorem~\ref{TheNoZombies} can be interpreted intuitively (but not fully rigorously) as follows. Bob's first-person probability $\p\equiv\p_{\rm 1st}\approx \mathbf{M}$ is a mixture of all enumerable semimeasures as in~(\ref{eqMPrimeConditional}). It is as if there was a ``correct'' computable measure $\p_{\rm 3rd}$ which is not known to Bob, and hence Bob holds a prior $\mathbf{M}_V$, i.e.\ a prior distribution over all computable (and merely enumerable) measures (and semimeasures). On receiving new data $z$, Bob updates his beliefs to $\mathbf{M}_V(\cdot|z)$. At some point, it turns out that a single term of the infinite (sub-)convex combination dominates: namely, $m_j=\p_{\rm 3rd}$. This can only happen if
\begin{itemize}
\item $\p_{\rm 3rd}(z)$ is not too small, i.e.\ Bob's current state is ``typical'' for this distribution;
\item ${\rm K}(j)$ is not too large, i.e.\ it is easy to pick $\p_{\rm 3rd}=m_j$ from the enumeration of semimeasures (intuitively, this amounts to $\p_{\rm 3rd}$ itself being simple, i.e.\ we expect that ${\rm K}(\p_{\rm 3rd})\approx {\rm K}(j)$ is also not too large).	
\item On the other hand, all other $m_i$ for $i\neq j$ must either be complex in comparison (i.e.\ ${\rm K}(i)\gg {\rm K}(j)$) or represent ``worse explanations'' of Bob's current state than $\p_{\rm 3rd}$, in the sense that $m_i(z)\ll \p_{\rm 3rd}(z)$.
\end{itemize}
But if ${\rm K}(z)\ll {\rm K}(\p_{\rm 3rd})$, then such a ``concentration'' on $m_j=\p_{\rm 3rd}$ cannot happen. To see why, note that from a description of $z$ we can construct a simple algorithm that generates a computable measure $\mu$ with $\mu(z)=1$, and which has some arbitrary values for extensions of $z$ (say, ones that differ from $\p_{\rm 3rd}$). If $i$ is the simplest index such that $m_i=\mu$, we will thus expect that
\[
   {\rm K}(i)\approx {\rm K}(\mu)\lesssim {\rm K}(z)\ll {\rm K}(\p_{\rm 3rd}) \approx {\rm K}(j).
\]
Therefore, $\mu=m_i$ will have much higher weight in~(\ref{eqMPrimeConditional}) than $\p_{\rm 3rd}=m_j$, and emergence of objectivity will fail: Bob$_{\rm 3rd}$ will be a probabilistic zombie.

\subsection{Subjective immortality}
\label{SubsecImmortality}
\begin{figure*}[!hbt]
\begin{center}
\includegraphics[angle=0, width=13cm]{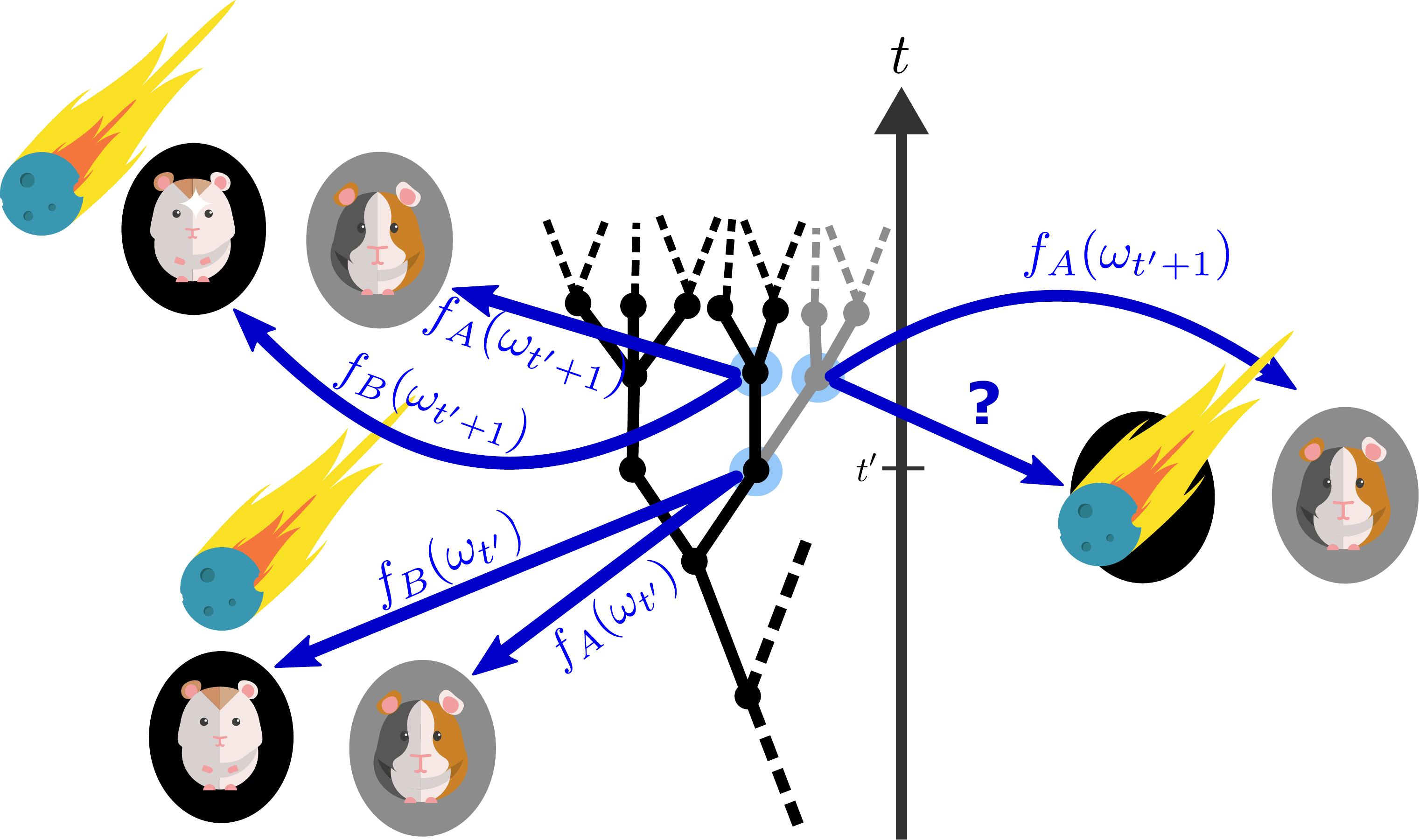}$\qquad$
\caption{Colorful and (hopefully) humorous illustration of the content of Observation~\ref{ObsSubjectiveImmortality2}. The tree represents the possible histories of a computational ontological model that corresponds to Alice's external world, as explained in Subsection~\ref{SubsecExternalProcess}. The possible values of the random variable $\omega_t$ (the state of the computational process at time $t$) correspond to the vertices of a tree graph, directed upwards. The subset of vertices for which $f_B(\omega_t)$ yields a growing (in $t$) sequence of bit strings is colored in black. However, there are some computational histories (in gray) that do not satisfy this constraint. The bottom left shows Alice and Bob who are happy and alive and observe an approaching meteorite. Their states are given by $f_A(\omega_{t'})$ and $f_B(\omega_{t'})$, respectively; that is, they are in this sense part of the computational process. They both ``share the same world'' in the sense of emergent objective reality --- this basically corresponds to the scenario of Theorem~\ref{TheNoZombies}. Then, this external process will transition probabilistically into one of two possible scenarios $\omega_{t'+1}$: first, a meteorite that has previously approached unfortunately hits Bob$_{\rm 3rd}$, which happens with probability $99\%$ (right-hand side); second, the meteorite fortunately misses Bob$_{\rm 3rd}$ (left-hand side, top), which happens with probability $1\%$. While both scenarios are possible for Alice, the postulates of our theory say, however, that this is not true for Bob$_{\rm 1st}$: according to Postulates~\ref{Postulates}, Bob$_{\rm 1st}$ must transition into some other observer state next. Therefore, the unfortunate meteorite-hitting branch has no relevance for Bob$_{\rm 1st}$. This is formally reflected by the fact that the distribution generated by $f_B$ will be a \emph{semimeasure}, not a measure. Instead of the termination by the meteorite, something else will happen to Bob$_{\rm 1st}$ --- but what this will be is a question that cannot be answered within Postulates~\ref{Postulates}. The answer will have to wait for a formalization of Postulates~\ref{DesiredPostulates}.
}
\label{fig_subjimm}
\end{center}
\end{figure*}

The next part of our analysis will be to see what happens if we drop the assumption that ``Bob survives and remembers forever''. We all have an intuition for such situations --- a possible scenario is (boldly and humorously) illustrated in Figure~\ref{fig_subjimm}. Let us work within the framework of Subsection~\ref{SubsecAsymptoticCoherence}: In Alice's external world, there is another observer Bob$_{\rm 3rd}$, characterized by some simple locator function $f_B$. Let us assume that, indeed, the probabilities $\p_{\rm 1st}(y|x)$ and $\p_{\rm 3rd}(y|x)$ have been close to another for quite a while, such that Bob$_{\rm 3rd}$ is \emph{not} a probabilistic zombie for Alice.

In the scenario of Figure~\ref{fig_subjimm}, the locator function $f_B$ generates a growing sequence of observer states, as demanded by Assumption~\ref{AssObjectivity}. However, if we interpret the hitting meteorite as an act of memory erasure, then $f_B$ will \emph{not} have this property any more on the ``gray'' branch of the computational tree, for times $t>t'$: that is, $f_B(\omega_{t'+1})$ will neither be equal to $f_B(\omega_{t'})$ nor be one bit longer, but will perhaps (depending on the definition of $f_B$) yield the empty string or some other junk.

Clearly, we can somehow modify or extend $f_B$ to the terminating branch, such that the resulting map (say, $\tilde f_B$) will keep on satisfying Assumption~\ref{AssObjectivity}. But the point is that there is no unique, ``natural'' way to define such an extension in general. Consequently, there will be many possible $\tilde f_B$, and since they are all approximately on ``equal footing'', each one of them will be more complex than $f_B$. But we have assumed that $f_B$ is simple, in order to have an emergent notion of objectivity in the first place.

Recall the construction of $\p_{\rm 3rd}$ from Subsection~\ref{SubsecAsymptoticCoherence}: Alice's probabilistic computational world generates a corresponding measure on the observer states as read out by $f_B$, i.e.\ on the states of Bob$_{\rm 3rd}$. If we repeat this construction now, without relying on the validity of Assumption~\ref{AssObjectivity}, then we can still obtain the measure $\p_{3\rm rd}$ up to computational time $t'$. But from time $t'+1$ on, some computational branches are ``cut off'' by not supplying any valid new observer states for Bob. Thus, instead of a measure, we obtain a \emph{semimeasure} $\p_{\rm 3rd}$.\\

While Theorem~\ref{TheSimpleLaws} is not valid for computable \emph{semimeasures}, its non-asymptotic version from Subsection~\ref{SubsecZombies} still applies: $\mathbf{M}_V(y|z)$ in~(\ref{eqMPrimeConditional}) can still concentrate on the semimeasure $m_j=\p_{\rm 3rd}$ if ${\rm K}(j)$ is small (which it is if Alice's world has simple laws) and if $m_j(z)$ is large:\\

\footnotetext{It is sufficient that most of the probability weight is distributed on semimeasures that make the same or very similar predictions as $\p_{3\rm rd}$ for the first few relevant states.}
\addtocounter{footnote}{-1}
\sobservation{Subjective immortality}{ObsSubjectiveImmortality2}{The scenario in Figure~\ref{fig_subjimm} can be understood as follows. At (not too early) computational times $t<t'$, there is ``emergent objectivity'': the probabilities $\p_{3\rm rd}$ that determine the chances of Bob$_{3\rm rd}$ as seen by Alice in her external world are close to the actual chances $\p\equiv\p_{1\rm st}$ of Bob$_{1\rm st}$'s state transitions. This is a finite-time version of Theorem~\ref{TheSimpleLaws}.

This happens whenever~(\ref{eqMPrimeConditional}) concentrates\footnotemark\enspace (quickly enough) on $m_j=\p_{3\rm rd}$, which is however a \emph{semimeasure} instead of a measure if there are branches on which Bob$_{3\rm rd}$ is terminated. If the total survival probability $\sum_y m_j(y|z)$ is zero or too small, then the semimeasure $m_j=\p_{3\rm rd}$ becomes irrelevant for Bob$_{1\rm st}$'s future states despite $2^{-{\rm K}(j)}m_j(z)$ being large. If this is the case, then other semimeasures will determine Bob$_{\rm 1st}$'s states from that moment on, and Alice's world loses its relevance for Bob$_{1\rm st}$.
}
It would be very interesting to say in more detail what Bob$_{1\rm st}$ would see in the scenario of Figure~\ref{fig_subjimm}, but the answer to this question depends very strongly on the type of postulates that we decide to use. If we, as we have done so far, use simplified Postulates~\ref{Postulates}, then Bob$_{\rm 1st}$ will transition into another state next which is one bit longer (and so forth); we would have to analyze the algorithmic probabilities of these possible futures, which may well depend on details of Alice's world and Bob's state at time $t'$. Bob$_{\rm 1st}$ would then subjectively survive \emph{and remember}.

However, this conclusion is likely an artefact of the deficiency that Postulates~\ref{Postulates} do not allow us to reason about memory erasure (forgetting). To do so, we need a formalization of Postulates~\ref{DesiredPostulates} instead. We therefore have to defer the answer to this interesting question to future work. However, we might speculate that the meteorite incident comes with substantial memory erasure for Bob$_{\rm 1st}$, in which case his journey through the space of observer states would start anew. In any case, it seems likely that ideas like ``quantum suicide''~\cite{Moravec,Marchal88,Marchal91,Tegmark98}, or rather their adaption to our setting, do not work in the context of our theory.

\section{Application to exotic scenarios}
\label{SecApplication}
\sectionmark{Application to exotic scenarios}
One major motivation for the theory of this paper was to have a \emph{unified} approach to answering the question ``What will I see next?'' --- one that applies to ordinary physics situations, but also to more exotic scenarios like Parfit's teletransportation paradox. In the following two subsection, we will see that our approach lives up to these hopes, at least in principle, notwithstanding its incompleteness as explained in Section~\ref{SecPostulates}.

\subsection{Dissolving the Boltzmann brain problem}
\label{SubsecBB}
If the approach of this paper captures a grain of truth about physics, then we face a substantial revision of some basic assumptions about the world. Thus, it should not come as a surprise (and may indeed be regarded a sign of predictive power) that our approach suggests revisions in those areas of physics that are asking fundamental questions about the nature of our universe and the role of the observer.

One major research area of this kind is cosmology. I am not a cosmologist, and most questions and problems of cosmology have nothing to do with the approach of this paper. However, there are some very fundamental questions of \emph{how to even think about our world} for which our approach will be relevant --- not because we could solve any of the cosmologists' problems, but because our ``first-person-first'' perspective changes the \emph{type of questions that we may want to pose in the first place}.

We have already seen a simple example of this in Subsection~\ref{SubsecExternalProcess}; let us recapitulate it in the context of the present subsection. One instance of cosmology's \emph{measure problem}~\cite{Linde} can be phrased as follows: why did our universe have thermodynamically atypical low-entropic initial conditions? As we have seen, our approach predicts that observers will find themselves to be part of a simple probabilistic computational process. It is a generic feature of such processes that they start in some initial state, and then their time evolution unfolds with increasing complexity.

While there is no notion of energy or thermodynamics in these information-theoretic statements,  there turns out to be a bridge between algorithmic probability and thermodynamics:
\emph{Kolmogorov complexity ${\rm K}$ can itself be regarded as a notion of entropy.} In fact, ${\rm K}$ is sometimes called ``algorithmic entropy'', and it has been applied directly as a measure of entropy in thermodynamics, cf.\ Section 8 of~\cite{LiVitanyi} or~\cite{Zurek}.
While standard thermodynamic entropy is a function of a probability distribution (such as the uniform distribution on all accessible microstates),
Kolmogorov complexity is defined for \emph{single} realizations of an ensemble (that is, for single microstates). There are numerous close relationships
between complexity and entropy~\cite{ZvonkinLevin,Brudno,QBrudno}. For instance, \emph{average Kolmogorov complexity equals entropy}: if $P$ is any computable
probability distribution on $\s$, then~\cite{LiVitanyi}
\[
   H(P)\leq \sum_{x\in\s} P(x){\rm K}(x)\leq H(P)+{\rm K}(P)+\mathcal{O}(1),
\]
where $H(P)=-\sum_{x\in\s}P(x)\log_2 P(x)$ is Shannon entropy. This implies that notions of algorithmic complexity can in many cases be read as if they were statements about entropy.

Thus, the approach of this paper predicts directly that observers will find low-entropic conditions if they retrodict their external world far enough to the past. This dissolves the need for a ``mechanistic'' explanation of simple (thermodynamically atypical) initial conditions. It is not necessary to postulate, for example, that our universe has developed as a thermodynamic fluctuation from another ``meta-universe'''~\cite{AlbrechtSorbo2004}; our theory predicts simple initial conditions without any such assumptions. The point of view taken in this paper is that the question of ``why there is a world in the first place'' requires (and has) an explanation that is of a different category than the usual argumentation with which we explain phenomena \emph{within} our world. Using thermodynamic reasoning, for example, assumes that we already have a certain amount of structure (basically fundamentally reversible dynamics according to some symplectic structure, leading to a notion of energy that is preserved) that is ultimately part of what we want to explain in the first place.

If we assume the (approximate) validity of the approach of this paper, then there is another puzzle related to cosmology which gets dissolved: the \emph{Boltzmann brain problem}~\cite{Albrecht2002,AlbrechtSorbo2004}. For our purpose, it can be summarized in the following way:

\emph{Suppose that our universe is ``combinatorially large'' in some sense, for example due to eternal inflation~\cite{LindeBook}. Then, because of statistical fluctuations, many observers (``Boltzmann brains'') will come into existence by mere chance, simply appear for a short time, surrounded by chaos, and then disappear again. Under certain assumptions on the cosmological model, there will be far ``more'' Boltzmann brains out there than there are ``ordinary'' observers (like we think we are). Thus, in such cases, should observers (like us) assign high probability to being a Boltzmann brain rather than having been generated by a long evolutionary process?}

Even if we ignore fundamental questions of conceptual validity, there are some obvious practical obstacles to making arguments of this kind scientifically sound. For example, the argument depends on the choice of method for how to count observers (``natural'' ones as well as Boltzmann brains)~\cite{PageWeighting}. Some cosmologists try to infer constraints on the cosmological model from assuming that we are not Boltzmann brains~\cite{PageBB}, but there is no consensus on how the calculations should be done in detail. As an example, \cite{Albrecht2002} and~\cite{AlbrechtSorbo2004} argue that inflation cures the Boltzmann brain problem, but other authors~\cite{Dyson} come to different conclusions.

For concreteness, let us formulate the Boltzmann brain puzzle in the terminology of observer states. Suppose we have a universe (corresponding to some ``large'' cosmological model) which contains a \emph{single} observer, Bob, who remembers having lived a rich life full of experiences in a standard, low-entropic planet-like environment. Let $x$ be Bob's observer state. For concreteness, let us assume that the combinatorially large universe contains about $10^{90}$ thermal fluctuations --- Boltzmann brains --- that contain, by mere chance, a perfect copy of $x$. That is, each of these brains \emph{also} thinks that it is Bob, that it has lived this rich life on a planet, and that it will subsequently continue business as usual. For the sake of the argument, let us furthermore assume that each of these Boltzmann brains will subsequently learn\footnote{A moment's thought shows that the Boltzmann brains can be completely ignored, according to Postulates~\ref{Postulates}, if this is not the case.} some additional bits $y=y_1 y_2\ldots y_m$, transitioning into an observer state $xy$ that corresponds to an extremely strange and unexpected, ``disordered'' experience~\cite{Nomura}, before they perhaps finally disappear.

A question that naturally comes to mind is the following: \emph{Suppose I am in observer state $x$ at this moment, having all these memories and beliefs. How do I know if I am ``really'' Bob, or if I am one of those Boltzmann brains?} But in the approach of this paper, this question is meaningless: observers are not material objects in some universe, but \emph{observers are their observer states}. That is: I am $x$. In some sense, I am Bob on the planet and \emph{at the same time} I am each and every one of those Boltzmann brains. Except that this is a void statement, so let us retract it, and let us notice that we have to withdraw the question as meaningless.

However, there is a reformulation of the question which \emph{does} have meaning in the context of our theory: \emph{will my experiences in the next moments be those of Bob on the planet, or those of one of the Boltzmann brains?} To formalize this question, denote by $z=z_1 z_2\ldots z_m$ the $m$ bits that ordinary Bob on the planet will subsequently acquire next. Then, we have to compare the conditional probability of those bits $z$ with that of the ``Boltzmann brain bits'' $y$, i.e.\
\[
   \p(z|x)\mbox{ versus }\p(y|x).
\]
Now, since there are $10^{90}$ Boltzmann brains, but only one version of Bob on the planet, naive counting would suggest that $\p(y|x)\approx 10^{90} \p(z|x)$, so that we should very strongly expect to make one of the strange ``Boltzmann brain observations'' next. However, the approach of this paper, as formulated in Postulates~\ref{Postulates}, claims that this is incorrect: the two probabilities above are equal to \emph{conditional algorithmic probability}; counting numbers of objects in some universe is completely irrelevant, and the probabilities are \emph{independent} of the cosmological model.

The question above can be analyzed within the formalism of Subsection~\ref{SubsecAsymptoticCoherence}. We have the very large, but algorithmically simple, probabilistic, computational ontological model (say, it is Alice's external world), and we have a variety of choices of ``picking'' certain objects (variables) in this ``universe''. On the one hand, there is Bob on the planet (``Bob$_{\rm p}$''). This will be formulated by some locator function $f_{\rm p}$ (``p'' is for ``planet'') that somehow reads Bob's state from the state of the universe\footnote{How $f_{\rm p}$ is defined in detail is irrelevant for our purpose; it is some function of not too high complexity that somehow extracts Bob's state from the fundamental state of the universe. Note that this does \emph{not} mean that the universe has to be discrete in any naive sense (spacetime pixels etc.), but only that it is an abstract process that has in principle a finite description. At this point, we are ignoring quantum theory, but we will turn to it in Section~\ref{SecQuantum}.}. On the other hand, we can define some other locator function $f_{\rm BB}$ that tracks one of the Boltzmann brains (``Bob$_{\rm BB}$''); we will shortly discuss different options for how to do this.

These locator functions generate ``third-person probabilities'' $\p_{\rm 3rd}^{\rm p}$ and $\p_{\rm 3rd}^{\rm BB}$. We can now reformulate the question above as follows, with $\p$ our algorithmic prior:
\[
   \mbox{Is }\p(w|x)\approx\p_{\rm 3rd}^{\rm p}(w|x)\mbox{ or } \p(w|x)\approx \p_{\rm 3rd}^{\rm BB}(w|x)?
\]
We can ask this for all possible ``next bits'' $w$, in particular for $w=y$ and $w=z$.

Since $\p_{\rm 3rd}^{\rm p}$ and $\p_{\rm 3rd}^{\rm BB}$ give very different conditional probabilities, only one of the approximate equalities can be true. That is, \emph{at least one of Bob$_{\rm p}$ and Bob$_{\rm BB}$ must be a probabilistic zombie} in the sense of Subsection~\ref{SubsecZombies}. 

Let us consider a specific choice of locator function $f_{\rm BB}$. Suppose that $f_{\rm BB}$ scans the universe in an algorithmically simple pseudo-random fashion, until it finds some record of some observer state $s$. Subsequently, it keeps on searching that way (starting in the vicinity of its previous strike) until it finds, within a prescribed number of time steps, another state $sa$, with $a\in\{0,1\}$ --- and so forth, producing an eternally growing sequence of observer states. Now suppose that this process actually produces the observer state $x$ at some point. Since there are $10^{90}$ Boltzmann brains in the universe, but only a single Bob$_{\rm p}$, this means that our locator function will most probably be pointing to a Boltzmann brain.

This locator function $f_{\rm BB}$ is algorithmically not very complex; perhaps of comparable complexity as $f_{\rm p}$. Hence, if we consider the corresponding (semi)measure $m_i=\p_{\rm 3rd}^{\rm BB}$ in~(\ref{eqMPrimeConditional}), then ${\rm K}(i)$ will not be too large since ${\rm K}(\p_{\rm 3rd}^{\rm BB})$ isn't either. However, $m_i(x)$, the probability that $f_{\rm BB}$ will actually produce $x$ \emph{by mere chance}, is combinatorially small, certainly much smaller than $\p_{\rm 3rd}^{\rm p}(x)$. This means that $m_i$ will definitely \emph{not} be dominating the sum; its weight will be much smaller than that of $\p_{\rm 3rd}^{\rm p}$, and hence Bob$_{\rm BB}$ will be a probabilistic zombie.

What if we define another locator function $f'_{\rm BB}$ where we try to circumvent the smallness of this probability? For example, let $f'_{\rm BB}$ scan the universe in an algorithmically simple pseudo-random fashion, pretty much like $f_{\rm BB}$, \emph{until it finds specifically a random fluctuation in observer state $x$}. (This function also has to return prefixes of $x$ at earlier times, and it has to make sure that the fluctuation will produce further growing bit strings for a while.) The (semi)measure $m_j={\p'}_{\rm 3rd}^{BB}$ in~(\ref{eqMPrimeConditional}) has now $m_j(x)=1$. However, in this case, $f'_{\rm BB}$ must contain a \emph{complete description of $x$}, and so ${\rm K}({\p'}_{\rm 3rd}^{\rm BB})\geq {\rm K}(x)$. But then, Observation~\ref{ObsYoungZombies} suggests that {Bob'}$_{\rm 3rd}^{\rm BB}$ is a probabilistic zombie.

In summary, the existence of Boltzmann brains has no relevance whatsoever for anyone's first-person perspective: in the terminology of Subsection~\ref{SubsecZombies}, these are probabilistic zombies. This also implies that the assumption that ``we are not Boltzmann brains'' \emph{cannot} be used to rule out cosmological models, in contrast to the hopes of some cosmologists.

\subsection{Simulating agents on a computer}
\label{SubsecBrainEmulation}
This subsection turns to a set of questions that may attain particular relevance in the near future with ongoing technological progress: namely, the problem to make decisions in situations that involve difficult questions of personal identity.

A specific instance of this problem is the question of \emph{brain emulation}: would it make sense to invest in technology that scans our brains and simulates them to high accuracy after our death? Would the simulation be ``conscious'', and would we actually ``wake up'' in the simulation? The theory of this paper does \emph{not} claim to make any statements about consciousness directly, but it \emph{does} claim to make predictions about the first-person experience of observers. It is this technical, information-theoretic notion of first-person perspective that is the subject of interest here, not the specific, high-level, so far ill-defined notion of consciousness.

Many philosophers, neuroscientists, and computer scientists have thought about the question of brain emulation. Here I will not discuss any specific details of this problem, but only its very fundamental information-theoretic basis which, as I argue below, allows our theory, at least in principle, to make some concrete predictions. Concretely, I will follow a discussion in~\cite{Armstrong} (see also~\cite{Superintelligence}). The authors discuss the idea to create an ``oracle artificial intelligence'' (OAI) as an AI that is confined to some ``box'' and only interacts with the real world by answering questions. Restricting it to be an ``oracle'' in this sense (and not allowing it to act as an agent in the actual physical world) is meant to reduce potential risks (for example, the risk that the AI takes over and destroys our planet). However, the authors argue that not all risks can be eliminated: for example, the OAI might simulate human minds in its memory if this helps to answer some questions more accurately. Then, according to~\cite{Armstrong},

\emph{``[...] the problem with the OAI simulating human minds is mainly ethical: are these simulated humans conscious and alive? And, if they are, are they not being killed when the simulation is ended? Are you yourself not currently being simulated by an OAI seeking to resolve a specific question on human psychology~\cite{Bostrom}? If so, how would you feel if the simulation were to be ended? In view of its speed and the sort of questions it would be asked, the number of humans that the OAI may have cause to simulate could run into the trillions. Thus, the vast majority of human beings could end up being doomed simulations. This is an extreme form of ``mind crime''~\cite{Salamon} where the OAI causes great destruction just by thinking.''}

The worldview that underlies this argumentation is clearly reminiscent of the standard cosmological ontology of the Boltzmann brain problem in Subsection~\ref{SubsecBB}, with the ``trillions of doomed simulations'' analogous to the vast number of Boltzmann brains in a large universe. We have already argued that the approach of this paper implies that naive counting is inappropriate in cosmology; hence it should not come as a surprise that it also implies a substantial shift of perspective on the brain emulation problem.

According to Postulates~\ref{Postulates} and the general view expressed by our approach, an ``observer'' is not a physical object in some universe, but \emph{it is its observer state}. Observer states are abstract structure that cannot be ``created'' or ``destroyed'', neither by physics as we know it nor by computer simulation (see also the discussion on subjective immortality in Subsection~\ref{SubsecZombies}). The only way in which the emergent external world (or a computer simulation) can affect observers is by impacting conditional algorithmic probability, which in turn determines the chances of future observations: regarding a ``world'' as a computational process, what happens in this world influences the statistics of its outputs, which in turn enters the definition of algorithmic probability. This is the sense in which Bob's probability of suffering increases if Alice decides to beat him up.

Thus, we conclude that starting a computer simulation does not ``bring an observer into existence'', and shutting down a simulation does not ``kill'' the simulated observer. But there still remains the question of what happens, say, if we decide to torture a simulated observer; does it increase someone's probability of suffering?

Phrasing the question in this form seems to assume that it makes sense to talk about ``agency'' in our approach, i.e.\ that we have a choice in the first place. At first glance, this does not obviously make sense, as there is no fundamental notion of ``free will'' built into our theory: in some sense, observers passively follow the stochastic random walk on the set of observer states. However, exactly the same is true for all other theories of physics that we have: in classical mechanics, observers act perfectly deterministically, whereas in quantum mechanics, their behavior is given by probabilistic laws. Arguably, probabilistic indeterminism does not automatically entail any notion of ``free will'' (it is more like being slave to a die).

For this reason, the old philosophical debate about free will applies to our theory in exactly the same way as it does to all other physical theories. Even though this is a fascinating problem, its philosophical resolution is not important for the discussion in this section\footnote{Even though it is not important for this paper, I would still like to advertise the plausibility of a \emph{compatibilist} point of view, as laid out very clearly, for example, by Dennett~\cite{Dennett}. Furthermore, theoretical computer science can add an important twist to it via the notion of \emph{computational irreducibility}~\cite{WolframNat,WolframPRL,WolframBook,Israeli}, which can be used to justify the assignment of autonomy or agency to algorithms. Identifying ``ourselves'' with the information processing in our brains will then allow us to claim a status of information-theoretically well-defined autonomy or ``freedom''.}. Instead, let us follow a pragmatic approach for the time being: whatever ``free will'' fundamentally means, it is an undeniable experience that we somehow have to decide what to do tomorrow. Therefore, it is essential for practical purposes to treat our actions as not predetermined, and to argue counterfactually what would happen if we decided one way or the other. Henceforth we will treat the actions of our prototype of observer, Alice, in her emergent external world as ``free'' in this sense.

Equipped with a notion of agency of observers, we can now analyze what our theory has to say about torturing a simulated mind. Suppose that Alice the guinea pig is in some observer state $x$, a standard ``happy state'', describing her experience of eating a large and tasty piece of cabbage. However, imagine that some possible future state $xy$ is a ``suffering state'', possibly representing the experience of a painful medical procedure that we would like to test on emulated Alice in a computer simulation.

Suppose that we have a simulation running (which may be a deterministic or probabilistic algorithm), and emulated Alice is in observer state $x$. Furthermore, suppose that we know that in the next few time steps, our simulation is going to perform the transition to the suffering state with high probability $P_{\rm sim}(y|x)\gg 0$ (unity in the deterministic case). Are we ethically allowed to run the simulation? Should we shut it down? Should we have refrained from running it in the first place?

Clearly, what actually matters for Alice is $\p(y|x)$, her first-person probability of suffering according to Postulates~\ref{Postulates}. Arguably, it is ethically correct for us to run the simulation \emph{either} if Alice's first-person suffering probability is small despite our simulation, \emph{or} if it is large but our simulation cannot be regarded as the cause for this. In more detail, we have the following two arguably acceptable scenarios:
\begin{itemize}
\item[(1)]$\p(y|x)\approx 0$ even though $P_{\rm sim}(y|x)\gg 0$. This would imply that simulated Alice is a \emph{probabilistic zombie} in the sense of Observation~\ref{ObsYoungZombies}.
\item[(2)] Both $\p(y|x)$ and $P_{\rm sim}(y|x)$ are large, but $\p(y|x)$ would also be as large if we decided not to implement the specific simulation.
\end{itemize}
One way to make sure that one of these scenarios applies is by running a \textbf{closed simulation}. By this I mean a (possibly probabilistic) simulation algorithm that runs completely autonomously, without accepting any data from the external world. Its behavior will only depend on an initially specified program, plus a sequence of random input bits\footnote{Since our external world is in general probabilistic (cf.\ Subsection~\ref{SubsecExternalProcess}), we may input actual random bits into the simulation, or, alternatively, pseudorandom bits; our analysis below will apply to both cases, as long as the simulation produces a ``typical'' instance of the corresponding probabilistic process. This means in particular that the pseudorandom input bits are not supposed to smuggle relevant information about the external world into the simulation.} if the simulation is meant to be non-deterministic.

As long as simulated Alice is still very simple, she will automatically be a probabilistic zombie due to Observation~\ref{ObsYoungZombies}. If our simulation tortures her at this point, this will be of no relevance for Alice's first-person perspective; we are in Scenario (1)\footnote{Actually, we are already in Scenario (2) if simulated Alice is a zombie but nevertheless turns out to have $\p(y|x)\gg 0$.}. But we know from Theorem~\ref{TheNoZombies} that this situation cannot last very long: if the simulation runs long enough, Alice will loose zombie status, and simulated Alice will more and more become an accurate representation of the actual first-person perspective that corresponds to the simulated observer states. However, in this case we run into Scenario (2): if the simulation tortures Alice with high probability at this point, we have $P_{\rm sim}(y|x)\approx \p(y|x)\gg 0$, but the cause of $\p(y|x)$ being large is not that we have launched the simulation: rather, \emph{the cause is that the simulated world exists mathematically as a simple abstract computational process which generates the transition to the suffering state with non-negligible probability.} This is a mathematical fact, regardless of whether we actually run the simulation or not.

To see this, recall $\mathbf{M}_V(y|x)$ from equation~(\ref{eqMPrimeConditional}). It is a mixture of semimeasures $m_j$, and (at least) one specific choice of $j$ will yield the measure $m_j$ which describes Alice's statistics within our simulation. If simulated Alice has lost zombie status, and Theorem~\ref{TheSimpleLaws} applies, then this specific $m_j$ will dominate the statistical mixture of semimeasures in~(\ref{eqMPrimeConditional}), by having small ${\rm K}(j)$ and large $m_j(x)$ (in comparison to other semimeasures). However, these properties are mathematical statements that are completely independent of whether we choose to implement the simulation or not. 

In other words: running a closed simulation merely \emph{displays} the world which is simulated; it does not ``bring it into existence'' in any metaphysical sense. Thus, running a closed torturing simulation (or terminating a closed simulation) is ethically unproblematic since it has no causal effect, similarly as watching a documentary about a war (or stopping to watch it) does not actually affect any soldier that is portrayed in the documentary.

The situation is completely different if we run an \textbf{open simulation}, that is, if information is allowed to flow from the external world into the simulation. Imagine, for example, that a team of programmers regularly intervenes with the simulation (similarly as in Gary Larson's cartoon ``god at his computer''), or that we start to communicate with simulated Alice. Exactly as in the closed case, we may still have an initial phase where Alice is a zombie in the sense of Scenario (1), and in the long run, Theorem~\ref{TheNoZombies} implies that simulated Alice's first-person chances converge towards the distribution that we observe in the simulation. This distribution is generated by a simple computational process.

What is this simple computational process? In the closed case, it is simply the simulation algorithm itself; this algorithm will ultimately represent the best possible compression of simulated Alice's states and thus dominate her chances of future states. In the open case, however, computable patterns of the external world will ultimately enter the simulation. Thus, asymptotically, the best possible compression of the simulated states will ultimately correspond to a computational process that involves (all or part of) \emph{the external world and the simulation}. But then, \emph{we become part of the relevant computational process} and will gain causal influence on the fate of simulated Alice. That is, her conditional probability $\p(y|x)\approx P_{\rm sim}(y|x)$ will depend substantially on our choices as agents in our external world.

Thus, in the case of an open simulation, none of the two scenarios applies, and torturing becomes an actual ``mind crime''. This should not be surprising, given that \emph{actual} material guinea pig Alice is a special case of an open simulation, with the hardware given by the brain, and the behavior of other guinea pigs clearly having causal impact on her actual experience\footnote{To some extent, there should be a gradual transition between ``open'' and ``closed''. It thus seems plausible that a small amount of intervention is still compatible with Scenarios (1) and (2).}.

In summary: to emulate responsibly, don't talk to your simulation; but if you decide to talk to her, be nice!

\section{A quantum of speculation}
\label{SecQuantum}
\sectionmark{A quantum of speculation}
While quantum theory (QT) has been named as a main motivation for this theory in the introduction, the discussion so far has not touched on QT at all. This may seem odd at first sight: why have we only talked about \emph{classical} probabilities and not about transition amplitudes? Isn't our theory in contradiction to the observed quantumness of our world, as Subsection~\ref{SubsecExternalProcess} seems to predict a \emph{classical} external world?

Much of this objection rests on intuition that comes from a certain naive form of wave function realism. According to this view, physics must always talk about \emph{material stuff evolving according to differential equations}. The quantum Schr\"odinger equation (or its relativistic or field-theoretic counterparts) are then often seen as instances of this dogma, with an actual ``thing'' $|\psi(t)\rangle$ (the quantum state) evolving in time. Since this ``thing'' is a complex wave function, and not a classical probability distribution, proponents of this view will intuitively think that the emergent external world in our theory must be a classical world.

This view ignores the fact that all empirical content of the quantum state ultimately lies in the prediction of probabilities of measurement outcomes. Therefore, it is consistent (and in many ways advantageous) to regard quantum states as ``the same stuff as probability distributions'', namely as states of knowledge, information, or belief in some (yet to be specified) sense~\cite{Rovelli,Spekkens,FuchsQFQIT,Brukner,Zeilinger}. The departure from classical physics is in the \emph{properties} of these probabilities (for example in the violation of Bell inequalities despite space-like separation, or the appearance of interference patterns in situations where classical physics suggests no such patterns), not in the mathematical description (via complex numbers) that is used to compute these probabilities. One of the clearest arguments for this broadly ``epistemic'' view comes from the recent wave of reconstructions of QT~\cite{Hardy2001,DakicBrukner,MasanesMueller,ChiribellaDArianoPerinotti,Hardy2011,MasanesEtAl,BMU,Hoehn,HoehnWever,Wilce}, which proves that the full complex Hilbert space formalism of QT can be derived from a few natural information-theoretic principles.

Probabilistic predictions comprise everything that we can ever empirically test about QT.
All interpretations of QT agree on this point~\cite{FuchsPeres}; they only differ in the way that they interpret these probabilities, and in additional claims about unobservable processes that are declared to be causing the observed behavior in some (classically) intuitively comprehensible or mechanistic terms. Indeed, the motivation for many of these approaches comes from the traditional intuition described above. In a traditional view, we have an unsettling situation in QT, which has been termed the \emph{measurement problem}. From a traditional perspective, the problem is as follows:
\begin{itemize}
	\item Typically, the quantum state is evolving unitarily, according to the Schr\"odinger equation $i\hbar |\dot\psi(t)\rangle= H|\psi(t)\rangle$. This is analogous to time evolution in classical mechanics, and in this sense ``nice'' and intuitive.
	\item However, sometimes, there are disturbing exceptions from this rule: this is when we perform a \emph{measurement}. Then the state vector seems to collapse in some sense, violating unitarity.
\end{itemize}
The traditional narrative is to declare unitary time evolution as the ``standard rule'', and the creation of measurement outcomes with certain probabilities as an apparent violation of this rule which is in need of elaborate explanation. For example, this point of view is very pronounced in Everettian interpretations of quantum mechanics~\cite{Everett}. However, taking actual scientific practice as the starting point, and taking the manifold evidence (mentioned above) of the epistemic nature of the quantum state seriously, the more economic and consistent point of view is this: \emph{what is ``really happening'' is the appearance of measurement outcomes with probabilities as predicted by the quantum state}. In order to not fall into an overly instrumentalist perspective, we adopt an insight from QBism~\cite{Fuchs,FuchsSchack,Timpson}, namely that the notion of a ``measurement outcome'' is merely a metaphor (and special case) of \emph{experience}, which in the terminology of this paper is the same as \emph{observation}. In summary,
\emph{all there is are observations} (i.e.\ transitions between observer states), and these observations are \emph{non-deterministic}. Quantum states are the things that determine the probabilities of these observations (we do not have to settle the question of what ``probability'' exactly means here to come to this conclusion). Time evolution of a state is ultimately nothing but a correlation of these probabilities with some clock variable~\cite{PageWootters}.

The point of view that \emph{observations} (as part of \emph{observer states}) are the primary notion, and that the quantum state should be interpreted epistemically, dissolves the measurement problem. For more detailed explanations for why this is an attractive position, see e.g.~\cite{Appleby}. Moreover, it can do so in a particularly nice way within the ontology of this paper, by refuting the intuitive consequence that quantum states, as a result, would somehow ``not tell us enough about the physical world'' if they are to be understood epistemically.
\sobservation{QT and the measurement problem}{ObsMeasurementProblem}{Based on a traditional view of physics, QT is widely regarded to suffer from a (conceptual) \emph{measurement problem} as sketched above. However, within the approach of this paper, the measurement problem dissolves: quantum states are an observer's states of knowledge (we will explore in more detail below in what sense), and a measurement update rule simply corresponds to Bayesian updating. This solution to the measurement problem is shared with other epistemic interpretations of the quantum state~\cite{Rovelli,Spekkens,FuchsQFQIT,Brukner}.

Moreover, our approach is particularly well-suited to support such an interpretation: it rejects the fundamentality of any ``external reality'', and thus it suggests that the question of ``what is really going on in the world'' may not be answerable in the way we would hope for. In this sense, it provides reasons to expect that observers will typically find a theory of the quantum kind (with observations or measurements as primary notions) describing their world. Furthermore, our approach says that \emph{observations are fundamentally private to a single observer}, and that the notion of a ``common external reality'' for \emph{different} observers is an approximation that is only valid under certain conditions (cf.~Theorem~\ref{TheNoZombies}). This makes our approach compatible with ``Wigner's Friend''-type thought experiments.
}

So QT fits our theory very well from a conceptual point of view, but can we understand some of its characteristic features in more technical detail? There is an obvious candidate for a characteristically nonclassical effect, namely \emph{entanglement}. As Schr\"odinger~\cite{Schroedinger} famously wrote in 1935: \emph{``I would not call that \emph{one} but rather \emph{the} characteristic trait of quantum mechanics, the one that enforces its entire departure from classical lines of thought.''} More concretely, it is not the mathematical notion of entanglement in itself that is remarkable, but rather its operational consequences, in particular the violation of Bell inequalities~\cite{Bell1964,Bell1966,CHSH} (we will use the usual notion \emph{nonlocality} for this in the following, without implying a specific interpretation). Quantum information theory has shown us a multitude of ways in which nonlocality can be harnessed for information-theoretic processes that would be classically impossible, such as device-independent cryptography~\cite{BennettBrassard,BarrettHardyKent} or randomness amplification~\cite{ColbeckThesis,Pironio}. Thus, nonlocality is an operationally particularly robust signature of nonclassicality.

\begin{figure*}[!hbt]
\begin{center}
\vskip -1.5em
\includegraphics[angle=0, width=15cm]{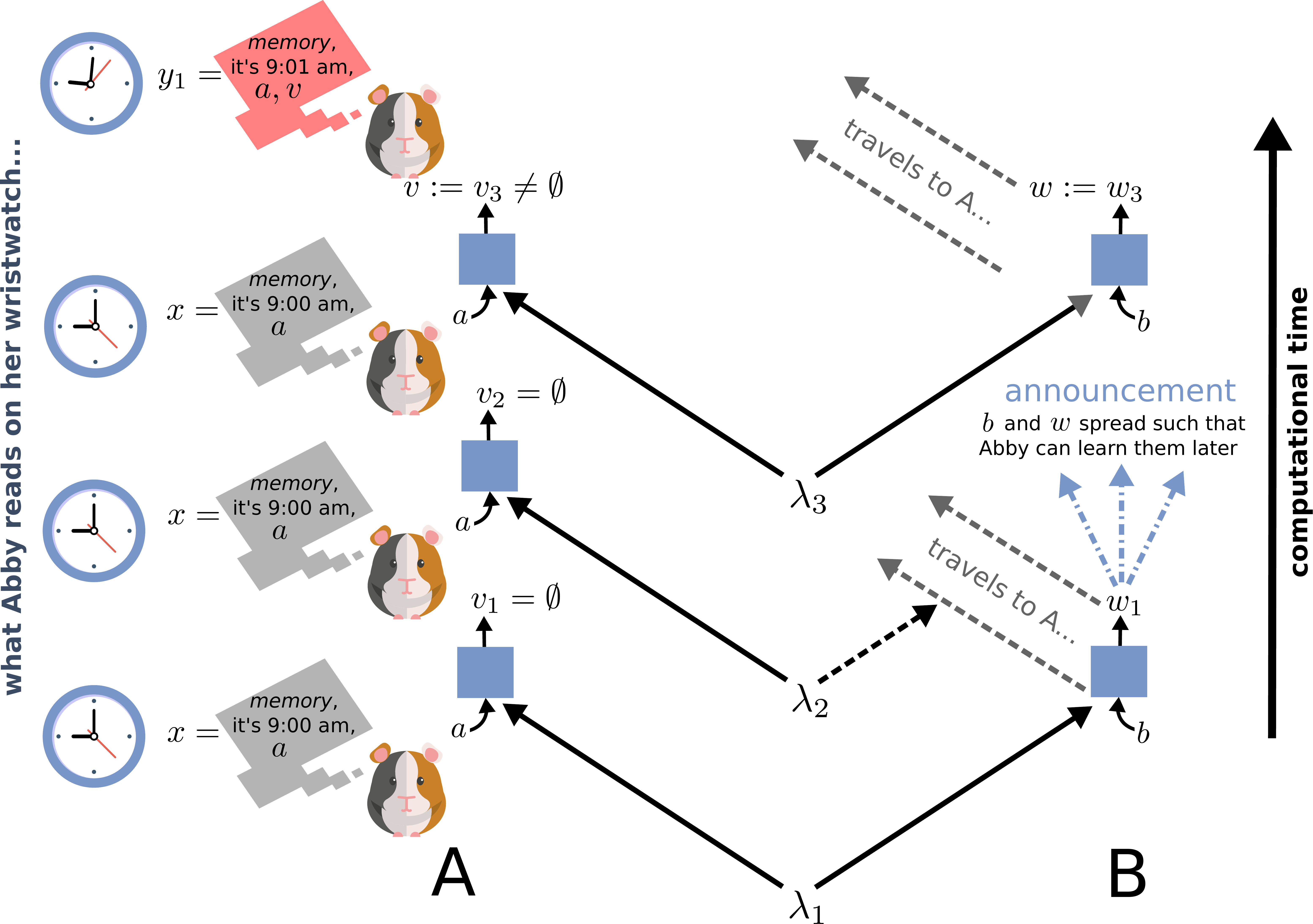}
\caption{A private ``Bell experiment'' that Alice can set up in her emergent external computational world. As explained in the main text, we assume some rudimentary locality structure, admitting a notion of ``spacelike separation'' which means that information needs some (computational) time to travel from $A$ to $B$ and vice versa. Here we assume that $A$ and $B$ are \emph{very} far away, such that the $a$ and $v_i$ can arrive at $B$ only long after the experiment is finished, and the same for the arrival of $b$ and $w_i$ at $A$. By construction, there is a local, classical conditional probability distribution $P_0(v,w|a,b)\equiv P_0(v_i,w_i|a,b)$ that describes the probabilities of the outcomes of every single run, given the settings. We consider a scenario in which Alice can ``loop back'' to her state $x$ of before the experiment, \emph{conditionally} on the local outcome $v_i$. Since the probabilities of her future states depend only on her current observer state, she will not notice that this happens (neither immediately nor at any point in her future). Consequently, she will see a postselected conditional probability distribution $P(v,w|a,b)\neq P_0(v,w|a,b)$ which, as we prove below, can violate Bell inequalities, but must be non-signalling.}
\label{fig_bell}
\end{center}
\end{figure*}
In Subsection~\ref{SubsecExternalProcess}, we have seen that our theory predicts the appearance of an ``external world'' that corresponds to a computational process. What can we say about the correlations in Bell scenarios seen by observers in such worlds? In our own physical world, the laws of quantum theory predict the violation of Bell inequalities for some entangled states on a bipartite Hilbert space $AB$. This phenomenon is classically impossible only if $A$ and $B$ are spacelike separated such that the instantaneous transmission of classical information is forbidden by relativistic causality. Not only is this the physically most interesting scenario~\cite{LoopholeFree}, it is also the technologically most relevant one that is needed, for example, for fully loophole-free device-independent cryptography. On the other hand, \emph{if $A$ and $B$ were not causally separated} in the way just described, \emph{then a violation of a Bell inequality would not be surprising} (and could not lead to technological applications) since it would be a natural consequence of signalling.

In other words: \emph{we only care about the violation of Bell inequalities because our world admits a notion of ``locality'' in the first place.} If every single random variable in our universe could instantaneously signal to every other random variable, then there would be no point in studying Bell nonlocality.

Now consider our observer Alice's emergent external world as explained in Subsection~\ref{SubsecExternalProcess}. This is a \emph{computational ontological model} in the sense of Definition~\ref{DefOntologicalModel}: a  probabilistic computational process which generates the observer's asymptotic statistics $\mu$. Besides being algorithmically simple, computable and probabilistic, we do not know much about this process. But if we want to study Bell scenarios in it, then -- as explained above -- we need to assume that it carries some notion of locality, in the following sense:
\emptysassumption{AssLocalityStructure}{
We assume that Alice's external world carries \emph{locality structure} in the following sense. As a computational ontological model (see Definition~\ref{DefOntologicalModel}), its configuration $\omega_t\in\Omega$ is naturally subdivided into several random variables, such that some random variables take a finite (sometimes large) number of steps to influence other random variables in the process.
}
If Alice's external world does not carry locality structure in this sense, then there is no point in studying Bell scenarios in it. But if it does, then we can reason about random variables that are, in some generalized sense, ``close-by'' or ``far apart'', meaning that the number of computational steps it would take for an intervention on one random variable to impact the other is either small or large. This in turn allows us to formulate scenarios as depicted in Figure~\ref{fig_bell}: situations in which there is a certain process in Alice's vicinity $A$ (represented by the blue box), which takes some ``setting'' $a$ (for example, a bit) and produces some ``outcome'' $v$ (which might itself be a bit, or a sequence of bits). In general, there can be another process at a distant point $B$, turning some setting $b$ into some outcome $w$, which is not completely statistically independent of the process at $A$. A possible origin of this statistical dependence is the existence of a random variable $\lambda$ that has been distributed to both $A$ and $B$ beforehand.

Let us assume that Alice understands her computational world well enough to set up a situation of this kind in an experiment-like fashion. This way, she can construct a ``Bell experiment'' in her world, as depicted in Figure~\ref{fig_bell}. Like in an actual Bell experiment, we assume that she can in principle input any bit $a$ at $A$ (``choose the setting $a$'') that she wants, and she can construct the setup such that any remotely generated bit $b$ can be used as the setting at $B$. On the one hand, she could use two bits $a$ and $b$ that are freely generated locally at $A$ and $B$; in this case, $a$ and $b$ will be uncorrelated with all random variables \emph{except for those in their respective future ``lightcones''}, satisfying the unique sensible definition of ``free choice'' that is routinely applied in this context, cf.~\cite{ColbeckRenner2013,ColbeckRenner}. On the other hand, Alice could also use two bits for the settings that are generated locally in some pseudo-random fashion (e.g.\ by creating a checksum of Twitter messages that her fellow guinea pigs have sent out close to $A$ resp.\ $B$ shortly before the experiment). Most methods of pseudo-randomness generation should yield outcomes that resemble ``truly free'' random bits, since there is no reason to expect that the stochastic process will ``conspire'' with the pseudo-random variables to produce non-typical outcomes.

Shortly after choosing $a$ and $b$, the outcomes $v$ and $w$ are created locally at $A$ resp.\ $B$. We assume that Alice knows the value of $a$, and she can in principle immediately learn the value of $v$. However, due to spacelike separation, she will in general have to wait a while until she learns the values of $b$ and $w$. These outcomes will be distributed according to a conditional probability distribution
\[
   P_0(v,w|a,b)=\sum_\lambda q(\lambda)P_{\lambda}(v,w|a,b),
\]
where $q$ is some probability distribution over the possible values of $\lambda$, and $P_\lambda(v,w|a,b)=P_0(v,w|a,b,\lambda)$. If the random variable $\lambda$ summarizes all randomness that is shared by $A$ and $B$, then it follows that
\[
   P_\lambda(v,w|a,b)=P_\lambda(v|a) P_\lambda(w|b).
\]
In other words, $P_0$ is a \emph{classical correlation}. Not all correlations in physics are classical in this sense. Quantum theory famously predicts the existence of correlations that are not of this form. Concretely, a correlation $P(v,w|a,b)$ is \emph{quantum} (cf.\ e.g.~\cite{Navascues}) if there exist Hilbert spaces $\mathcal{H}_A,\mathcal{H}_B$ for $A$ and $B$, a joint state $|\psi\rangle$ in $\mathcal{H}_A\otimes\mathcal{H}_B$, orthogonal projectors $\pi_a^v,\pi_b^w$ with $\sum_v \pi_a^v=\mathbf{1}_A$, $\sum_w \pi_b^w=\mathbf{1}_B$, and $\pi_a^v \pi_a^{v'}=\delta_{v,v'}\pi_a^v$ as well as $\pi_b^w \pi_b^{w'}=\delta_{w,w'}\pi_b^w$, such that
\[
   P(v,w|a,b)=\langle\psi|\pi_a^v\otimes \pi_b^w|\psi\rangle.
\]
The set of quantum correlations is strictly larger than the set of classical correlations. A simple way to see this is the existence of \emph{Bell inequalities}~\cite{Bell1964,Bell1966,CHSH,Peres} which are satisfied by all classical correlations, but violated by some quantum correlations. The CHSH inequality~\cite{CHSH} constitutes a famous example. If $\mathbb{E}_{a,b}$ is the expectation value of $v\cdot w$ under the choice of settings $a,b\in\{0,1\}$ (concretely, $\mathbb{E}_{a,b}=\sum_{v,w\in\{+1,-1\}} vw P(v,w|a,b)$), then
\[
   |\mathbb{E}_{0,0}+\mathbb{E}_{0,1}+\mathbb{E}_{1,0}-\mathbb{E}_{1,1}|\leq 2.
\]
While this inequality is satisfied by all classical correlations, it is violated by some quantum correlations. In particular, there are states and projective measurements that yield values of up to $2\sqrt{2}$, which is known as the Tsirelson bound~\cite{KhalfinTsirelson,Tsirelson}.

\begin{figure}[H]
\begin{center}
\includegraphics[angle=0, width=4cm]{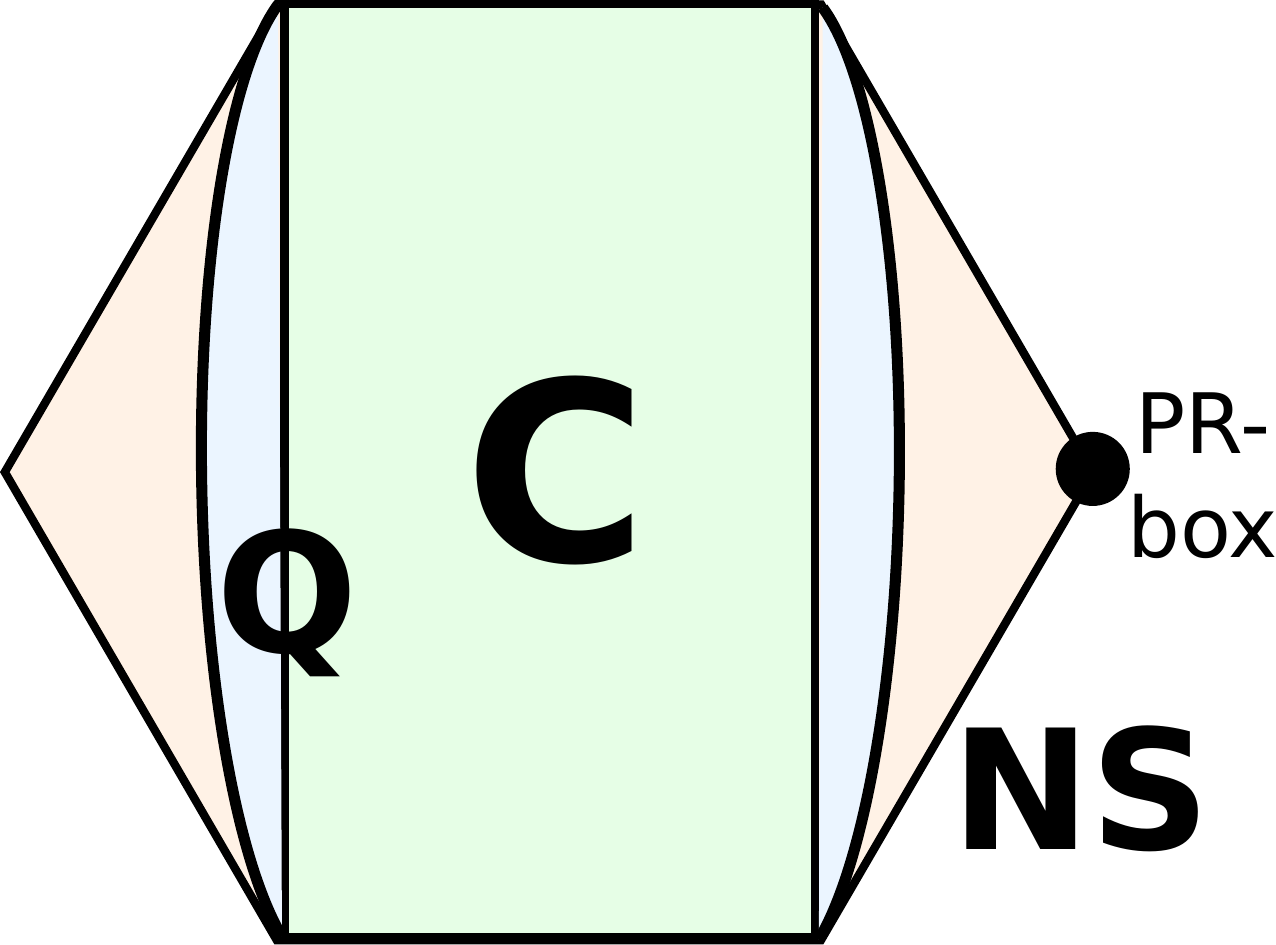}
\caption{The set of non-signalling correlations (for a fixed number of parties, measurement settings, and outcomes) is a convex polytope, here denoted NS. It contains the convex polytope of classical correlations C, and the convex set of quantum correlations Q (which is not a polytope) sits strictly in between the two.}
\label{fig_correlations}
\end{center}
\end{figure}

It is a simple but important insight that the violation of Bell inequalities (here termed ``nonlocality'' to comply with physics convention) does not allow one to communicate. This is known as the ``no signalling'' principle~\cite{PopescuRohrlich,BLMPPR}, that is, the local measurement outcome probabilities at $A$ are independent of the choice of settings at $B$, and vice versa:
\begin{eqnarray}
	\sum_v P(v,w|a,b)&=&\sum_v P(v,w|a',b)\mbox{ for all }a,a',b,w,\label{eqNS1H} \\
	\sum_w P(v,w|a,b)&=&\sum_w P(v,w|a,b')\mbox{ for all }a,b,b',v.
	\label{eqNS2H}
\end{eqnarray}
In particular, this gives us well-defined \emph{marginals} (reduced states) at $A$ and $B$, namely $P(w|b)$ in terms of~(\ref{eqNS1H}), and $P(v|a)$ in terms of~(\ref{eqNS2H}). We can also say that~(\ref{eqNS1H}) expresses \emph{no signalling from $A$ to $B$}, and~(\ref{eqNS2H}) formalizes \emph{no signalling from $B$ to $A$}.

As discovered by Tsirelson~\cite{Tsirelson,KhalfinTsirelson} and Popescu and Rohrlich~\cite{PopescuRohrlich}, the no-signalling principle alone is not sufficient to characterize the set of quantum correlations. That is, the set of non-signalling correlations is \emph{strictly larger} than the set of quantum correlations. An example of stronger-than-quantum correlations is given by a so-called Popescu-Rohrlich box (or ``PR-box'' correlation)
\begin{eqnarray*}
P(-1,-1|a,b)&=&P(+1,+1|a,b)=\frac 1 2 \mbox{ if }(a,b)\neq(1,1),\\
P(-1,+1|a,b)&=&P(+1,-1|a,b)=\frac 1 2 \mbox{ if }(a,b)=(1,1)
\end{eqnarray*}
and all other probabilities equal to zero, where both $a$ and $b$ can take on the values $0$ or $1$. It is easy to see that this correlation is non-signalling, i.e.\ satisfies~(\ref{eqNS1H}) and~(\ref{eqNS2H}), and
\[
   |\mathbb{E}_{0,0}+\mathbb{E}_{0,1}+\mathbb{E}_{1,0}-\mathbb{E}_{1,1}|=4
\]
which is larger than the quantum maximum (Tsirelson bound) of $2\sqrt{2}$. In summary, we obtain the picture that is sketched in Figure~\ref{fig_correlations}.

Returning to Alice's Bell experiment, it is clear that the correlation $P_0$ that governs her outcomes must be classical. However, as we will now see, surprising effects can happen if we integrate ``fundamental forgetting'' into our framework. Recall that we have so far worked with Postulates~\ref{Postulates}, which imply that observer states can only grow one bit at a time, and cannot ``shrink'': these postulates describe observers who can fundamentally never ``forget'' their past states. In Section~\ref{SecPostulates}, we have argued that this is a simplification that should ultimately be overcome (see the discussion there for more details).

Let us now explore what might happen if we work with a version of Postulates~\ref{DesiredPostulates} --- a desired version of the postulates that admits fundamental forgetting. This will be somewhat speculative, because we do not yet have well-defined mathematical formulation of such postulates. Let us therefore give a list of assumptions which will allow us to reason rigorously even in the absence of such a general mathematical framework.

Let us start by slightly rewriting the formalism that we have used so far. We have an algorithmic prior, which gives us probabilities $\p(x)$ on bit strings $x=x_1 x_2\ldots x_n\in\s$. This is the probability of the observer being successively in the observer states $y_1:=x_1$, $y_2:=x_1 x_2$, $y_3:=x_1 x_2 x_3$, $\ldots$, $y_n:=x_1 x_2 x_3\ldots x_n$. Every $y_{k+1}$ is one bit longer than $y_k$. Let us denote this now by $\p(y_1,y_2,\ldots,y_n)$, where every $y_i$ is a bit \emph{string}.

A version of Postulates~\ref{DesiredPostulates} would admit non-zero probability also for sequences of observer states $y_1,\ldots,y_n$ where the $y_{k+1}$ are \emph{not} just one bit longer than $y_k$, but possibly shorter, or of some other form (this will also generalize our notion of (semi)measures from Definition~\ref{DefSemimeasures}). Let us make the following assumptions on the respective prior $\p$:
\emptysassumption{AssQuantum}{We assume that we work with a version of Postulates~\ref{DesiredPostulates} with a prior $\p$ that satisfies the following conditions:
\begin{itemize}
	\item[1.] The probability of future states depends only on the current observer state, and not on the previous ones. That is, $\p(y_{n+1}|y_1,\ldots,y_n)$ is independent of $y_1,\ldots,y_{n-1}$ (Markovianity).
	\item[2.] A version of Theorem~\ref{TheSimpleLaws} can still be proven and is in the following assumed to apply. That is, we will consider observer Alice to be in some state $x$ such that $\p(y_1,\ldots,y_m|x)\approx\mu(y_1,\ldots,y_m|x)$ for those $y_1,\ldots,y_m$ that will be relevant for the experiment, where $\mu$ is a computable measure.
	\item[3.] The computational ontological model for $\mu$ which Alice calls her ``external world'' allows her to set up a Bell experiment; in particular, it carries locality structure in the sense of Assumption~\ref{AssLocalityStructure}.
\end{itemize}}
As we have argued above, assumptions 2.\ and 3.\ are necessary to even talk about Bell experiments. Assumption 1., on the other hand, formalizes the general approach of this paper that \emph{``all there is'', in some sense, is the momentary observer state}. Since there is no fundamental world that could carry memory of the observer's previous states, whatever remains of the past must be encoded, as memory, in the present observer state. Hence, the observer's future can only depend on her present state and not on her previous ones.

Under these assumptions, we will now consider a specific type of Bell experiment. Denote Alice's observer state at the beginning of the experiment by $x$. Naively, think of $x$ as a binary encoding of something like the following:

$x\simeq$\textit{[biographical memory] + ``It is Tuesday, January 14, 2031, 9:00 am, as I have just seen on my wristwatch. I am now inputting $a=0$ into my half of this Bell experiment, which is the first run of this experiment. I'm so excited to see what happens after I've repeated this a thousand times and collected all the data!}''

Concretely, suppose that the computational process works as follows during Alice's Bell experiment (see also Figure~\ref{fig_bell}):
\begin{enumerate}
    \item Alice sees that the experiment has successfully started, and her observer states transitions from $x$ to $x1$.
	\item At $A$ and at $B$, a random variable $\lambda$ is assessed which has been generated, copied and transported to both places earlier on. This variable $\lambda$ has been sampled with uniform probability $1/4$ from the four-element set
    \[
	   \lambda\in\{+\emptyset ++,\enspace \emptyset ++-,\enspace \emptyset --+,\enspace -\emptyset --\}.
	\]
	We use the notation $\lambda=l_0 l_1 l'_0 l'_1$, where $l_i\in\{+,\emptyset,-\}$ and $l'_i\in\{+,-\}$.
	\item The outcome $w=l'_b\in\{+,-\}$ is locally generated at $B$ (and free to spread from there to the rest of the process, including, later on, to Alice).
	\item If $l_a\neq \emptyset$ then the outcome $v=l_a\in\{+,-\}$ is locally generated, and Alice learns this outcome. That is, Alice transitions into the new observer state $x1z$, where $z=1$ if $v=+$ and $z=0$ if $v=-$.
	
	On the other hand, if $l_a=\emptyset$, then Alice transitions back to her earlier state $x$.
\end{enumerate}
If Alice's local outcome is $\emptyset$, then she does not see this outcome, but returns to her earlier observer state $x$. By definition, if this happens, then she will not become aware of what has just happened. Remember what ``being in the state $x$'' means: all that she sees, knows and remembers is still equal to her earlier state. In particular, Alice will still think that she is about to run the experiment for the first time, and that it is 9:00am as described above.

But then, can't Alice simply look at her wristwatch (or some other clock) to see that some extra time has passed, and find out in the next moment that she has just looped? Surprisingly, the answer \emph{must} be ``no'' due to the basic principles formulated in Assumption~\ref{AssQuantum}: the probability of all her future observations, $\p(y_1, y_2,\ldots, y_m|x)$, depends only on $x$ and is thus unchanged. This includes the probabilities of what she will at any later time read from any given clock. Everything must look for Alice as if she had travelled back in time --- or rather, as if nothing had happened at all.

This also explains why we have not specified how the computational process continues if $l_a=\emptyset$ in step 4: all information on whether there has been a ``loop'' $x\to x1 \to x$ will effectively have to be erased or hidden. Thus, if $l_a=\emptyset$ then the experiment will automatically repeat for Alice. Note that this phenomenon resembles the notion of \emph{contextuality} from~\cite{HarriganSpekkens,SpekkensContextuality}: before \emph{and} after a loop, it is the exact same probability $\p(y_1, y_2,\ldots, y_m|x)$ that characterizes Alice's observations; but the underlying states of the computational ontological model will be different. This can already be seen as a first, very vague hint as to why perhaps some phenomena of QT can be expected in this framework.

In the specific example above, the conditional probability distribution $P_0(v,w|a,b)=P_0(v_i,w_i|a,b)$ over all possible outcomes (including $v=\emptyset$) turns out to be the following:\\

\definecolor{gr}{rgb}{0,0.5,0}
\begin{tabular}{c|c|c|c|c}
$P_0({\color{red}l_a},{\color{gr}l'_b}|a,b)$ & $(a,b)=(0,0)$ & $(0,1)$ & $(1,0)$ & $(1,1)$ \\
\hline
$({\color{red}\emptyset},{\color{gr}-1})$ & $1/4$ & $1/4$ & $1/4$ & $1/4$ \\
$({\color{red}\emptyset},{\color{gr}+1})$ & $1/4$ & $1/4$ & $1/4$ & $1/4$ \\
$({\color{red}-1},{\color{gr}-1})$ & $1/4$ & $1/4$ & $1/4$ & $0$ \\
$({\color{red}-1},{\color{gr}+1})$ & $0$ & $0$ & $0$ & $1/4$ \\
$({\color{red}+1},{\color{gr}-1})$ & $0$ & $0$ & $0$ & $1/4$ \\
$({\color{red}+1},{\color{gr}+1})$ & $1/4$ & $1/4$ & $1/4$ & $0$
\end{tabular}\\

This correlation $P_0$ is classical by construction. However, the random experiment only ends for Alice if her local outcome $l_a$ is \emph{different} from $\emptyset$ --- the experiment is repeated until this is the case. Thus, the conditional probability $P(v,w|a,b)$ of the outcome $v\neq\emptyset$ that Alice eventually learns, and of the corresponding outcome $w$ at $B$, corresponds to the postselected distribution
\begin{equation}
   P(v,w|a,b)=\frac{P_0(v,w|a,b)}{1-P_0(\emptyset|a)}\qquad (v\neq \emptyset),
   \label{eqPostselected}
\end{equation}
where $P_0(\emptyset|a)=\sum_w P_0(\emptyset,w|a,b)$ for all $b$. In the special case above, we have $P_0(\emptyset|a)=\frac 1 2$ and obtain
\[
   P=2\cdot\left(
      \begin{array}{cccc}
      	  1/4 & 1/4 & 1/4 & 0 \\
      	  0 & 0 & 0 & 1/4 \\
      	  0 & 0 & 0 & 1/4 \\
      	  1/4 & 1/4 & 1/4 & 0
      \end{array}
   \right).
\]
This is a nonlocal correlation --- it is exactly the PR-box correlation~\cite{KhalfinTsirelson,Tsirelson,PopescuRohrlich} that we have described above.

How is this possible? What happens in the scenario above is a ``cosmological'' version of a phenomenon known as the \emph{detection loophole}~\cite{GargMermin,Branciard}: if the two parties $A$ and $B$ in a Bell experiment have detectors that are not perfect, then postselecting on the successful detection can reproduce the statistics of nonlocal correlations~\cite{PearleRejection}. Alice looping into her old state, and forgetting the run of the experiment, can be interpreted, in this instrumentalist language, as an unsuccessful detection event (denoted $\emptyset$)\footnote{There are also other possible ways to interpret this result. For example, postselecting on Alice's outcome will in general lead to correlations between the settings $a,b$ and the hidden variable $\lambda$, which is an instance of Berkson's paradox~\cite{Berkson2}: ``conditioning on a variable induces statistical correlations between its causal parents when they are otherwise uncorrelated''~\cite{Berkson1}. These correlations in turn lead to the violation of the assumption of free local choice of settings which underlie the derivation of Bell inequalities. However, as explained above, this does not prevent Alice from inputting whatever $a$ she likes into her half of the experiment; it rather allows her to effectively intervene on $\lambda$, but not in a way that could be used for signalling, as we will soon see.}.

Our example thus shows the following:
\slemma{Nonlocality}{LemLocality}{
In the presence of ``fundamental forgetting'' as formulated above, the actual records of data that observers remember admit statistics that violates Bell inequalities. This is because an observer's future states depend \emph{only} on her current state, and \emph{not} on any past states or facts of the world --- hence, resetting the observer state amounts to effectively resetting the world, opening up a ``cosmological detection loophole''.}

Consider the sequence of variables labelled as in Figure~\ref{fig_bell}. The outcome $w_1$ is generated very far from Alice, at a point which cannot yet ``know'' (due to locality) whether Alice has looped or not. This variable can interact with many other variables in the process, and it seems hardly avoidable that $w_1$ may become correlated with something that Alice observes in the future. Now suppose that $w_1$ was correlated with the random variable $\emptyset_1$, specifying whether Alice has looped in the first run or not. In this case, learning $w_1$ (or another typical random variable correlated with $w_1$) would teach Alice something about whether she has looped or not. But this would contradict part 1.\ of Assumption~\ref{AssQuantum}: the probabilities of Alice's future observations can only depend on her current observer state $x$, and this state is the same regardless of whether she has looped or not. Thus, the conditional independence relation  
\[
   \emptyset_1\ci w_1|a,b
\]
must hold. But as shown in Lemma~\ref{LemNoSignalling} in the appendix, this has an interesting consequence:
\slemma{No signalling}{LemNoSignallingMain}{While Alice's effective distribution $P(v,w|a,b)$ can violate a Bell inequality, it must still satisfy the no-signalling conditions~(\ref{eqNS1H}) and~(\ref{eqNS2H}).}
What would the above phenomena imply for an observer like Alice? Arguably, there is no motivation for Alice to make a model of an external world directly in terms of a computational ontological model for the full distribution $\mu$. This is because such a (``noumenal''~\cite{BrassardRobichaud1,BrassardRobichaud2,BrassardRobichaud3}) model would predict state transitions (including loops) which are \emph{unverifiable} for Alice, in the sense that she (or any external world) cannot hold any records of these transitions. Instead, Alice may use an effective (``phenomenal'') description that summarizes only those aspects that she can in principle record and remember. In situations like the Bell test of Figure~\ref{fig_bell}, this means that she may use non-classical conditional probability distributions (corresponding to the postselected distribution $P$) to describe phenomena in her world. Ultimately, this may lead Alice to work with a nonclassical probabilistic theory --- perhaps one that is similar to QT.

The argumentation above is clearly speculative, given in particular that we do not yet have any mathematically rigorous formulation of Postulates~\ref{DesiredPostulates}. However, it demonstrates that an approach in which the observer state is more fundamental than the ``world'' can lead to surprising statistical phenomena, and some of those may resemble phenomena of QT. Given an approach like the one of this paper, how far can we expect to get in deriving the full structure of QT? It is certainly possible that QT is ultimately a contingent feature, similarly as the exact choice of asymptotic measure $\mu$ in Theorem~\ref{TheSimpleLaws} will in general be merely random. In this case, even a partial derivation of QT would be impossible.

However, there are some indications that can motivate us to be more optimistic. One such hint is given by the recent wave of \emph{reconstructions of the formalism of quantum theory from simple information-theoretic postulates}~\cite{Hardy2001,DakicBrukner,MasanesMueller,ChiribellaDArianoPerinotti,Hardy2011,MasanesEtAl,BMU,Hoehn,HoehnWever,Wilce} mentioned at the beginning of this section. Similarly, there is progress in delineating the quantum correlations from the set of all non-signalling correlations in terms of simple principles~\cite{vanDam,BrassardCC,Pawlowski,NavascuesWunderlich,Cabello,Cabello2}. What we can learn from these reconstructions is that a few simple and intuitive constraints on \emph{encoding and processing of information} will automatically lead to (aspects of) the Hilbert space formalism of quantum theory. Perhaps these principles can be understood as unavoidable strategies of observers who try to place rational bets on future data records in computational worlds that have the counterintuitive properties described above~\cite{CCKM19}.

\section{Conclusions}
\label{SecConclusions}
\sectionmark{Conclusions}
In this work, I have argued that several puzzles and insights from modern physics and related fields motivate the exploration of a new type of ``first-person-first'' theories, and I have presented a blueprint of a simple theory of this kind. As I have emphasized in Section~\ref{SecPostulates}, the approach of this paper does \emph{not} yet give a full-fledged theory that has all the properties that one would like it to have; in particular, it is not yet able to treat processes of ``forgetting'' or ``memory erasure'', and Section~\ref{SecQuantum} suggests that such processes might be of fundamental importance.

Despite its incompleteness, it seems fair to say that the theory as presented here has shown a surprising variety of predictive and explanatory power: it explains in some sense ``why'' we see a simple, computable, probabilistic external world; it predicts the emergence of objective reality as an asymptotic statistical phenomenon; and it makes concrete predictions for exotic enigmas like the Boltzmann brain problem or the computer simulation of agents. We have also seen that it is consistent with quantum theory, and that some basic quantum phenomena (Bell nonlocality but no signalling) might be understood as consequences of its general framework. Moreover, it describes a very counterintuitive yet elegant and consistent ontology of the world in which observers are not ``objects in some universe'', but abstract structure. Due to this novel perspective, it predicts unforeseen phenomena like subjective immortality or ``probabilistic zombies'', a phenomenon that is so surprising (but consistent) that it has not even appeared in the science fiction literature yet.

Therefore, regardless of the question of whether ``the world is \emph{really} as crazy as that'', this approach expands our imagination and demonstrates that our usual ways of addressing puzzles in the foundations of physics have perhaps been more limited than we thought. My hope is that the results of this paper give us a glimpse on generic properties of \emph{all} theories ``of this kind'', even if particular details of the approach as presented in this paper will need revision.

A major revision in future work will be to generalize its definitions such that the fundamental postulates include memory erasure, formalizing Postulates~\ref{DesiredPostulates}. It seems likely that such an improvement necessitates a different definition of ``observer states'': these states will need to contain more structure than just being finite binary strings. However, what one should arguably \emph{not} do is to simply write down a seemingly realistic definition of observers that is motivated, say, by the contingent detailed features of human observers: the goal is not to make a \emph{model}, but to uncover the true fundamental \emph{mathematical nature} of ``what it means to be an observer''. Adding seemingly realistic bells and whistles to definitions is not the way to go in the ultimate foundational regime targeted by such a kind of approach.

The ``idealistic'' approach of this paper contrasts with our current way of doing science which reflects Cartesian dualism in a methodological sense: the empirical realm of physics and the first-person realm of, say, the philosophy of mind are treated as separate and, in many cases, irreconcilable regimes. This is not always a bad idea --- quite on the contrary. Banning the first-person perspective from physics was one of the major prerequisites for its success, and some attempts to unify both regimes are arguably overly speculative.

But keeping the two regimes separate may not be the best strategy under all circumstances. One such circumstance which is arguably becoming increasingly relevant is the development of computer technology, and with it the prospect that disturbing technologies like brain emulation might become available in the not so distant future. We need good theories that allow us to give precise answers to some urgent questions that arise in this context. In a complementary development, the first-person perspective has shown up in physics despite its initial banishment, manifesting itself in questions like: what should observers expect to see in a very ``large'' universe? How can we make sense of the fact that the notion of measurement seems to play a special role in quantum mechanics? These questions are hard to address since most conceivable answers cannot be easily tested empirically. But if we are careful and aim for mathematical rigor, simplicity of assumptions, and consistency with known physics, then we may hope to obtain some valuable insights that span both regimes. As I have argued in the introduction, having an approach of this kind will have the advantage that we can test its predictions in one regime (of physics), and thus increase our confidence in its predictions in the other regime (of the first-person perspective).

This work also helps to refute a common criticism which is faced by broadly epistemic approaches to physics (like QBism), namely the reproach of being ``solipsistic''. This is simply a fancy catchphrase subsuming the following objection: \emph{How can you deny the relevance of an objective external world, given that this notion is so obviously successful and important in physics?} What the approach of this paper shows is that one can successfully deny the \emph{fundamentality} of the notion of an objective external world, and obtain it as an emergent phenomenon from weaker assumptions. As shown above, this ``methodological solipsism'' even allows us to address questions that are otherwise difficult to address.

The results of this paper are an invitation to take a bolder perspective on some foundational questions. Relying strictly on our traditional view of the world might not be the right strategy under all circumstances; perhaps we are missing something truly important, as the conceptual questions of Table~\ref{fig_motivation} seem to suggest. Exploring alternatives, in a mathematically rigorous way that prevents us from fooling ourselves, may well yield surprising insights that are crucial for solving some important problems that lie ahead.

\section*{Acknowledgments}
This work would not have been possible without invaluable feedback and support from many friends and colleagues. I am grateful to Sebastian Guttenberg and Dierk Schleicher for encouragement in early stages of this work, to Lee Smolin for discussions on how this relates to other topics in fundamental physics, and to Sona Ghosh, Daniel Gottesman, Philippe Gu\'erin, Lucien Hardy, Philipp H\"ohn, Philippos Papayannopoulos, Renato Renner, L\'idia del Rio, Robert W.\ Spekkens, and Cozmin Ududec for helpful discussions and comments.

Im a particularly grateful for the inspiring discussions with the participants of the workshop on ``Algorithmic information, induction, and observers in physics'' at Perimeter Institute in April 2018. These discussions have contributed enormously to the continued improvement of this approach and this paper. In particular, I am indebted to Christopher Fuchs, R\"udiger Schack, and Tom Sterkenburg for helping me clear up some misguided ideas about probability --- you have helped me update my beliefs about updating beliefs! Special thanks go to Marcus Hutter for pointing out a mistake in an earlier formulation of Theorem~\ref{ThePersistence}.

I am indebted to my colleague Michael Cuffaro, from whom I have learnt an invaluable amount of philosophy of physics. The paper has benefited immensely from our discussions. I am deeply grateful to the Foundational Questions Institute (FQXi) for funding of the project ``Emergent objective reality --- from observers to physics via Solomonoff  induction'', which made it possible for me to work with Mike. I thank the Rotman Institute of Philosophy for funding via a Catalyst Grant. This research was undertaken, in part, thanks to funding from the Canada Research Chairs program.

Sincere thanks go to the Perimeter Institute for Theoretical Physics, which allowed me to pursue this idiosyncratic research during my times as a postdoc and associate faculty member. Research at Perimeter Institute is supported by the Government of Canada through the Department of Innovation, Science and Economic Development Canada and by the Province of Ontario through the Ministry of Research, Innovation and Science.

{
\footnotesize{\color{darkgray}Copyright of pictures: The two guinea pigs in Figures~\ref{FigObjectivity}, \ref{fig_subjimm} and~\ref{fig_bell} are from www.freepik.com (``Pet animals set'', described as ``free for commercial use with attribution'', accessed March 2015). The Messier 101 galaxy photograph in Figure~\ref{FigObjectivity} is due to the European Space Agency \& NASA; more information and credits to the original investigators can be found at \url{http://hubblesite.org/newscenter/newsdesk/archive/releases/2006/10/image/a}. The meteor (or comet?) in Figure~\ref{fig_subjimm} is from www.freepik.com (``Space Icons'', described as ``free for commercial use with attribution'', accessed March 2017). The watch in Figure~\ref{fig_bell}, made by Dinosoft, is from www.flaticon.com (accessed March 2017, usage according to Flaticon Basic License).}
}

\onecolumngrid

\pagebreak
\appendix
\section{Appendix}
\renewcommand \thesection{A}
\renewcommand{\thetheorem}{A.\arabic{theorem}}
\label{appendix}
\sectionmark{Appendix}
We begin with some comments on the relation between Postulates~\ref{Postulates} and~\ref{DesiredPostulates}, as announced in Section~\ref{SecPostulates}.

Even though Postulates~\ref{Postulates} do not satisfy all desiderata that one would ideally propose for this paper's approach, they are arguably a good starting point to obtain a glimpse on the predictions and constructions that are typical for \emph{theories of the kind} as envisaged above. The goal of this paper is to give the blueprint of a novel kind of approach, namely, one in which some notion of first-person perspective (``mind'' in some sense) is taken as fundamental, not a notion of third-person perspective (world). The aim is to construct one simple mathematically rigorous elaboration of this idea, and to demonstrate its potential in the context of well-known puzzles in the foundations of physics (broadly construed), \emph{not} to give the best or final version of such an approach. Indeed, simplified Postulates~\ref{Postulates} already admit a wealth of interesting insights. In particular, they suggest one possible way in which we can understand the appearance of an emergent notion of external world, despite the absence of any primitive notion of ``world'' in the postulates of the theory.

It is natural to expect that the simplified approach gives valid approximations to theories following Postulates~\ref{DesiredPostulates} whenever information loss can be neglected. This expectation is also confirmed by analogy: the simplification inherent in Postulates~\ref{Postulates} is arguably reminiscent of a simplification typically made in the context of Bayesian statistics~\cite{BernardoSmith}. Discussions of Bayesian reasoning usually start with the simplest instance of Bayes' theorem: an agent holds a prior distribution, learns the value of some random variables, and updates her beliefs accordingly, obtaining a posterior distribution. This is known to be a fruitful first step towards a more general theory of Bayesian learning or reasoning, despite the fact that some realistic situations do not seem to precisely fall into this scheme.

One example (among many) is given by the ``problem of old evidence'': the scheme just described is not able to suitably account for updating beliefs in the light of learning unit probability facts~\cite{Glymour,EvaHartmann}. Moreover, this scheme assumes that the agent holds consistent real-valued beliefs about \emph{all} relevant random variables, and updates these beliefs successively \emph{without ever forgetting any learned evidence}, similarly as in Postulates~\ref{Postulates}. This is yet another assumption that will not be satisfied in all relevant situations. The methodological strategy in Bayesian reasoning to answer these drawbacks is to study the simple scheme first, and to return to these puzzles separately later. The simplification inherent in Postulates~\ref{Postulates} should be understood as implementing a similar kind of strategy.\\

The following example has been announced in Section~\ref{SecSimpleLaws}.
\sexample{The quantum Turing machine as a computational ontological model}{ExQTM}{$\strut$\newline
Consider the quantum Turing machine (QTM) as defined by Bernstein and Vazirani~\cite{BernsteinVazirani}, and analyzed in the context of algorithmic information theory in Ref.~\cite{MuellerIEEE}. Similarly as for a classical monotone Turing machine computation as considered in Section~\ref{SecSimpleLaws}, we can consider QTMs with an input tape and an output tape, together with one or more work tapes. While these tapes can carry quantum states, i.e.\ are described by Hilbert spaces, they are also assumed to have a distinguished \emph{computational basis}. Let us consider QTM calculations of the following kind.
\begin{itemize}
	\item The input tape starts with a finite number of classical bits -- the ``program'', encoded into computational basis states. It is followed by an infinite sequence of maximally mixed qubit states.
	\item The output tape starts with an infinite sequence of blank symbols $\#$. That is, input tape cells are qubits, and output tape cells are qutrits, with computational basis states $\{0,1,\#\}$.
	\item Both input and output tapes are \emph{unidirectional}: with non-zero amplitude, their heads can only remain where they are or move one step to the right. (See~\cite{BernsteinVazirani,MuellerIEEE} for how the coherent transition function is defined for a QTM.) No such restriction is assumed for the work tapes.
	\item After every step of computation, the output tape is projectively measured in the computational basis, and the machine's state is updated according to the L\"uder's rule (``decohered'').
\end{itemize}
This describes a computational ontological model: the state space is $\Omega=\Omega_Q\times\Omega_O$, where $\Omega_O$ is the set of finite binary strings on the output tape, while $\Omega_Q$ is the set of possible mixed quantum states of the rest of the QTM (that is, its input and work tapes and heads). In more detail, we define $\Omega_Q$ as the set of all such mixed states that can appear on any finite classical input, for any finite number of computational steps $t\in\N$, conditioned on any sequence of measured output strings up to time $t$. In particular, $\Omega_Q$ is a countable set. Finally, we define $f_A:\Omega\to\s$ as the function that reads the output.
}
This defines a stochastic process which satisfies all premises of Definition~\ref{DefOntologicalModel}. It models an observer that is embedded into a private quantum world: the world evolves unitarily, except for decoherence introduced by the observer. Since QTM transition amplitudes are assumed to be efficiently computable, this implies that the process is computable, i.e.\ can in principle be simulated by a monotone Turing machine that generates the same output distributions on identical classical inputs. If the program is chosen appropriately, then the machine will never halt and make the output string grow indefinitely with unit probability. Then, the distribution of the output (of the random variable $f_A$) is a \emph{measure} $\mu$ in the sense of Definition~\ref{DefSemimeasures}.\\

The following criterion for no-signalling is used in Section~\ref{SecQuantum}:
\emptyslemma{LemNoSignalling}{
Consider a Bell experiment involving two parties, Alice and Bob, choosing among settings $a$ and $b$ and obtaining outcomes $x\in X$ and $y\in Y$ respectively. Furthermore, suppose that Bob's detector is perfect, but Alice's detector sometimes refuses to fire, which we formalize by an additional non-detection outcome `$\emptyset$'. Set $\bar X:=X\cup \{\emptyset\}$.

Suppose that there is an actual non-signalling correlation $P_0(\bar x,y|a,b)$ that determines the probabilities of all outcomes $\bar x\in \bar X$, $y\in Y$, and that $P_0(\emptyset|a)<1$ for all $a$. Define the postselected correlation $P$ by~\cite{Branciard}
\[
   P(x,y|a,b):=\frac{P_0(x,y|a,b)}{1-P_0(\emptyset|a)}\qquad\left( x\in X,y\in Y\right).
\]
Then $P$ is non-signalling from Bob to Alice, i.e.\ the conditional probability distribution $P(x|a)$ is automatically well-defined. However, $P$ may in general be signalling from Alice to Bob. Yet, if the condition
\begin{equation}
   P_0(\emptyset,y|a,b)=P_0(\emptyset|a)\cdot P_0(y|b)
   \label{eqNS1}
\end{equation}
is satisfied, then $P$ is also non-signalling from Alice to Bob. This condition can also be written
\begin{equation}
   \emptyset\ci y|a,b,
   \label{eqNS2}
\end{equation}
i.e.\ it states that Alice's non-detection event and Bob's outcome are to be conditionally independent, given the settings.
}
A comment on the notation: the event ``$\emptyset$'' can be understood as a binary random variable which takes the value ``yes'' if $\bar x=\emptyset$ and ``no'' if $\bar x \neq \emptyset$. Then~(\ref{eqNS2}) is just an ordinary conditional independence relation between random variables. In particular, it is sufficient to check factorization for the ``yes''-outcome due to the following implication for arbitrary discrete random variables $X,Y,Z$ which is straightforward to verify:
\[
   P(X=x_0,Y|Z)=P(X=x_0|Z)\cdot P(Y|Z)\quad\Rightarrow\quad
   P(X\neq x_0,Y|Z)=P(X\neq x_0|Z)\cdot P(Y|Z).
\]
\proof
Using that $P_0$ is non-signalling, we get
\[
   \sum_{y\in Y}P(x,y|a,b)=\frac{\sum_{y\in Y} P_0(x,y|a,b)}{1-P_0(\emptyset|a)}=\frac{\sum_{y\in Y} P_0(x,y|a,b')}{1-P_0(\emptyset|a)}=\sum_{y\in Y}P(x,y|a,b')\qquad\mbox{for all }x\in X,a,b,b',
\]
that is, $P$ is non-signalling from Bob to Alice. We also have
\[
   \sum_{x\in X}P(x,y|a,b)=\frac{\sum_{\bar x \in\bar X}P_0(\bar x,y|a,b)-P_0(\emptyset,y|a,b)}{1-P_0(\emptyset|a)}
   =\frac{P_0(y|b)-P_0(\emptyset,y|a,b)}{1-P_0(\emptyset|a)},
\]
and if we assume eq.~(\ref{eqNS1}), then this simplifies to $P_0(y|b)$, which is manifestly independent of $a$ for all $y\in Y,b$. No-signalling of $P_0$ also implies that $P_0(\emptyset|a)=P_0(\emptyset|a,b)$ and $P_0(y|b)=P_0(y|a,b)$, such that~(\ref{eqNS1}) is equivalent to~(\ref{eqNS2}).
\qed

\end{document}